\documentclass[3p]{elsarticle}
\makeatletter
\def\ps@pprintTitle{%
 \let\@oddhead\@empty
 \let\@evenhead\@empty
 \def\@oddfoot{\centerline{\thepage}}%
 \let\@evenfoot\@oddfoot}
\makeatother
\usepackage{lineno,hyperref}
\usepackage{amsmath}
\usepackage{float} 
\usepackage{esvect} 
\usepackage{color,soul} 
\usepackage{subcaption} 
\usepackage{color} 
\usepackage[]{footmisc}

\modulolinenumbers[5]
\biboptions{numbers,sort&compress}

\begin{document}

\begin{frontmatter}

\title{A multi-scale model for electrokinetic transport in networks of micro-scale and nano-scale pores}

\author{Shima Alizadeh\textsuperscript{1,2}}
\author{Ali Mani \textsuperscript{1,2,}\footnote[1]{To whom correspondence should be addressed.    Email: alimani@stanford.edu (Ali Mani)}}
\address{1 Department of Mechanical Engineering, Stanford University, 
Stanford, California 94305, USA}
\address{2 Center for Turbulence Research, Stanford University, 
Stanford, California 94305, USA}

\begin{abstract}
We present an efficient and robust numerical model for simulation of electrokinetic phenomena in porous networks over a wide range of applications including energy conversion, desalination, and lab-on-a-chip systems. Coupling between fluid flow and ion transport in these networks is governed by the Poisson-Nernst-Planck-Stokes equations. These equations describe a wide range of transport phenomena that can interact in complex and highly nonlinear ways in networks involving multiple pores with variable properties. Capturing these phenomena by direct simulation of the governing equations in multiple dimensions is prohibitively expensive. We present here a reduced order computational model that treats a network of many pores via solutions to 1D equations. Assuming that each pore in the network is long and thin, we derive a 1D model describing the transport in pore's longitudinal direction. We take into account the non-uniformity of potential and ion concentration profiles across the pore cross-section in the form of area-averaged coefficients in different flux terms representing fluid flow, electric current, and ion fluxes. Distinct advantages of the present framework include: a fully conservative discretization, fully bounded tabulated area-averaged coefficients without any singularity in the limit of infinitely thick electric double layers (EDLs), a flux discretization that exactly preserves equilibrium conditions, and extension to general network of pores with multiple intersections. By considering a hierarchy of canonical problems with increasing complexity, we demonstrate that the developed framework can capture a wide range of phenomena. Example demonstrations include, prediction of osmotic pressure built up in thin pores subject to concentration gradient, propagation of deionization shocks and induced recirculations for intersecting pores with varying properties. 
\end{abstract}


\end{frontmatter}

\section{Introduction}

\begin{figure}[H]
        \centering
        \begin{subfigure}[H]{0.38\textwidth}
                \includegraphics[width=\textwidth]{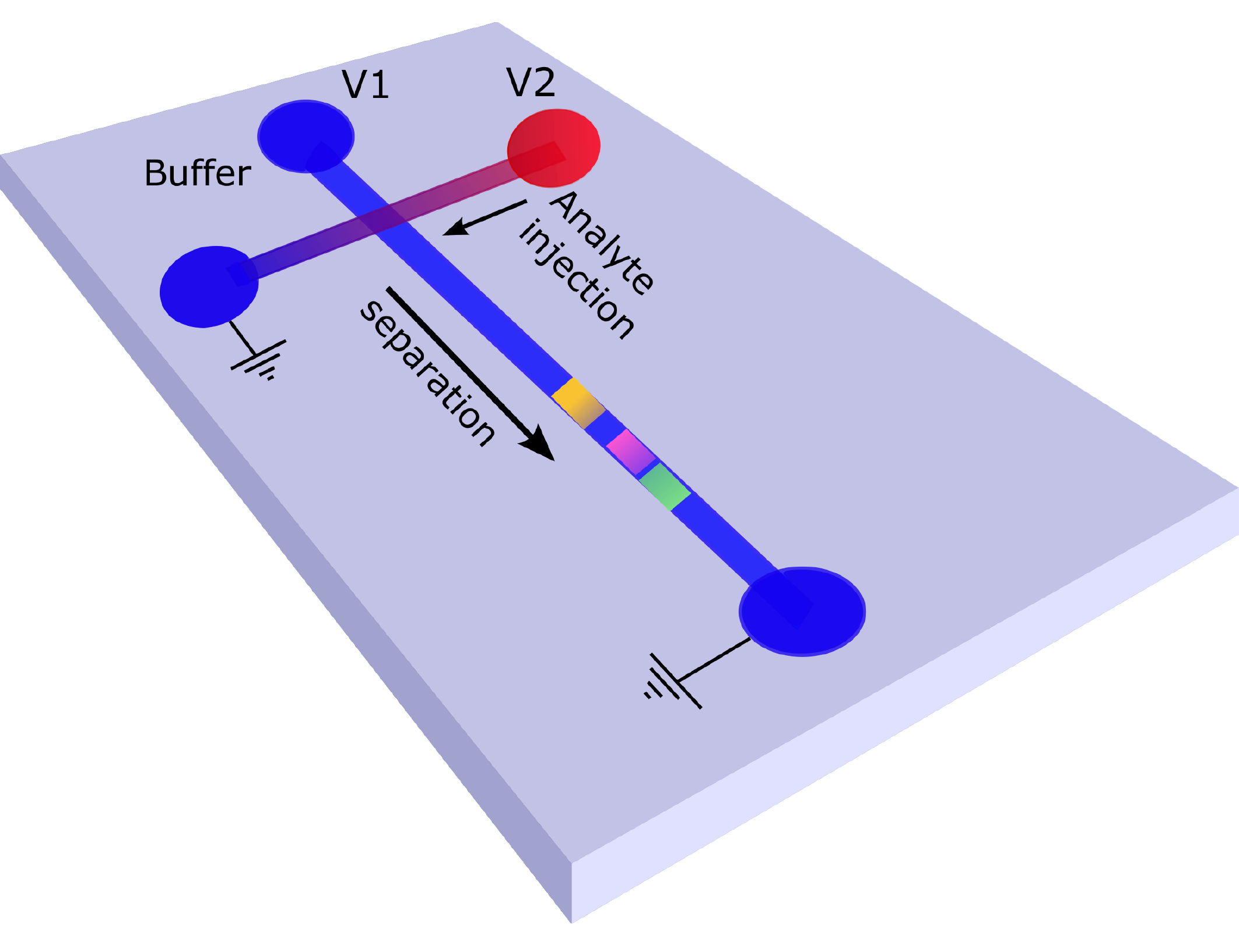}
                \caption{}
                \label{fig:chip}
        \end{subfigure}%
        ~ 
        \begin{subfigure}[H]{0.5\textwidth}
                \includegraphics[width=\textwidth]{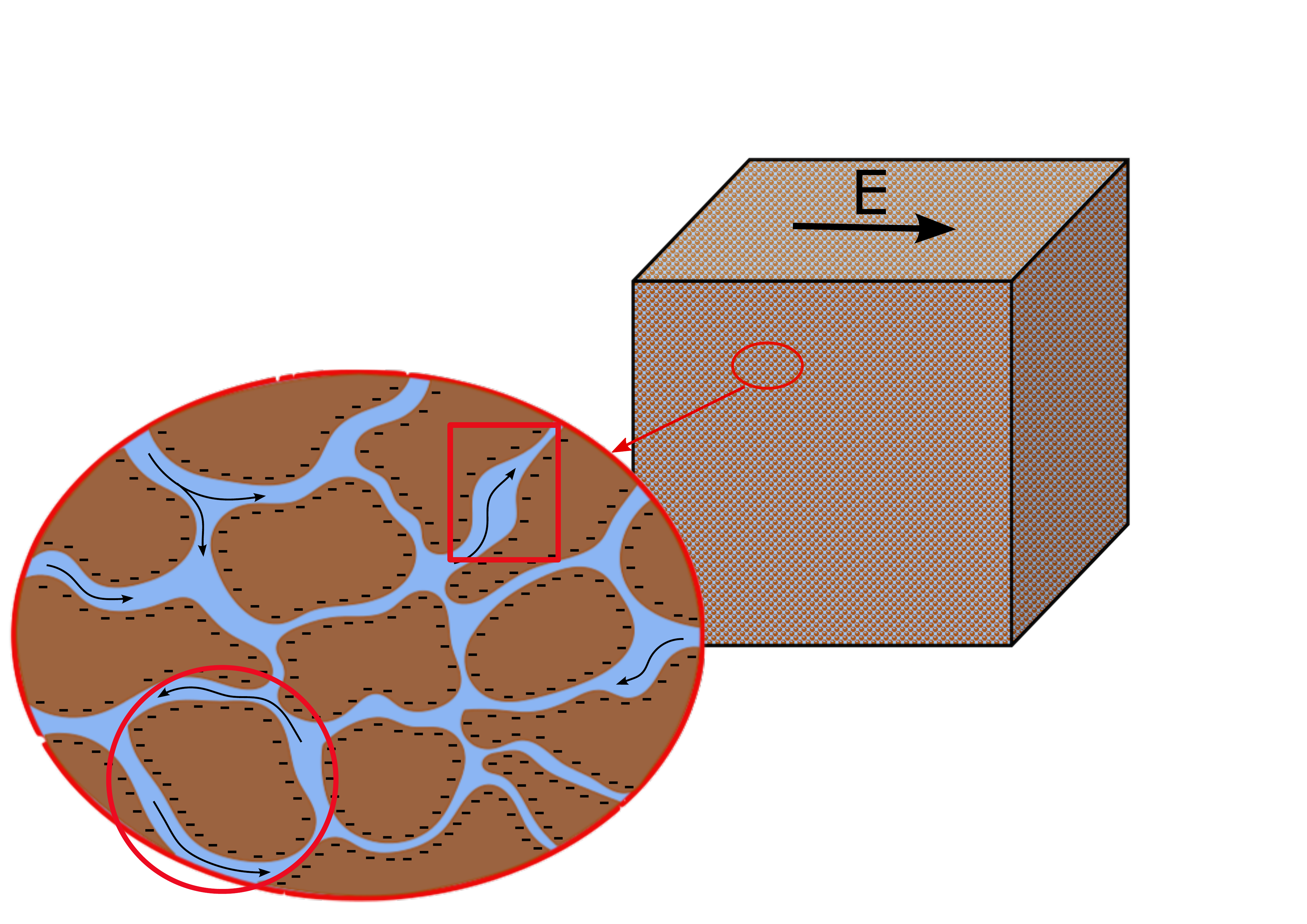}
                \caption{}
                \label{fig:porous}
        \end{subfigure}
        ~ 
          \begin{subfigure}[H]{0.27\textwidth}
                \includegraphics[width=\textwidth]{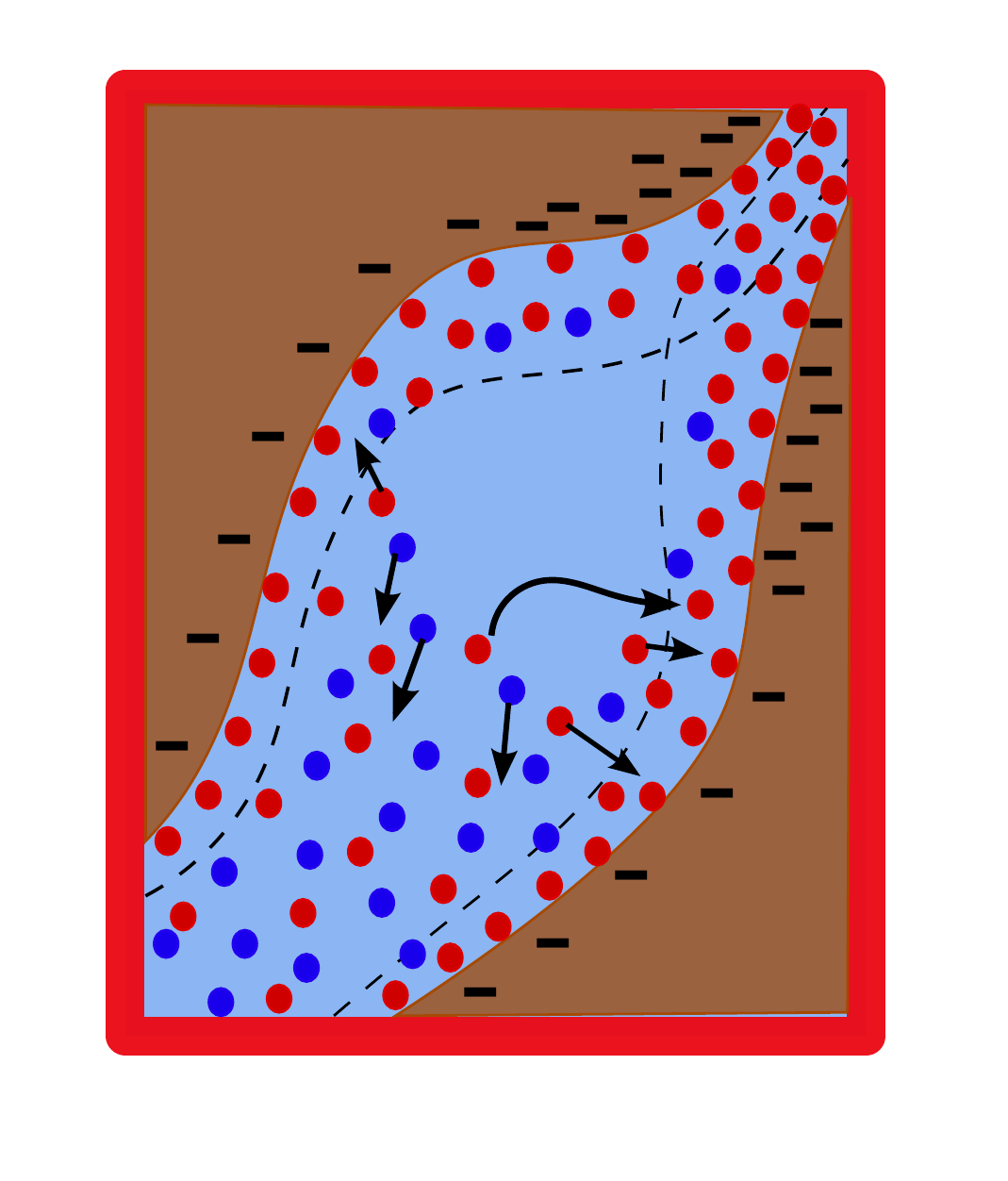}
                \caption{}
                \label{fig:icp_shock}
        \end{subfigure}
        \begin{subfigure}[H]{0.3\textwidth}
                \includegraphics[width=\textwidth]{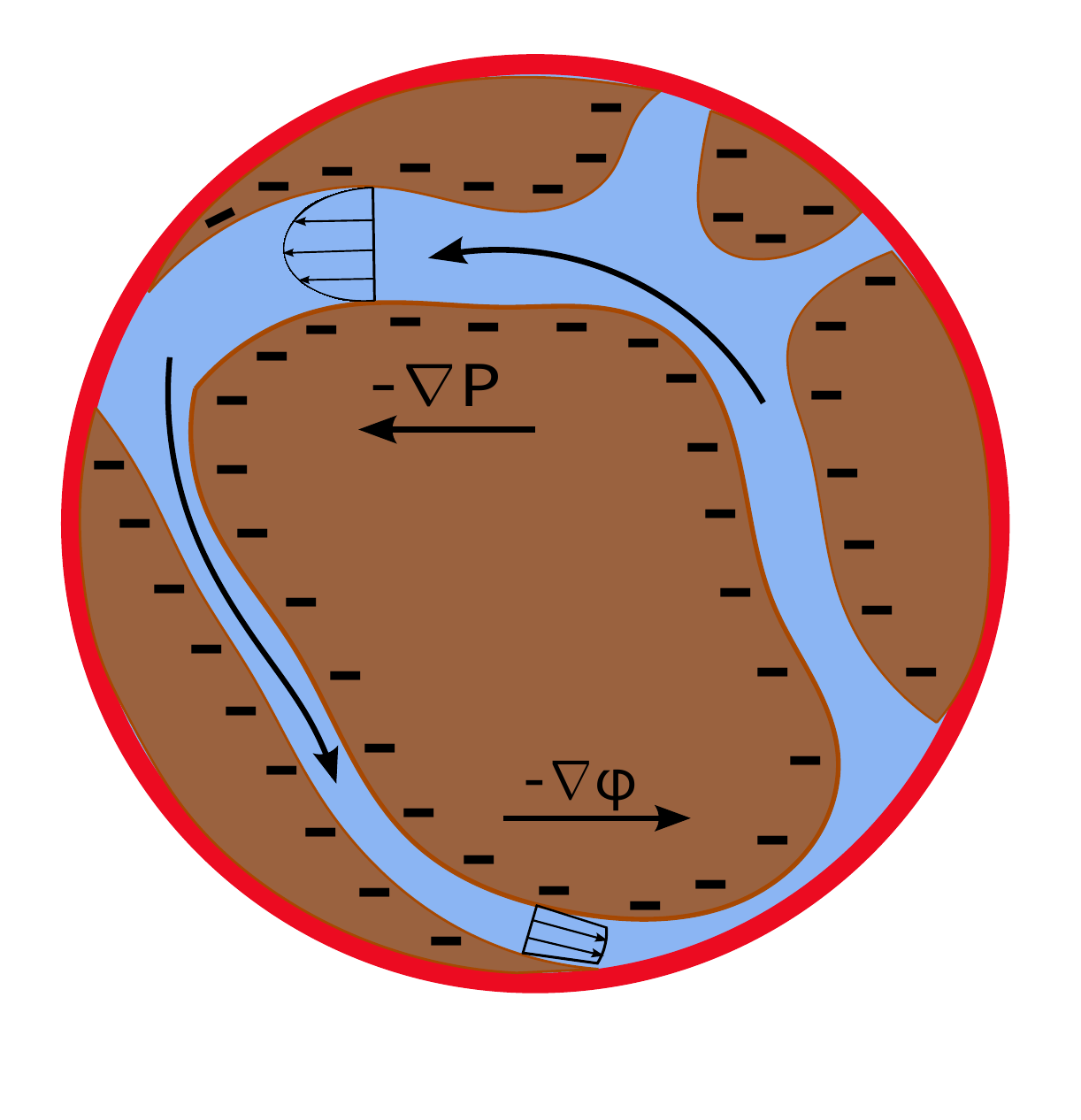}
                \caption{}
                \label{fig:recirculation}
        \end{subfigure}
         \caption{(a) Schematic of a lab-on-a-chip system operating based on electrokentic injection of analyte into a buffer. (b) Schematic of electrokinetic flows in a porous material which consists of many micro-scale and nano-scale pores exposed to an external electric field (E). (c) Schematic of the propagation of a deionization shock initially generated at the interface of two pores in series. (d) Internal recirculation in parallel pores. A backward pressure driven flow dominates in the micro-pore with lower hydraulic resistance, while the electro-osmotic flow becomes dominant in the nano-pore, where the surface effect is more significant.}
                \label{fig:pore_networks}
 \end{figure}

Electrokinetic phenomena in microstructures have implications in different technological applications and have been attracting significant attention. These microstructures can be either fabricated with specified geometry, such as in microfluidics (Figure (\ref{fig:chip})), or involve random porous media (Figure (\ref{fig:porous})) made of either conducting or non-conducting materials. In this paper we collectively refer to these systems as \textit{network of pores}.   Microfluidic lab-on-a-chip devices are widely used to analyze biological entities and perform processes including mixing, separation \cite{hibara, penn1, griggiths, marshall2011}, and biomolecular detection \cite{han, tegenfedt, penn3}. On a larger scale, electrokinetics in porous media is shown to have practical relevance for a wide variety of applications from filtration, desalination \cite{ko2012}, and fluid pumping \cite{suss2011}, to energy storage \cite{biesheuvel2011, mathias2012, kharkats, linden, tanner, virkar, elmekawy} and highly efficienct energy conversion \cite{dirk2012,hoffmann2013}. A number of these applications involve transport nonlinearities induced when pores with different geometrical and electro-chemical properties are connected \cite{wang2005, holtzel2007, ramirez2006}. Below we describe aspects of physical complexities associated with a network of pores.\\

Ion Concentration Polarization (ICP) is a well known phenomenon \cite{levich1947} that is described by induced concentration gradient in systems where ion selectivity varies in the direction of transport. In the context of porous networks, ICP can occur when pores of varying cross sections are connected in series  \cite{park2006, plecis2005, schoch2005, schoch2006, mishchuk1995, dukhin1969, ehlert2007, mani2009, manimartin2011} or T-junctions \cite{probstein2003, horno1989, rubinstein2000, pu2004} as well as in pore-membrane junctions \cite{holtzel2007, ehlert2007, kwak2011, kim2010, kim2012, deng2013, dydek2011}. Wang et al. \cite{wang2005} reported the use of ICP to generate high-focusing ratios of an ionic protein near a microchannel-nanochannel junction. By examining different systems of pores, Kim et al. \cite{kim2009,kim2010} have demonstrated that ICP dynamics can be exploited for a variety of applications including energy efficient desalination systems and biomolecule preconcentration devices.\\

Mani, Zangle, and Santiago  \cite{mani2009, zangle2009, zangle2010} performed an extensive numerical-experimental study on the propagation of ICP enrichment and depletion zones at galvanostatic (constant current) and potentiostatic (constant potential) conditions. They showed that under certain circumstances, ICP enrichment and depletion regions could spread as shocks (Figure (\ref{fig:icp_shock})), which are driven by both electromigration and diffusion. Mani and Bazant \cite{manimartin2011} referred to the propagation of depletion zone as deionization shock, and explained in detail the underlying physics of the shock and its stability in pores with slow variation of surface charge and geometry. They illustrated how the nonlinear dynamics of deionization shock result from exchanges between bulk and surface conduction carried by the excess charge in the EDLs. Other related studies have analyzed the overlimiting current through porous material subject to ICP \cite{deng2013}. It has been demonstrated that in these systems, a novel mechanism for overlimiting current exists due to surface conduction which is activated after deionization shocks propagate through pores\cite{nielsen2014,dydek2011,dydek2013}. However, coupling between surface conduction and deionization shocks in networks of many pores is yet to be understood.\\

Additional complexity arises when pores are connected in parallel in a network. As we will demonstrate, such connections can lead to flow loops and enhanced mixing when pores with different properties are connected to each others (Figure (\ref{fig:recirculation})).\\

All of the aforementioned effects occur on the scale of pore length or intersection distance in a network, and therefore, traditional homogenization models fail to capture such phenomena. On the other hand, full numerical simulation of electrokinetic transport in three-dimensional porous media is prohibitively expensive and infeasible for large networks of pores. One technique to overcome the present barriers is the utilization of area-integrated models, which are developed based on the principle of local thermodynamic equilibrium in the pore cross-section dimension. \cite{mani2009, nielsen2014, Peters2016, yaroshchuk2010, yaroshchuk2011, harvie2012, biscombe_circuit2012, biscombe2014, biscombe2012}. This approach reduces the transport equations to a one-dimensional model for each pore, which relates the cross-section averaged fluxes including fluid velocity $\bar{u}$, electric current density $\bar{i}$, and ion fluxes $\bar{j}^{\pm}$ to the local driving forces with prefactors. The prefactors are area-averaged quantities that take into account the non-uniformity of the concentration profile across EDLs. For long and thin pores, the driving forces are proportional to gradients of driving potential functions describing a combination of pressure, concentration, and electrostatic fields. As we will see, these driving potentials can be constructed such that they only depend on longitudinal coordinates in each pore while remaining uniform in cross-sectional planes. Such potential functions, which are most suitable for model reduction, are also referred to as virtual quantities in the literature \cite{Peters2016, yaroshchuk2010}. The driving forces and their potential functions can be expressed in different ways. In our modeling, we introduce three forces in the form of gradients in virtual total pressure ($P_0$), virtual electro-chemical potential for counterions (in our case, $\mu^+$  given that there is negative surface charge), and virtual concentration ($C_0$). Later in this section, we will illustrate the reason of this choice by declaring the key advantages of these driving potentials over the other forms represented in the literature. Detailed derivation of each force will be discussed in Sections \ref{sec:c0} and \ref{sec:p0mu}. As shown in Figure (\ref{fig:onsager}), we sum the effects of three driving forces for each cross-section averaged flux term and write a set of quasi-linear equations as follows:           
\begin{figure}[h]
\makebox[\textwidth][c]{\includegraphics[width=0.86\textwidth]{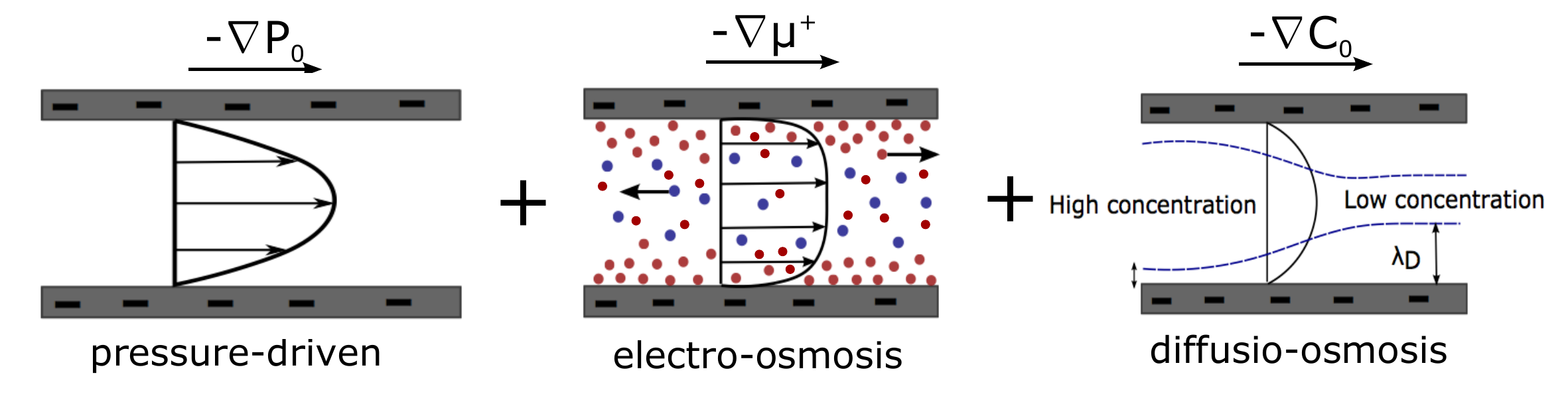}}
\caption{Linear combination of three mechanisms driving fluid flow based on the gradients in pressure, electro-chemical potential of counterions, and concentration. In addition to fluid flow, the same mechanisms can lead to the generation of ion and charge transport.}
\label{fig:onsager}
\end{figure}

\begin{equation}
\bar{u} \quad = \quad \bar{g}_{11} \nabla P_0 \quad + \quad \bar{g}_{12}\nabla \mu^+ \quad + \quad \bar{g}_{13}\nabla C_0,
\label{eq:1u}
\end{equation}
\begin{equation}
\bar{i} \quad = \quad \bar{g}_{21}\nabla P_0 \quad + \quad \bar{g}_{22}\nabla \mu^+ \quad + \quad \bar{g}_{23}\nabla C_0,
\label{eq:1i}
\end{equation}
\begin{equation}
\bar{j}^- \quad = \quad \bar{g}_{31}\nabla P_0 \quad + \quad \bar{g}_{32}\nabla \mu^+ \quad + \quad \bar{g}_{33}\nabla C_0.
\label{eq:1j}
\end{equation}

The pre-factors $\bar{g}_{ij}$ are area-averaged coefficients, which can be computed from the solution of the Poisson-Boltzmann equation across each pore cross-section. As will be discussed in Section \ref{sec:area integrated transport},  in our modeling we assume that for each pore the area-averaged charge is equal to that on the solid walls (total cross-section is electroneutral). This allows tracking of only one ion in a binary electrolyte system; the area-averaged concentration of the other ion can be computed from the electroneutrality condition. We can rewrite equations (\ref{eq:1u})-(\ref{eq:1j}) in a matrix form and develop a right hand side coefficient matrix as follows:
\begin{equation}
\left[ \begin{array}{c} \bar{u}  \\ \bar{i} \\ \bar{j}^- \end{array} \right] =
\begin{bmatrix} \bar{g}_{11} & \bar{g}_{12} & \bar{g}_{13}\\
			\bar{g}_{21} & \bar{g}_{22} & \bar{g}_{23}\\
			\bar{g}_{31} & \bar{g}_{32} & \bar{g}_{33}\end{bmatrix}
\left[ \begin{array}{c} \nabla P_0 \\ \nabla \mu^+ \\ \nabla C_0 \end{array} \right]. \label{eq:onsager}
\end{equation}

For the sets of driving potentials and fluxes employed here, the coefficient matrix, $\bar{g}_{ij}$ satisfies the Onsager symmetry, a property derived based on the assumption of the reversibility of microscopic equations of motions \cite{onsager1931a, onsager1931b, monroe2009}. The property of linear response of the fluxes to the gradient of driving potentials, which will be derived in detail below, has been utilized and discussed widely in electro-chemistry literature \cite{dirk2012, hoffmann2013, gross1968, mazur1951, overbeek1953, ajdari2001, brunet2004, bahga2010, behrens2001, mathias2011, jensen2011}. One should note that equation (\ref{eq:onsager}) is only valid for local gradients at each axial position along the pore. In general it does not hold for large distances due to the nonlinear response of the system given $\bar{g}_{ij}$ is a function of dynamically evolving fields with time. For each axial position, the determinant of the corresponding matrix is always positive. In other words, the coefficient matrix is positive definite, which verifies that the developed model satisfies the second law of thermodynamics \cite{Peters2016} stating that the global entropy increases during an irreversible process.\\

Despite the useful insights obtained from the previous reduced order models which have similar forms to equation (\ref{eq:onsager}), they suffer from several limitations, which make them unsuitable for computational implementation and analysis of complex pore networks. Many of these models have been developed at steady state \cite{yaroshchuk2011,biscombe2012,nielsen2014,manimartin2011}, or in the limit of thin EDL \cite{manimartin2011,nielsen2014}. Many are geometry dependent and are applicable only to simple network topologies \cite{mani2009, manimartin2011, nielsen2014, harvie2012, biscombe_circuit2012}. In this work, we have derived a new reduced order transport model accompanied with a novel computational discretization that offers several advantages for computational modeling of porous networks:
\begin{enumerate}
\item The developed reduced order model covers the entire range of EDL to pore size ratios. Large pores and high concentrations lead to thin EDL limit, and small pores and low concentration result in highly overlapped EDLs. Furthermore, the developed model includes an asymptotic treatment for the limit of zero concentration (infinite EDL thickness). This entire regime is relevant to electrokinetic transport in porous systems since deionization shocks often result in a wide range of variations in ion concentration. 
\item The matrix coefficients in the developed model are bounded over the entire range of concentrations and pore sizes. 
\item The formulation setup in the developed model uses driving potentials that remain bounded in the limit of very small (and large) concentration. 
\item The discretization scheme is fully conservative without any numerical leakage of ion concentrations. We should note that the variables that form the driving potentials (e.g. virtual concentration) are not conservative by nature. However other field variables, such as area-averaged concentration, are physically conserved. To maintain all of the advantages described above, our formulation uses evolution equations for conservative fields as primary unknowns; the driving potentials are closed in terms of conservative quantities through solutions to the Poisson-Boltzmann equation. 
\item Our formulation is discretely well balanced. In other words, in the absence of any driving force, equilibrium solution is recovered regardless of spatial mesh spacing. This property has been achieved by utilizing virtual quantities as driving potentials since they are invariants of thermodynamic equilibrium states. 
\item The derived methodology is capable of handling discrete jumps in pore size, or multi-pore intersections. Handling such configurations is challenging since pressure, concentration, and electrostatic potential can discretely jump across such interfaces. However, the virtual potentials remain continuous, and this property has been utilized in the present model.  
\end{enumerate}

The outline of the paper can be summarized as follows: we begin by describing the model problem and the governing equations. After nondimensionalizing the equations, we introduce main assumptions of the model and motivate deriving reduced order models for thin and long pores. Then, we present the derivation of model reduction in which the thermodynamic equilibrium is assumed in thin directions to pre-specify the field profiles in cross-sectional directions. We then introduce the 1D model governing the axial transport of ions at each pore and provide a comprehensive discussion of different flux terms. We explain the coupling of solutions for different pores by proper application of conservation laws at pore intersections. In the result section, we first present the results obtained by using tabulated coefficients to investigate the influence of pore size and surface charge on the osmotic phenomenon in a single pore. We then demonstrate our model validation by comparing our results with those from direct numerical simulations of two pore systems: (1) a single micropore confined by a cation-selective membrane, and (2) a microchannel-nanochannel-microchannel setting exposed to an external pressure gradient. We also discuss the temporal and spatial evolution of electrokinetic behaviors emerging in networks of micro-pores and nano-pores in series, parallel, and H-junction configurations.

\section{Model problem}
\begin{figure}[h]
\makebox[\textwidth][c]{\includegraphics[width=0.7\textwidth]{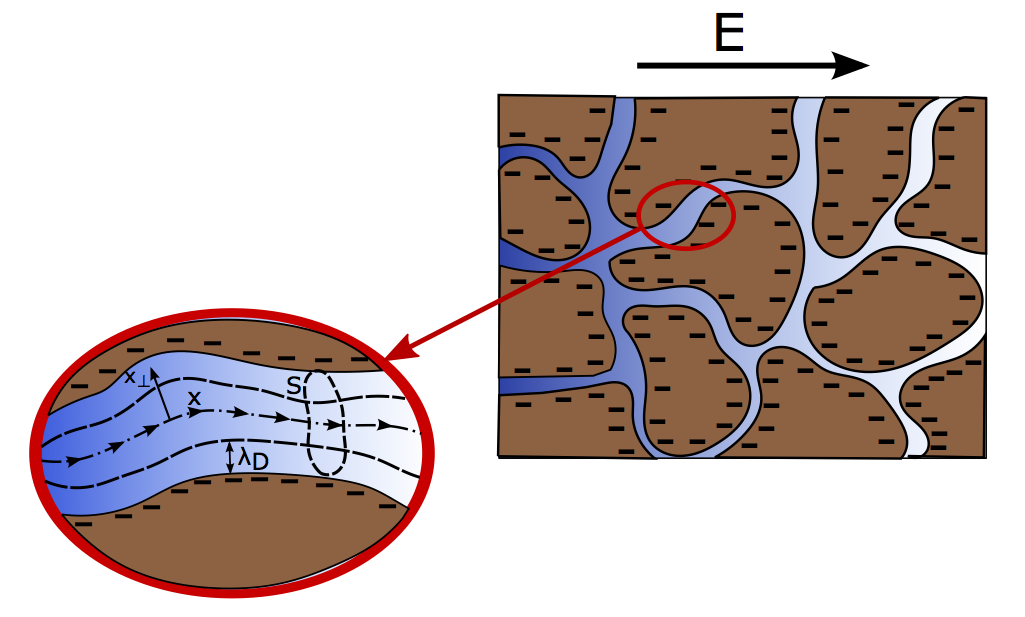}}
\caption{Sketch of an electrokinetic porous structure exposed to an external electric field. The coordinate is set along the pore axis and in the direction normal to the axis. The Debye length has been shown next to the charged surfaces, where the counterions are accumulated to screen the surface charge.}
\label{fig:pore_coord}
\end{figure}

We model an electrokinetic porous structure as a network of micro-scale and nano-scale pores with small aspect ratio; that is they are assumed to be long and thin pores. Figure (\ref{fig:pore_coord}) shows an example of a porous network. For each pore, the x-axis is assigned in the axial direction. Each pore can have a general area cross-section ($S$), and the area and the surface charge density ($\sigma$) can vary smoothly in the pore longitudinal direction. For simplicity, we assume that the surface charge density is fixed in time and the pore is filled with a symmetric binary electrolyte with diffusivity D and valence z. However, extension of the model to time-variable surface charge, such as in conducting material or in systems with surface reaction, and inclusion of multi-species is straightforward. As will be explained in detail, pores may lead to terminal reservoirs, where the salt concentration is known, or be connected together via intersections that are modeled as ``internal reservoirs'' whose concentration is to be dynamically determined. The pore surface charge is screened by counterions through the electric double layer (EDL), denoted by $\lambda_D$ in Figure (\ref{fig:pore_coord}). 

\begin{table}[h]
  \caption{Nomenclature}
  \label{tbl:variables}
  \resizebox{\textwidth}{!}{\begin{tabular}{llll}
    \hline
    dimensionless quantity                   & description                                                        & dimensional parameter  & description\\
    \hline
    $C$                              & concentration of ion species             &     $V_T$                   & thermal voltage \\
    $C_0$                          & virtual concentration                         &	$L$                       & axial length scale\\
     P                                 & hydrodynamic pressure			&  	$R$         & pore diameter\\
    $P_0$                          & pressure (hydrodynamic plus osmosis)     & $\nabla$              & gradient operator\\
    $\phi$                          & electric potential 			&    $D$                            & diffusivity coefficient\\
    $\psi$                          & surface induced electric potential 	&    $k_B$                         & Boltzmann constant\\
    $\phi_0$                      & virtual electric potential    		        &     $T$                 & absolute temperature \\
    $\mu^+$                      & electro-chemical potential                 &    $z$                       & atomic valence \\
    $\textbf{U}$                 & velocity vector                                   &    $e$                       & elementary charge\\ 
    $u$                              & axial velocity component                  &   $\varepsilon$             & electric permittivity\\
    $\textbf{U}_{\perp}$    & velocity components in the wall normal direction                                            &   $t_{\text{diff}}$                & diffusion time scale\\
    $g^p$                         &  the pre-factor of pressure driven flow                     &   $\sigma$                     & surface charge density \\
     $g^e$                         & the pre-factor of electro-osmotic flow                    &  $\lambda_{\text{ref}}$             & reference Debye length \\
    $g^c$                         & the pre-factor of diffusion-osmotic flow       &    $h_p$                          & local effective pore size\\
    $\lambda_D$             & Debye length normalized by $h_p$	             &    $C_{\text{ref}}$     & reference concentration\\		
      $\lambda_0$             & normalized Debye length w.r.t. local  $C_0$ &  $u_{\text{diff}}$    &   diffusion velocity scale\\
      $\lambda^*$              &  normalized Debye length w.r.t. local $\overline{C^-}$  &   &  \\
      $a$                        & pore aspect ratio   &    &  \\
      $\kappa$                                   & electro-hydrodynamic coupling constant  &  & \\
      $S$                            & pore cross sectional area   &   &\\ 
      $S_r$                      & surface area of internal reservoir   &          &\\
      $V_r$        & the volume of internal reservoir    &   & \\  
      $\rho$                         & volumetric net charge density    &    &  \\  
      $\sigma^*$    &   normalized surface charge density   &   & \\                  
      $C_s = -2 \lambda_D^2 \sigma^*$            & excess concentration of counterions in     &  	&    \\
      &                 the EDL to screen local surface charge                   				  &    &    \\
      $A_2, B_2, C_2$                   & pre-factors of gradients in conservation of current  &   &\\
      $A_1, B_1, C_1$                   &pre-factors of gradients in conservation of fluid flow  &    &\\ 
      $F_x$                         & ion flux along the pore axis &     & \\ 
      $F_{\perp}$                & ion flux component in normal direction to pore axis   &     & \\
      $I$                   & electric current  &   &\\
      $Q$                   & background solvent flow rate   &     &  \\
      $\bar{g}$, $\overline{g^p}$, $\overline{g^e}$, $\overline{g^c}$, $\overline{g^{p-}}$, $\overline{g^{e-}}$, $\overline{g^{c-}}$ ,      &   area averaged coefficients   &   &  \\
        $\overline{g^{e+}}$, $\overline{g^{pe}}$, and $\overline{g^{pc}}$, $\bar{f}$ &     &    &  \\
       $\pi_i$     &     pressure boundary value for a reservoir    &    &  \\
        $\theta_i$                     & electro-chemical potential boundary value for a reservoir    &   &  \\                                                                             
  \hline
    subscript                   & description                                                        & subscript               & description\\
    \hline
    L                                  &  left reservoir				                         & R			 	&  right reservoir\\
    r                                &  reservoir quantity					        & diff			        &  diffusion-related quantity\\
   $\perp$			      & in the wall normal direction				& ref	                         & reference quantity\\
    \hline
    superscript                   & description                                                        & superscript               & description\\
    \hline
    $\sim$                          &  dimensional field variable				        &  c   &    diffusio-osmotic variable \\
     ---                                &  area averaged quantity					& $(1)$,$(2)$,$(3)$	& solution of conservation equations for different boundary values\\
    $'$			       &  perturbation from mean value				&  $\pm$			&  cation and anion\\
    p			&    pressure-driven variable				&      &\\
    e			&    electro-osmotic variable				&		&\\
    \hline
  \end{tabular}}
\end{table}

\subsection{Governing equations}
We have employed coupled  Poisson-Nerst-Planck-Stokes equations for the mathematical modeling of the ion transport through the pore systems. Below we use the symbol ``$\sim$'' to denote dimensional field variables including space, time, and vector and scalar fields.  All constants regardless of their dimensionality and dimensionless field variables are denoted without $\sim$. The evolution of ion concentration fields is represented by transport equations utilizing the Nernst-Planck flux form:
\begin{equation}
\frac{\partial\tilde{C}^{\pm}}{\partial \tilde{t}} + 
\tilde{\nabla}.(\tilde{\textbf{U}} \tilde{C}^{\pm})=
 \tilde{\nabla}. \{ D \tilde{\nabla}\tilde{C}^{\pm} \pm \frac{\text{z} e D}{k_B T}\tilde{C}^{\pm} \tilde{\nabla}\tilde{\phi} \},
 \label{eq:np_dim}
\end{equation}
where $\tilde{C^\pm}$ are concentrations for positive and negative ions and $\tilde{\textbf{U}}$ is the velocity field vector. The parameters $D$, z, $e$, $k_B$, and $T$ refer to ion diffusivity, ion valence, elementary charge, Boltzmann constant, and absolute temperature, respectively. Here we have employed Einstein's relation to write the ion electromigration mobility in terms of its diffusivity. We consider zero ion flux normal to the pore surfaces:
\begin{equation}
\tilde{\textbf{U}}_{\perp} \tilde{C}^{\pm} - D \tilde{\nabla}_{\perp}\tilde{C}^{\pm} \mp \frac{\text{z} e D}{k_B T}\tilde{C}^{\pm} \tilde{\nabla}_{\perp}\tilde{\phi} = 0,
\end{equation}
where $\tilde{\textbf{U}}_{\perp}$ is the velocity component normal to walls and $\tilde{\nabla}_{\perp}$ is the dimensional gradient operator in wall-normal directions.\\
The electric potential field, $\tilde{\phi}$ is governed by the Poisson equation:
\begin{equation}
\tilde{\nabla}^2 \tilde{\phi}=-\frac{\tilde{\rho}}{\varepsilon}, 
\label{eq:poisson_dim}
\end{equation}
where  $\varepsilon$ is the $\text{electric permittivity}$ of the solution and $\tilde{\rho}$ is the local charge density equal to:
\begin{equation}
\tilde{\rho}= \text{z} e( \tilde{C}^+ - \tilde{C}^- ).
\end{equation} 
Using Gauss's law, we write the boundary condition for the electric field on the pore wall as follows:
\begin{equation}
\frac{\partial \tilde{\phi}}{\partial \tilde{x}_{\perp}}|_{\text{wall}} = \frac{\sigma}{\varepsilon},
\end{equation}
where $\sigma$ is the known surface charge density and assumed to be constant in time and distributed uniformly over the pore perimeter at all axial positions, but it can vary along the pore axis. 
We consider an incompressible fluid flow at low Reynolds number regime governed by the Stokes equations:
\begin{equation}
\tilde{\nabla}. \tilde{\textbf{U}}=0,
\label{eq:continuity_dim}
\end{equation}

\begin{equation}
-\tilde{\nabla} \tilde{P}+ \mu \tilde{\nabla} ^2 \tilde{\textbf{U}} -\text{z} e ( \tilde{C}^+ - \tilde{C}^-) \tilde{\nabla} \tilde{\phi}=0,
\label{eq:stokes_dim}
\end{equation}
where $\mu$ is the fluid dynamic viscosity, and $\tilde{P}$ represents the hydrodynamic pressure in the domain. For all velocity components, no-slip and no-penetration conditions are imposed at pore walls. We will discuss the boundary treatment in the axial directions in Section \ref{sec:network}. 

\subsection{Dimensionless formulation}

\begin{table}[htb]
  \caption{Normalization factors used for directional non-dimensionalization}
  \label{tbl:normal_factors}
  \centering
  \begin{tabular}{lll}
    \hline
    Variable  & Notation  & Normalization\\
    \hline
    Axial coordinate     & $x$   & $L$ \\
    Transverse coordinates   & $x_{\perp}$   & $h_p$\\
    Time     & $t$    & $t_{\text{diff}}=L^2/D$\\
    Axial velocity component & $u$  & $u_{\text{diff}}=D/L$\\
    Tansverse velocity components & $\textbf{U}_{\perp}$ & $u_{\text{diff}}h_p/L$\\
    Ion concentrations & $C^{\pm}$ & $C_{\text{ref}}$\\
    Hydrodynamic pressure & P & $C_{\text{ref}} k_B T$\\
    Electric potential & $\phi$ & $V_T=k_BT/ze$\\
    \hline
  \end{tabular}
\end{table}

In an approach analogous to lubrication theory, we have used a directional normalization to account for different length scales present in pore systems. Table (\ref{tbl:normal_factors}) summarizes the normalization factors used to nondimensionalize the governing equations. Concentration is nondimensionalized using $C_{\text{ref}}$, which is typically the reservoir concentration. We have used reference osmotic pressure and thermal voltage to normalize the hydrodynamic pressure and  electric potential respectively. Moreover, diffusion time, $t_{\text{diff}}=L^2/D$ and diffusion velocity, $u_{\text{diff}}=D/L$ are used to normalize time and the axial component of velocity. The axial coordinate is nondimensionalized by the characteristic pore length, L, and for normal directions we have used local effective pore size, $h_p$, defined as follows:
\begin{equation}
h_p\equiv \frac{\text{cross-sectional area}}{\text{cross-sectional perimeter}}.
\end{equation}
According to this definition, $h_p$ depends on the local area cross section of the pore and hence can change along the pore axis. For a rectangular pore with width much larger than the pore thickness, $h_p=\frac{h}{2}$, and for a circular cross-section with diameter $d$, $h_p=\frac{d}{4}$. Using this parameter, we take $u_{\text{diff}}h_p/L$ as the normalization factor for velocity components in transverse directions.
\\
The dimensionless parameters present in the normalized equations and boundary conditions are pore aspect ratio, $a$,  normalized Debye length, $\lambda_D$, electrohydrodynamic coupling constant, $\kappa$, and normalized surface charge density defined below: 
\begin{equation}
a = \frac{h_p}{L} \quad , \quad \lambda_D=\frac{\lambda_{\text{ref}}}{h_p} \quad , \quad \kappa = \frac{\varepsilon}{\mu D}V_T^2 \quad , \quad \sigma^*= \frac{h_p \text{z} e \sigma}{k_B T \varepsilon}
\end{equation}
where $\lambda_{\text{ref}}=\sqrt{\frac{\varepsilon k_B T}{2 C_{\text{ref}} \text{z}^2 e^2}}$ is the reference Debye length and $V_T=\frac{k_B T}{ze}$ is the thermal voltage. Note that $a$ and $\lambda_D$ can vary along the pore axes, as both are defined based on $h_p$. Using the normalization factors introduced in Table(\ref{tbl:normal_factors}), we obtain the non-dimensional equations as follows:
\begin{equation}
a^2 \frac{\partial C^{\pm}}{\partial t} + a^2 \nabla.(\textbf{U} C^{\pm})=
a^2 \frac{\partial}{\partial x}\lbrace \frac{\partial C^{\pm}}{\partial x} \pm C^{\pm} \frac{\partial \phi}{\partial x}\rbrace + \nabla_{\perp}.\lbrace
\nabla_{\perp} C^{\pm} \pm C^{\pm} \nabla_{\perp} \phi
\rbrace,
\label{eq:np}
\end{equation}

\begin{equation}
- (a^2 \frac{\partial ^2 \phi}{\partial x^2} + \nabla^2_{\perp} \phi)=\frac{1}{2\lambda_D ^2} (C^+- C^-),
\label{eq:poisson}
\end{equation}

\begin{equation}
\nabla.\textbf{U}=0,
\label{eq:continuity}
\end{equation}

\begin{equation}
a^2 \frac{\partial ^2 u}{\partial x^2} + \nabla^2_{\perp}u = \frac{\kappa}{2 \lambda_D^2} \lbrace \frac{\partial P}{\partial x} + (C^+ - C^-)\frac{\partial \phi}{\partial x} \rbrace 
\label{eq:x-mom},
\end{equation}

\begin{equation}
a^2 (a \frac{\partial ^2(a \textbf{U}_{\perp})}{\partial x^2} + \nabla^2_{\perp} \textbf{U}_{\perp})= \frac{\kappa}{2 \lambda_D^2} \lbrace 
 \nabla_{\perp}P+ (C^+ - C^-)  \nabla_{\perp}\phi 
 \rbrace.
 \label{eq:y-mom}
\end{equation}

$\nabla_{\perp}$ is the non-dimensional gradient operator in the transverse directions. Equations (\ref{eq:x-mom}) and (\ref{eq:y-mom}) are the momentum transport equations in axial and traverse directions respectively.
\\
The dimensionless boundary conditions at pore walls for equations (\ref{eq:np}), (\ref{eq:poisson}), (\ref{eq:x-mom}), and (\ref{eq:y-mom}) are respectively as follows:
\begin{equation}
\nabla_{\perp}C^{\pm} \pm C^{\pm} \nabla_{\perp}\phi = 0,
\end{equation}
\begin{equation}
\frac{\partial \phi}{\partial x_{\perp}}\vert_{\text{wall}}=\sigma^*,
\label{eq:sigma_s}
\end{equation}
\begin{equation}
\textbf{U} = 0.
\label{eq:u_boundary}
\end{equation}

\section{Reduced order model} \label{sec:reduced order model}
We consider pores with small values of aspect ratio ($a \ll1$), which implies long and thin pores. From a physical point of view, this means that the dynamics are fast in thin directions, and thus it is reasonable to assume equilibrium condition in these directions. Mathematically, this assumption leads to a perturbation analysis through which the terms in equations (\ref{eq:np}) to (\ref{eq:u_boundary}) are prioritized based on powers of $a$ in the preceding pre-factors.\\

In this section, we introduce the numerical modeling of a single pore in the limit of small aspect ratio and derive a set of reduced order equations in this setting. In Section \ref{sec:network}, we explain how the solutions from different pores can be coupled at the pore intersections using the proper conservation laws.

\subsection{Cross-sectional equilibrium profiles} \label{sec:c0}
In the limit of small $a$, equation (\ref{eq:np}) is simplified to:
\begin{equation}
\nabla_{\perp}. \lbrace \nabla_{\perp}C^{\pm} \pm C^{\pm}\nabla_{\perp}\phi \rbrace =0.
\label{eq:simplified_C}
\end{equation}
Given the zero ion flux normal to the wall (for now we have ignored chemical reactions among ions and surface groups), the normal flux for each ion species is zero. Thus, one can conclude:
\begin{equation}
\nabla_{\perp}C^{\pm} \pm C^{\pm}\nabla_{\perp}\phi =0.
\label{eq:const_normal_flux}
\end{equation}
Using the definition of electro-chemical potential for co-ion and counterion ($\mu^{\pm}=\text{ln}C^{\pm} \pm \phi$), we can interpret equation (\ref{eq:const_normal_flux}) as constant electro-chemical potentials in the wall normal direction. 
The ion concentration fields then follow the Boltzmann distribution:
\begin{equation} 
C^{\pm}= C^{\pm}_0(x) \exp(\mp \phi),
\label{eq:boltzmann_C}
\end{equation}
where the integration constants $C_0^{\pm}$ are the ion concentrations where $\phi$ is zero and are only functions of the axial coordinate. To write both solutions in relation (\ref{eq:boltzmann_C}) with the same constant, we define the following parameters:
\begin{equation}
C_0(x)=\sqrt{C^+_0(x) C^-_0(x)} = \sqrt{C^+(x, x_{\perp}) C^-(x, x_{\perp})}, 
\label{eq:c0}
\end{equation}
\begin{equation}
\phi(x, x_{\perp})= \phi_0(x) + \psi(x, x_{\perp}),
\label{eq:psi_def}
\end{equation}
where $\phi_0$ is defined as follows:
\begin{equation}
\exp(\phi_0(x))=\sqrt{\frac{C^+_0}{C^-_0}} = \sqrt{\frac{C^+}{C^-}} \exp(\phi).
\label{eq:phi0}
\end{equation}
In other words, $\sqrt{C^+C^-}$ and $\sqrt{C^+/C^-}\exp(\phi)$ are two invariants of the solution introduced in equation (\ref{eq:boltzmann_C}) at each cross-section. Using these definitions, relation (\ref{eq:boltzmann_C}) can be written as:
\begin{equation}
C^{\pm}= C_0(x) \exp(\mp \psi).
\label{eq:final_boltzmann}
\end{equation}
The physical interpretation of $C_0$ and $\phi_0$ is straightforward in the limit of thin EDLs: $C_0$ represents concentration of either species in the electroneutral fluid outside the EDLs (e.g., pore centerline). Likewise, $\phi_0$ represents the electrostatic potential outside the EDLs. The field $\psi$ represents the equilibrium electrostatic potential due to the EDLs. For cases that EDLs are not thin, $C_0$ and $\phi_0$ respectively represent the concentration and electrostatic potential in a virtual electroneutral reservoir which is in equilibrium with the local cross-section (thus, they are called virtual quantities). As we will see, these quantities will form the driving potentials whose gradients will be the driving forces for axial fluxes. In the limit of small $a$, we can rewrite equation (\ref{eq:poisson}) using equation (\ref{eq:final_boltzmann}) as: 
\begin{equation}
\nabla^2_{\perp}\psi =\frac{1}{\lambda_0^2} \sinh(\psi),
\label{eq:psi_eqn}
\end{equation}
where $\lambda_0 =  \frac{1}{h_p} \sqrt{\frac{\varepsilon k_B T}{2 \tilde{C}_0 \text{z}^2 e^2}}=\frac{\lambda_D}{\sqrt{C_0(x)}}$ is the normalized Debye length based on the local $C_0$. The boundary condition represented in equation (\ref{eq:sigma_s}) is also simplified to:
\begin{equation}
\frac{\partial \psi}{\partial x_{\perp}}\vert_{\text{wall}}=\sigma^*(x).
\label{eq:psi_bc}
\end{equation}
Equation (\ref{eq:psi_eqn}) can be solved for equilibrium potential profiles and corresponding equilibrium concentrations via (\ref{eq:final_boltzmann}) for a wide range of $\sigma^*$ and $\lambda_0$. These profiles are smooth functions of parameters $\sigma^*$ and $\lambda_0$. Therefore, one can numerically tabulate the results of these  solutions using a relatively coarse mesh in the input parameter space. The tabulation strategy will be discussed in detail in Section \ref{sec:table}. Furthermore, we note that the asymptotic solutions from equations (\ref{eq:final_boltzmann})-(\ref{eq:psi_bc}) lead to concentration fields whose integral in each cross-sectional plane are electroneutral when the surface charge is accounted for. This can be readily examined by applying Green's theorem to the integral of equation (\ref{eq:psi_eqn}) and using the boundary condition (\ref{eq:psi_bc}). This asymptotic electroneutrality will be utilized when developing the axial transport model in Section \ref{sec:area integrated transport}.

\subsection{Virtual total pressure and electro-chemical potential} \label{sec:p0mu}
In our model, two of the driving potentials are virtual total pressure, $P_0$ and virtual counterion electro-chemical potential, $\mu^+$, instead of hydrodynamic pressure, $P$, and virtual electric potential, $\phi_0$. This choice has been beneficial for the following reasons: (1) $P_0$ and $\mu^+$ remain continuous at the pore interfaces, where the pore area cross-sections have jumps, whereas electric potential and hydrodynamic pressure may have significant jumps at these locations. (2) $\mu^+$ remains finite in regions of very low concentration, e.g. behind deionization shocks, while $\phi_0$ becomes unbounded. (3) As we will explain in Section \ref{sec:table}, these variables result in bounded area-averaged coefficients that are used for computing ion flux terms. Using bounded coefficients is advantageous to deal with singularities in regions of low and high concentration.\\

\textbf{Virtual total pressure}: We can solve for cross-sectional variations of pressure using equation (\ref{eq:y-mom}) for small $a$. This yields:
\begin{equation}
\nabla_{\perp}P= -( C^+ - C^-) \nabla_{\perp}\phi = 2C_0\nabla_{\perp}(\cosh(\psi)),
\end{equation}
which results in the following solution after integrating from the point with $\psi=0$ to a transverse location, $y$ with potential $\psi$: 
\begin{equation}
 P=2 C_0 \cosh(\psi) + P_0(x).
 \label{eq:finalP}
\end{equation}
In other words, $P_0=P-2C_0\text{cosh}(\psi)$ is another cross-sectional invariant. This quantity physically represents the superposition of hydrodynamic pressure and osmotic pressure.\\

\textbf{Virtual electro-chemical potential}: Using equations (\ref{eq:c0}) through (\ref{eq:phi0}), we can show that electro-chemical potentials for ion species are only functions of x:
\begin{equation}
 \mu^{\pm}=\ln(C_0 \text{exp}(\mp \psi)) \pm (\psi+\phi_0)=\ln(C_0(x)) \pm \phi_0(x).
 \label{eq:finalMu}
\end{equation}
As we will see in Section \ref{sec:velocity profiles}, this allows us to integrate the axial momentum equation and obtain the velocity profile as a sum of the terms including axial gradients of driving potentials multiplied by transverse-dependent fields. These transverse-dependent fields can be determined by using the equilibrium profiles discussed in Section \ref{sec:c0}. 

\subsection{Cross-sectional velocity profile} \label{sec:velocity profiles}
\begin{figure}[H]
\makebox[\textwidth][c]{\includegraphics[width=0.6\textwidth]{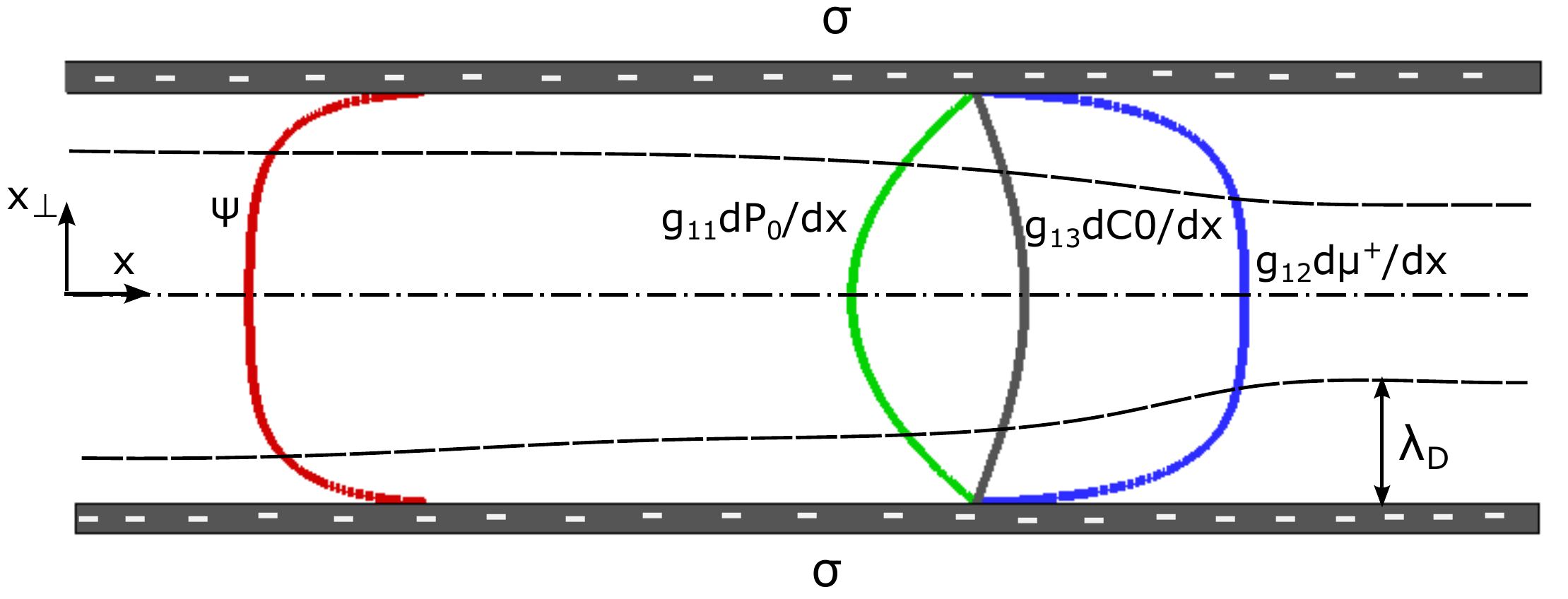}}
\caption{A sample solution of equilibrium electrostatic potential, $\psi$, inside a micro-pore for some values of $\sigma^*$ and $\lambda_0$. The thickness of EDL varies along the pore axis due to the axial concentration gradient. In another axial position, three axial velocity profiles have been depicted, which result from the local gradients of $P_0$, $\mu^+$, and $C_0$. The net axial velocity profile is obtained by the linear superposition of the three terms: $u = g_{11} \frac{d P_0(x)}{d x}+g_{12} \frac{d \mu^+(x)}{d x}+g_{13} \frac{d C_0(x)}{d x} $. According to the notation on the figure and equation (\ref{eq:axial_vel_sol}), one can realize that $g_{11}=\frac{\kappa}{2\lambda_D^2}g^p$, $g_{12}=\kappa g^e$, and $g_{13}=\frac{\kappa}{2\lambda_D^2}g^c$.}
\label{fig:sample_equlb_soln}
\end{figure}

Using relations (\ref{eq:finalP}) and (\ref{eq:finalMu}) in the axial momentum equation (\ref{eq:x-mom}) in the limit of small $a$, we can rewrite this equation as follows:
\begin{equation}
 \nabla^2_{\perp}u=\frac{\kappa}{2\lambda_D^2}\{
\frac{d P_0}{d x} + (C^+-C^-)\frac{d \mu^+}{dx} + 2\exp(\psi) \frac{dC_0}{dx}\}.
\label{eq:axial_u_eq}
\end{equation}
Due to the linearity of equation (\ref{eq:axial_u_eq}), we take advantage of superposition principle and write the solution as follows:
\begin{equation}
u=\frac{\kappa}{2 \lambda_D^2} \{ g^p(x_{\perp}) \frac{d P_0(x)}{d x} +2\lambda_D^2 g^e(x_{\perp}) \frac{d\mu^+(x)}{d x}+g^c(x_{\perp}) \frac{d C_0(x)}{d x}\}.
\label{eq:axial_vel_sol}
\end{equation}
Equation (\ref{eq:axial_vel_sol}) has three terms describing the fluid motions generated by three mechanisms, which are pressure driven flow, electro-osmosis, and diffusio-osmosis respectively. $g^p$, $g^e$, and $g^c$ are the transverse-dependent part of the solutions and obtained by solving the following differential equations:
\begin{equation}
 \nabla^2_{\perp}g^p=1 ,\qquad g^p\vert_{\text{wall}}=0,
 \label{eq:p_grad}
\end{equation}
\begin{equation}
 \nabla^2_{\perp}g^e=- \nabla^2_{\perp}\psi ,\qquad g^e \vert_{\text{wall}}=0,
 \label{eq:phi_grad}
\end{equation}
\begin{equation}
 \nabla^2_{\perp}g^c=2\exp(\psi) ,\qquad g^c\vert_{\text{wall}}=0.
 \label{eq:C_grad}
\end{equation}
Equation (\ref{eq:p_grad}) results in the classical Poiseuille flow. Equation (\ref{eq:phi_grad}) and (\ref{eq:C_grad}) can be solved using the equilibrium profiles of $\psi$, which are available from solutions to equation (\ref{eq:psi_eqn}) for different values of $\sigma^*$ and $\lambda_0$. Therefore, in addition to the profiles of $\psi$ and $C^\pm$ discussed in Section \ref{sec:c0}, one can access the velocity profiles, $g^p$, $g^e$, and $g^c$ for a given $\sigma^*$ and $\lambda_0$. Figure (\ref{fig:sample_equlb_soln}) demonstrates examples of three components of velocity profile as well as equilibrium potential distribution, $\psi$.

\subsection{Area-integrated transport equation} \label{sec:area integrated transport}
To obtain the one dimensional transport equation, first we rewrite equation (\ref{eq:np}) as below:
\begin{equation}
\frac{\partial C^{\pm}}{\partial t} + \frac{\partial F_x^{\pm}}{\partial x} + \nabla_{\perp}\textbf{F}_{\perp}^{\pm} = 0,
\label{eq:np_newform}
\end{equation}
where, 
\begin{equation}
F_x^{\pm} = uC^{\pm} - \frac{\partial C^{\pm}}{\partial x} \mp C^{\pm}\frac{\partial \phi}{\partial x},
\end{equation}
\begin{equation}
\textbf{F}_{\perp}^{\pm} = \textbf{U}_{\perp}C^{\pm} - \frac{1}{a^2} \{\nabla_{\perp}C^{\pm} \pm C^{\pm} \nabla_{\perp} \phi \}.
\end{equation}
Integrating equation (\ref{eq:np_newform}) over the area cross-section, $S$, we obtain each term in the area-integrated transport equation as follows:
\begin{equation*}
\int_S \frac{\partial C^{\pm}}{\partial t} ds= S\frac{\partial \overline{C^{\pm}}}{\partial t},
\end{equation*}
\begin{equation}
\int_S \frac{\partial F_x}{\partial x}ds= \frac{\partial}{\partial x} S\{\overline{uC^{\pm}} - \frac{\partial \overline{C}^{\pm}}{\partial x}  \mp \overline{C^{\pm}\frac{\partial \phi}{\partial x}}\},
\end{equation}
\begin{equation*}
\int_S \nabla_{\perp}.\textbf{F}_{\perp} ds =\oint_{\text{wall}}
\textbf{F}_{\perp}.\textbf{n}dl =0.
\end{equation*}
Note that the second term is integrated assuming that the axial variation of cross-sections is a smooth function of $x$ inside each pore (based on Leibniz integral rule). Writing the equation with respect to electro-chemical potential of positive ions, we obtain the area-integrated transport equations for positive and negative ions respectively as:
\begin{equation}
S\frac{\partial \overline{C^+}}{\partial t} +\frac{\partial}{\partial x} S\{ \overline{uC^+} - \overline{C^+}\frac{d \mu^+}{d x}\}=0,
\label{eq:area_pos_ion}
\end{equation}  
\begin{equation}
S\frac{\partial \overline{C^-}}{\partial t} +\frac{\partial}{\partial x} S\{\overline{uC^-} -2\frac{\overline{C^-}}{C_0} \frac{d C_0}{dx} + \overline{C^-}\frac{d \mu^+}{d x}\}=0.
\label{eq:area_neg_ion}
\end{equation} 
These equations are partial differential equations (PDEs) describing the temporal and axial evolution of area-averaged ion concentrations. In practice, one can choose to solve only one of these equations and use electroneutrality, as discussed in Section \ref{sec:c0}, to relate $\overline{C^+}$ to $\overline{C^-}$.  This can be obtained by applying Green's theorem and area integration of equation (\ref{eq:psi_eqn}) and using boundary condition (\ref{eq:psi_bc}):
\begin{equation}
\overline{C^+} - \overline{C^-} = C_s \equiv -2 \sigma^* \lambda_D^2
\label{eq:nutrality}
\end{equation}
The right hand side of equation (\ref{eq:nutrality}) refers to the excess concentration of counterions required to screen the local surface charge, denoted by $C_s$ in our formulation. This is another dimensionless parameter normalized by $C_{\text{ref}}$, which indicates the significance of surface conduction. Higher values of $C_s$ means that there is a strong surface conduction effect.\\

In our model, we have chosen to solve (\ref{eq:area_neg_ion}) for the concentration of coions (in this case, anions) and use the electroneutrality condition (\ref{eq:nutrality}) to obtain the counterion concentration. In continue, we describe how to close different area-averaged flux terms in equation (\ref{eq:area_neg_ion}) and determine the virtual total pressure and virtual electro-chemical potential fields.\\
\\
We relate $\overline{C^-}$ to $C_0$ by the area-averaged parameter, $\bar{g}(x)$:
\begin{equation}
\overline{C^-} = C_0\bar{g},
\label{eq:cbar_c0}
\end{equation}
where $\bar{g}$ can be computed using the available equilibrium potential profiles from the solution to equation (\ref{eq:psi_eqn}) as follows:
\begin{equation}
\bar{g} = \frac{1}{S}\int_S \text{exp}(\psi)ds.
\label{eq:fbar}
\end{equation}
Using equation (\ref{eq:fbar}), we can compute $\bar{g}$ and tabulate it for different values of $\sigma^*$ and $\lambda_0$. To determine the axial advective flux $\overline{uC^-}$, we utilize the Reynolds decomposition strategy for both $C^-$ and $u$:
\begin{equation}
C^- = \overline{C^-}(x) + C^{-\prime}(x,x_{\perp}) \quad , \quad u = \bar{u}(x)+u^{\prime}(x,x_{\perp}).
\label{eq:reynolds decomposition}
\end{equation}
Using equation (\ref{eq:axial_vel_sol}), we derive $\bar{u}$ and $u^{\prime}$ as follows:
\begin{equation}
\bar{u}= \frac{\kappa}{2 \lambda_D^2} \{ \overline{g^p}\frac{d P_0}{d x} + 2 \lambda_D^2 \overline{g^e} \frac{d \mu^+}{d x} + \overline{g^c} \frac{d C_0}{d x} \},
\label{eq:ubar}
\end{equation}
\begin{equation}
u^{\prime}= \frac{\kappa}{2 \lambda_D^2} \{ g^{p\prime}\frac{d P_0}{d x} + 2 \lambda_D^2 g^{e\prime} \frac{d \mu^+}{d x} + g^{c\prime} \frac{d C_0}{d x} \}.
\label{eq:uprime}
\end{equation}
$\overline{g^p}$ is computed using the Poiseuille velocity profile. For a 2D channel $\overline{g^p}=-1/3$ and for a cylindrical pore it is equal to $-1/2$. $\overline{g^e}$ and $\overline{g^c}$ are also obtained by area averaging the solutions of equations (\ref{eq:phi_grad}) and (\ref{eq:C_grad}). The profiles with the superscript prime (such as $g^{p \prime}$) can be simply obtained by subtracting the area-averaged values from the corresponding profiles. Therefore, by means of equations (\ref{eq:reynolds decomposition})-(\ref{eq:uprime}), the advective flux in (\ref{eq:area_neg_ion}) is written as:
\begin{equation}
\overline{uC^-}= \overline{u}\overline{C^-} + \frac{\kappa}{2 \lambda_D^2} \{ \overline{g^{p\prime} C^{-\prime}} \frac{d P_0}{d x}  + 2 \lambda_D^2 \overline{g^{e\prime}C^{-\prime}} \frac{d \mu^+}{d x}
+ \overline{g^{c\prime} C^{-\prime}} \frac{d C_0}{d x} \}.
\label{eq:UC-}
\end{equation}
To close the last three terms, we normalize the area-averaged coefficients with $C_s$, and introduce the area-averaged coefficients $\overline{g^{p-}}$, $\overline{g^{e-}}$, and $\overline{g^{c-}}$:
\begin{equation}
\overline{uC^-}=\bar{u}\overline{C^-} + C_s \frac{\kappa}{2 \lambda_D^2} \{ \overline{g^{p-}} \frac{dP_0}{dx} + 2 \lambda_D^2 \overline{g^{e-}}  \frac{d\mu^+}{dx} + \overline{g^{c-}}  \frac{d C_0}{dx} \},
\label{eq:neg_advection}
\end{equation}
where,
\begin{equation}
\overline{g^{p-}} =({\frac{\overline{g^{p \prime}C^{-\prime}}}{2\vert \sigma^*\vert \lambda_D^2}}), 
\quad
 \overline{g^{e-}} =({\frac{\overline{g^{e\prime}C^{-\prime}}}{2\vert \sigma^*\vert \lambda_D^2}}),
\quad
\overline{g^{c-}} = ({\frac{\overline{g^{c\prime}C^{-\prime}}}{2\vert \sigma^*\vert \lambda_D^2}}).
\label{eq:neg_averaged_terms}
\end{equation}
The coefficients defined in (\ref{eq:neg_averaged_terms}) are all functions of $\sigma^*$ and $\lambda_0$, as one can compute $\frac{C^{-\prime}}{2\vert \sigma^* \vert \lambda_D^2}$ using the equilibrium profiles as follows:
\begin{equation}
\frac{C^{-\prime}}{2\vert \sigma^* \vert \lambda_D^2} = \frac{(\text{exp}(\psi)-\overline{f})}{ 2\vert \sigma^* \vert \lambda_0^2}.
 \end{equation}
Normalization with $C_s$ is appropriate since deviation of concentration profiles from the mean is controlled by EDL whose total charge scales with $C_s$. As shown later, this normalization leads to bounded  coefficients even when the bulk concentration becomes very small. Knowing the local values of $\sigma^*$ and $\lambda_0$, one can first solve equations (\ref{eq:final_boltzmann}) to (\ref{eq:psi_bc}) to obtain $\psi(x_{\perp})$ and $C^-(x_{\perp})/C_0$, and then solve equations (\ref{eq:p_grad}) to (\ref{eq:C_grad}) to obtain $g^p(x_{\perp})$, $g^e(x_{\perp})$, and $g^c(x_{\perp})$. After computing $\bar{g}$ and applying Reynolds decomposition, one can compute the coefficients introduced in equation (\ref{eq:neg_averaged_terms}).\\
The same procedure is applied for the concentration of positive ions. Using the electro-neutrality assumption, we decompose $C^+$ as:
\begin{equation}
C^+=\overline{C^-}+ C_s + C^{+\prime}.
\label{eq:cation}
\end{equation}
The final result for the area-averaged advection of cation concentration is obtained as follows:
\begin{equation}
\overline{uC^+}=\overline{u}(\overline{C}+C_s)+ C_s \frac{\kappa}{2 \lambda_D^2} \{ \overline{g^{p+}} \frac{dP_0}{dx}+ 2 \lambda_D^2 \overline{g^{e+}} \frac{d\mu^+}{dx} +\overline{g^{c+}} \frac{d C_0}{dx} \},
\label{eq:pos_adv}
\end{equation}
where,
\begin{equation}
\overline{g^{p+}}=(\overline{\frac{g^{p \prime}C^{+\prime}}{2\vert \sigma^*\vert \lambda_D^2}}),
\quad
\overline{g^{e+}}=(\overline{\frac{g^{e\prime}C^{+\prime}}{2\vert \sigma^*\vert \lambda_D^2}}),
\quad
\overline{g^{c+}}=(\overline{\frac{g^{c\prime}C^{+\prime}}{2\vert \sigma^*\vert \lambda_D^2}}),
\label{eq:areapositiveterms}
\end{equation}
We can determine $\frac{C^{+\prime}}{2\vert \sigma^* \vert \lambda^2}$ using the equilibrium solutions as follows:
\begin{equation}
 \frac{C^{+\prime}}{2\vert \sigma^* \vert\lambda^2}=(\frac{\text{exp}(-\psi)-\overline{f}}{2\vert \sigma^* \vert \lambda_0^2}-1).
\end{equation}
In addition to area-averaged coefficients, we need to determine $P_0$ and $\mu^+$ to fully close equation (\ref{eq:area_neg_ion}). To this end, we apply conservation laws for fluid mass and net electric charge in the axial direction inside each pore. The area-integrated continuity of fluid flow, $\frac{d }{dx}(S\bar{u})=0$ yields:
\begin{equation}
 \frac{d}{d x}(A_1(x) \frac{d P_0}{d x}) + \frac{d}{d x}(B_1(x) \frac{d \mu^+}{d x})= \frac{d}{d x} (C_1(x)\frac{d C_0}{d x}),
\label{eq:cont}
\end{equation}
where,
\begin{equation*}
 A_1(x)=S\frac{\kappa}{2 \lambda_D^2} \overline{g^p} \quad , \quad B_1(x)= S\kappa \overline{g^e} \quad , \quad
 C_1(x)= - S\frac{\kappa }{2 \lambda_D^2}\overline{g^c}.
\end{equation*}
Conservation of electric charge can be stated as $\frac{d }{dx}\{S\text{z}(\overline{F^+_x} -\overline{F^-_x} )\}=0$, which can be written in a similar form to equation (\ref{eq:cont}): 
\begin{equation}
\frac{d}{d x}(A_2(x) \frac{d P_0}{d x}) + \frac{d}{d x}(B_2(x) \frac{d\mu^+}{d x})= \frac{d}{d x} (C_2(x) \frac{d C_0}{d x}),
\label{eq:current}
\end{equation}
where,
\begin{equation*}
 A_2(x) = SC_s \frac{\kappa}{2\lambda_D^2 } ( \overline{g^p} + \overline{g^{p+}} - \overline{g^{p-}}),
 \end{equation*}
 
\begin{equation*}
B_2(x) = -S \{ (2\overline{C^-}+C_s)+ C_s \kappa (\overline{g_e} + \overline{g^{e+}}-\overline{g^{e-}} ) \} ,
\end{equation*}

\begin{equation*}
 C_2(x)= - S \{ C_s \frac{\kappa}{2\lambda_D^2}(\overline{g^c} + \overline{g^{c+}} - \overline{g^{c-}}) + 2\bar{g} \}.
\end{equation*}
Inside each pore, we solve equations (\ref{eq:cont}) and (\ref{eq:current}) to obtain $P_0$ and $\mu^+$ fields at each time step. All x-variable coefficients in these equations are determined using the local values of $\overline{C^-}$ and 9 area-averaged coefficients obtained from the equilibrium solutions. The numerical strategy used to solve this set of equations are discussed in detail in Section \ref{sec:network}.\\

To summarize, one needs to compute 9 area-averaged coefficients that are $\bar{g}$, $\overline{g^e}$, $\overline{g^c}$, $\overline{g^{p-}}$, $\overline{g^{e-}}$, $\overline{g^{c-}}$, $\overline{g^{p+}}$, $\overline{g^{e+}}$, and $\overline{g^{c+}}$. These coefficients are only functions of $\sigma^*$ and $\lambda_0$. Therefore, to reduce the problem to 1D, one needs to store only these 9 numbers with respect to dimensionless surface charge and dimensionless Debye length. These coefficients are determined using the equilibrium solutions to the cross-sectional profiles of electrostatic potential ($\psi$), concentration, and velocity fields.  Knowing the values of the area-averaged coefficients, one can use the following order of procedures to time-advance all quantities: 
\begin{enumerate}
\item The field $\overline{C^-}(x)$ is already known from previous time step. One can compute $C_0(x)$ from equation (\ref{eq:cbar_c0}).
\item Having $C_0(x)$, we can solve equations (\ref{eq:cont}) and (\ref{eq:current}) coupled with each other to find $P_0$ and $\mu^+$. 
\item $\bar{u}$ can be computed by substitution into equation (\ref{eq:ubar}) 
\item Mean advective flux, $\overline{uC^-}$ can be computed by substitution into equation (\ref{eq:neg_advection}) 
\item The substitution of the above quantities in (\ref{eq:area_neg_ion}) allows time-advancement for the field $\overline{C^-}$.  
\end{enumerate} 
 
\subsection{Tabulation strategy for area-averaged coefficients} \label{sec:table}
The procedure described above uses $\lambda_0(C_0)$ as the primary input for the computation of the area-averaged coefficients. However, the developed evolution equation (\ref{eq:area_neg_ion}) updates $\overline{C^-}$ as the primary field variable and not $C_0$. In the thin EDL limit, $C_0$ is equal to $\overline{C^-}$, but in general we have $\overline{C^-} < C_0$. In order to compute $\lambda_0$ one needs to close $C_0$ in terms of $\overline{C^-}$. This relation is provided in equation (\ref{eq:cbar_c0}), which sets an iterative procedure as described below. First, we define an explicitly available dimensionless Debye length based on the local $\overline{C^-}$: 
\begin{equation}
\lambda^* = \frac{1}{h_p} \sqrt{\frac{\varepsilon k_B T}{2 \tilde{\overline{C^-}} \text{z}^2 e^2}} = \frac{\lambda_D}{\sqrt{\overline{C ^-}}}.
\end{equation}

The idea is to tabulate all of the coefficients against $\sigma^*$ and $\lambda^*$ by bypassing $\lambda_0$. For any given $\sigma^*$ and $\lambda^*$, we iteratively guess the value of $\lambda_0$. Then, by using the procedure described in Section \ref{sec:c0} and computing $\psi$ across the pore cross-section, one can use equation (\ref{eq:fbar}) to obtain $\bar{g}(\sigma^*, \lambda_0)$ and validate the guessed $\lambda_0$ by testing whether substitution in equation (44) leads to the initially known $\overline{C^-}$. Equivalently, one can test whether the following relation holds:
\begin{equation}
\lambda_0 = \lambda^*\sqrt{\bar{g}}.
\label{eq:lam_0_s}
\end{equation}
If this equation does not hold, we update the guessed $\lambda_0$ by the value obtained from (\ref{eq:lam_0_s}). This iterative procedure is repeated until $\lambda_0$ and corresponding $\psi(x_{\perp})$ profile are converged to their correct values. After obtaining the converged solution, we can then compute area-averaged coefficients and tabulate them with respect to $\sigma^*$ and $\lambda^*$. Once the tabulated coefficients are saved, no iteration procedure would be needed for the main simulations, which time-advance $\overline{C^-}(x)$. The summary of our tabulation strategy is as follows:\\

Equation (\ref{eq:psi_eqn}) is solved for a wide range of parameters $\sigma^*$ and $\lambda^*$. Each value of $\sigma^*$ sets a new boundary condition for equation (\ref{eq:psi_eqn}). For each value of $\lambda^*$ an iterative procedure is applied to determine the correct value of $\lambda_0$ and corresponding equilibrium solution of equation (\ref{eq:psi_eqn}). The explanation of the numerical procedure to solve this equation is presented in \ref{app:poisson solver}. Once the converged solution of (\ref{eq:psi_eqn}) is obtained, we compute $g^p(x_{\perp})$, $g^e(x_{\perp})$, and $g^c(x_{\perp})$ profiles by solving equations (\ref{eq:p_grad}) to (\ref{eq:C_grad}) numerically. Having obtained the electric potential distribution, $\bar{g}$ is computed using equation (\ref{eq:fbar}). $\overline{g^p}$, $\overline{g^e}$, and $\overline{g^c}$ are determined by integrating $g^p$, $g^e$, and $g^c$ profiles over pore area cross-section. $\overline{g^{p-}}$, $\overline{g^{e-}}$, $\overline{g^{c-}}$ are computed by equations (\ref{eq:neg_averaged_terms}) and the last three coefficients, $\overline{g^{p+}}$, $\overline{g^{e+}}$, and $\overline{g^{c+}}$ are obtained using equations (\ref{eq:areapositiveterms}). The computed quantities are eventually tabulated based on the corresponding values of $\sigma^*$ and $\lambda^*$.\\

\begin{figure}[H]
\makebox[\textwidth][c]{\includegraphics[width=0.99\textwidth]{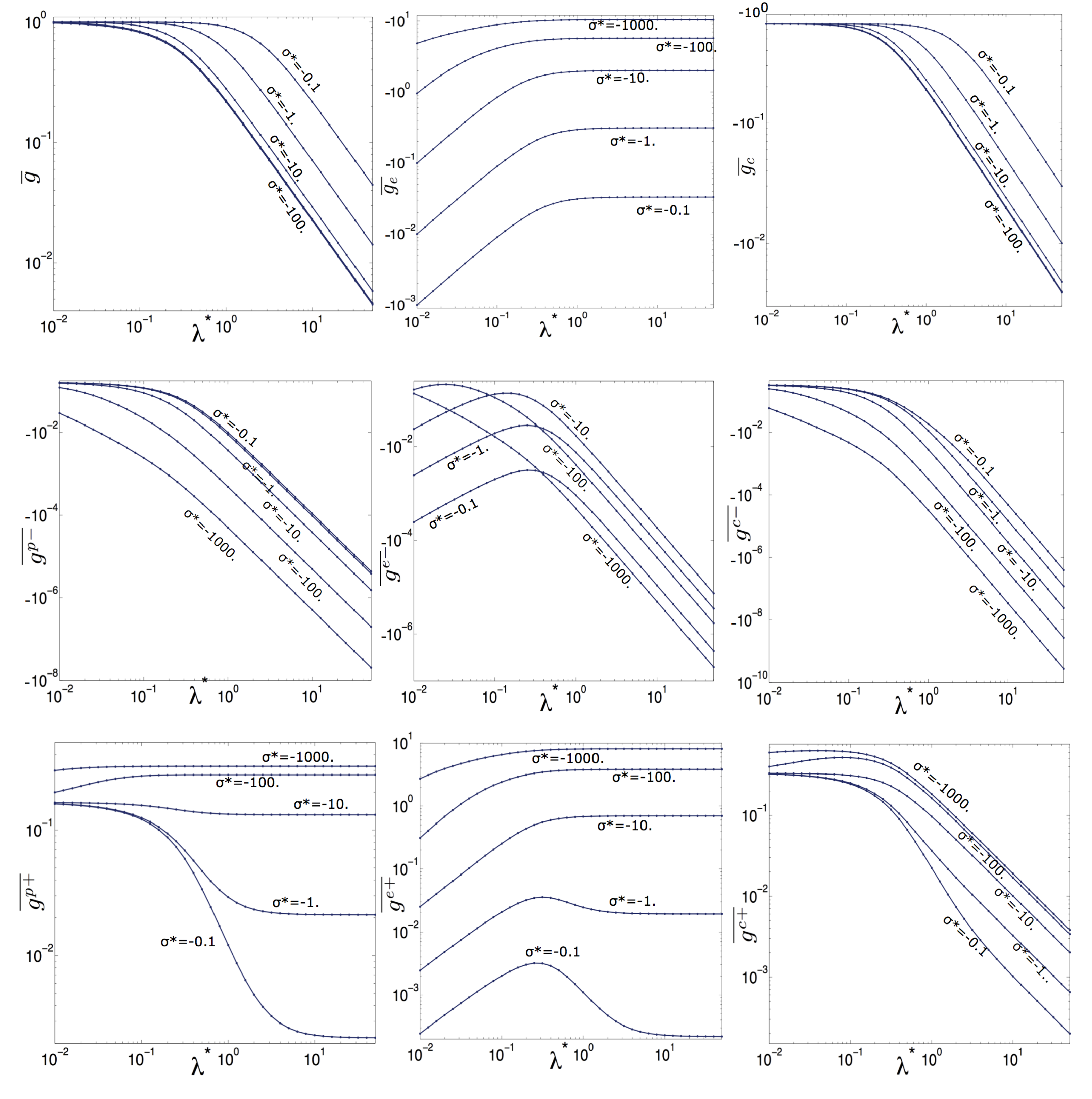}}
\caption{The variation of 9 area-averaged coefficients vs. $\lambda^*$ in the log-log scale for different values of $\sigma^*$. The coefficients were computed for a pore with rectangular cross-section. For such a geometry, $\overline{g^p} = -1/3$, which is computed analytically using the pressure-driven velocity profile obtained from equation (\ref{eq:p_grad}). All plots are generated based on the absolute values of all coefficients, and their actual signs are indicated on the vertical axes.}
\label{fig:table_coeff}
\end{figure}

Figure (\ref{fig:table_coeff}) represents the area-averaged coefficients computed for a rectangular cross-section channel and tabulated with respect to $\sigma^*$ and $\lambda^*$. Note that we have plotted the absolute values of all coefficients vs. $\lambda^*$ for different values of $\sigma^*$ and indicated the actual signs of the coefficients on the vertical axes.  As mentioned before, our formulation leads to bounded area-averaged coefficients for both thick and thin EDL limits. This helps eliminate the singularities of the models presented in the literature in the limit of low concentration ( large $\lambda^*$) \cite{mani2009, harvie2012}.\\

Before moving to the next section, we provide a short summary on the three dimensionless Debye lengths used in our model, which are $\lambda_D$, $\lambda^*$, and $\lambda_0$. All of these represent Debye lengths normalized by the local pore size, $h_p$, and thus are dependent on pore axial coordinates. $\lambda_D$ is the dimensionless Debye length defined based on $C_{\text{ref}}$ and thus, it is time independent. $\lambda^*$ is the Debye length defined based on the local $\overline{C^-}$, and together with $\sigma^*$, is the key input parameter to read the developed tables shown in Figure (\ref{fig:table_coeff}). $\lambda_0$ is the Debye length defined based on the local virtual concentration, $C_0$ and appears in equation (\ref{eq:psi_eqn}). This parameter was useful to lay out the derivation process. However, in the simulation algorithm, after the area-averaged coefficients are obtained, this parameter is not used explicitly. Overall, the important dimensionless input parameters for a single pore are $\lambda_D$, $\sigma^*$, $\kappa$, and the dimensionless driving forces which are externally applied; the presented leading order solution is independent of the parameter $a$.

\section{Extension to network of pores} \label{sec:network}
We introduced the reduced order models and explained how to solve for area-averaged concentration, virtual total pressure, and virtual electro-chemical potential using equations (\ref{eq:area_neg_ion}), (\ref{eq:cont}), and (\ref{eq:current}).  For a single pore connected to external reservoirs, these equations are naturally accompanied by Dirichlet boundary conditions that specify the values of $C_0$, $P_0$, and $\mu^+$. However, when there is a network of connecting pores and/or reservoirs with unknown pressure or potential (such as a dead-end pore with zero net flow), the boundary values are to be determined via additional procedures discussed below.\\

\begin{figure}[h]
\makebox[\textwidth][c]{\includegraphics[width=0.6\textwidth]{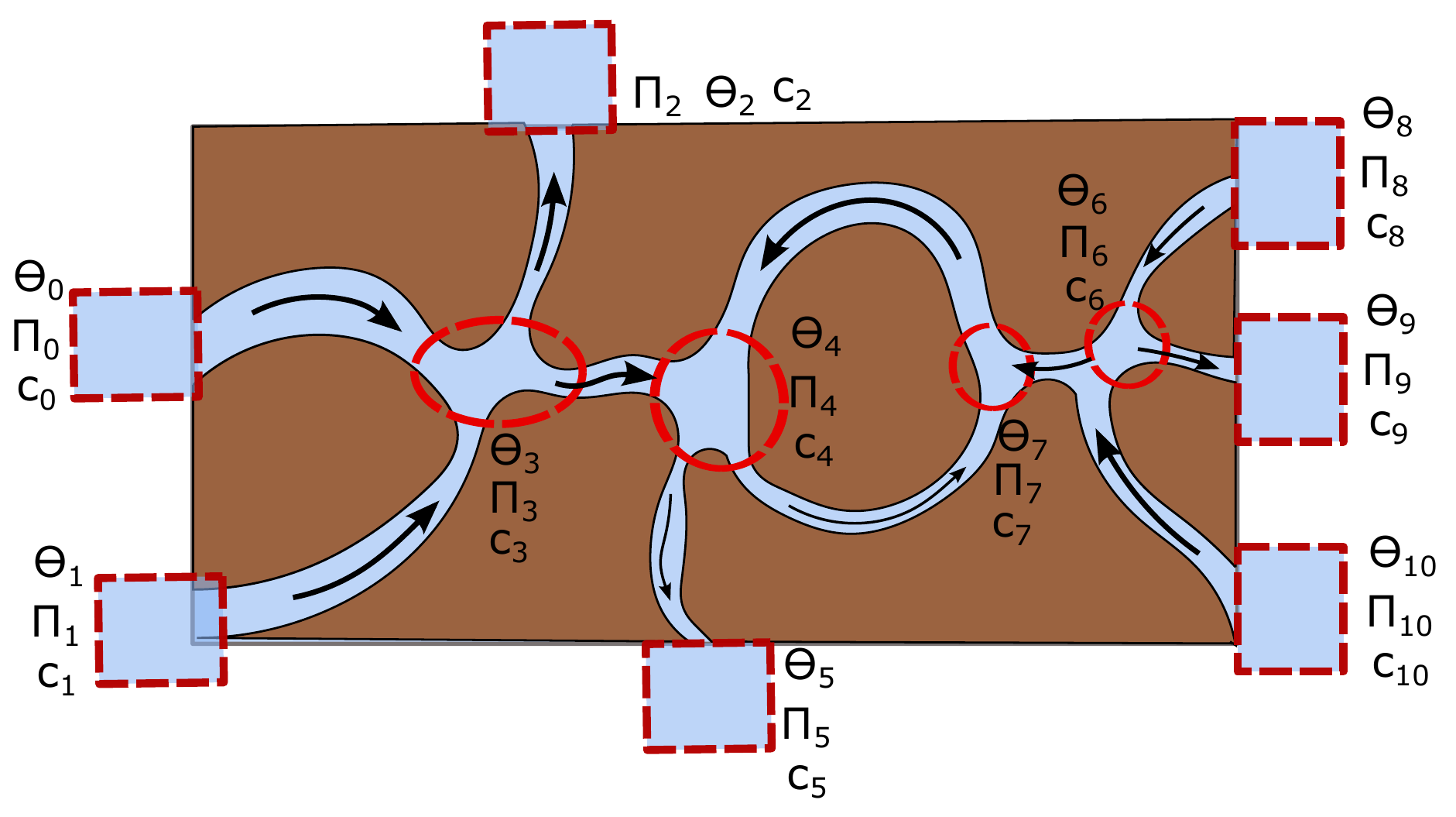}}
\caption{Schematic of an arbitrary network of pores and connecting reservoirs. The terminal reservoirs are shown in rectangles and internal reservoirs are marked by red ellipses. Depending on the characteristics of the reservoirs, their virtual total pressure ($\pi_i$), electro-chemical potential ($\theta_i$), or/and concentration ($c_i$) may be unknowns.}
\label{fig:network_ext}
\end{figure}

We model any porous structure as a network of small aspect ratio pores that are connected via small internal reservoirs. This can be viewed as a graph whose edges represent the pores and the nodes are the connecting reservoirs. We classify the reservoirs into two groups of terminal reservoirs and internal reservoirs. Terminal reservoirs refer to the end reservoirs with known concentration, however, the value of pressure and potential may be unknown (instead the flow rate and/or current may be specified). Internal reservoirs are the interfacial elements connecting two or more adjacent pores. In these reservoirs, the values of $\overline{C^-}$, $P_0$, and $\mu^+$ are all unknown. Figure(\ref{fig:network_ext}) shows a schematic of a pore network with different types of reservoirs. The terminal reservoirs are shown with rectangles and internal reservoirs are highlighted by red ellipses. Note that pores can still have variable cross-sections and surface charge densities along their axes. 
In order to determine $P_0$ and $\mu^+$ fields in the network, equations (\ref{eq:cont}) and (\ref{eq:current}) for each pore must be accompanied with proper boundary conditions in the connecting reservoirs. The procedure below describes how conservation laws at the reservoirs are utilized to determine the boundary conditions.\\

To obtain the boundary conditions, first we solve equations (\ref{eq:cont}) and (\ref{eq:current}) inside each pore for three different conditions mentioned in Table (\ref{tbl:PMu_solns}). The superposition of these three solutions can satisfy any arbitrary boundary condition for that pore.
\begin{table}[h]
  \caption{Three B.C. for solving equations (\ref{eq:cont}) and (\ref{eq:current})}
  \label{tbl:PMu_solns}
  \centering
  \begin{tabular}{llll}
    \hline
    Case  & Left B.C.  & Right B.C. & right hand side (RHS) terms\\
    \hline
    1        & $P_{0L} = \mu^+_L = 0$   & $P_{0R}=  \mu^+_R = 0$   & $\text{RHS}_1 = \frac{d}{dx}C_1(x)\frac{d C_0}{d x}$\\
              &                                           &                                            & $\text{RHS}_2 = \frac{d}{dx}C_2(x)\frac{d C_0}{d x}$\\
              &                                           &                                            &                                                                                     \\
    2        & $P_{0L} = \mu^+_L = 0$   & $P_{0R}= 1$, $ \mu^+_R = 0$   &  $\text{RHS}_1 = \text{RHS}_2 = 0$\\
              &                                           &                                                    &                                                          \\
    3        & $P_{0L} = \mu^+_L = 0$   & $P_{0R}= 0$, $ \mu^+_R = 1 $  &  $ \text{RHS}_1 = \text{RHS}_2 = 0$\\
    \hline
  \end{tabular}
\end{table}

The first set of boundary conditions results in a solution, denoted by $P_0^{(1)}(x)$ and $\mu^{+(1)}(x)$, which are the fields due to internal diffusio-osmotic effect present as source terms in RHS of equations (\ref{eq:cont}) and (\ref{eq:current}) for each pore. Case 2 is the solution induced by the nonzero pressure driven force, denoted by $P_0^{(2)}(x)$ and $\mu^{+(2)}(x)$. Case 3 represents the fields $P_0^{(3)}(x)$ and $\mu^{+(3)}(x)$ generated by the nonzero electro-chemical potential force. Using the linearity of equations (\ref{eq:cont}) and (\ref{eq:current}) with respect to $P_0$ and $\mu^+$, the general solution is written by the linear combination of the three solutions introduced above:
\begin{equation}
P_0 = P_0^{(1)} + \pi_L + (\pi_R - \pi_L)P_0^{(2)} + (\theta_R - \theta_L)P_0^{(3)},
\label{eq:p0}
\end{equation}
\begin{equation}
\mu^+ = \mu^{+(1)} + \theta_L + (\pi_R - \pi_L)\mu^{+(2)} + (\theta_R - \theta_L)\mu^{+(3)},
\label{eq:mu}
\end{equation}
where  $\pi_L$, $\pi_R$, $\theta_L$, and $\theta_R$ are correct, yet unknown, boundary conditions indicating $P_0$ and $\mu^+$ in the left and right reservoirs respectively. Substituting relations (\ref{eq:p0}) and (\ref{eq:mu}) in the area-integrated flow rate $Q = S\bar{u}$ and area integrated current $I = S\text{z}(\overline{F^+_x} -\overline{F^-_x} )$ yields:
\begin{equation}
Q = Q^{(1)} + (\pi_R - \pi_R)Q^{(2)} + (\theta_R - \theta_L)Q^{(3)},
\label{eq:Q}
\end{equation}
\begin{equation}
I = I^{(1)} + (\pi_R - \pi_R)I^{(2)} + (\theta_R - \theta_L)I^{(3)},
\label{eq:I}
\end{equation}
where $Q^{(1)}$ and $I^{(1)}$ are the flow rate and the electric current resulting from fields, $P_0^{(1)}$ and $\mu^{+(1)}$. $Q^{(2)}$ and $I^{(2)}$ are generated by $P_0^{(2)}$ and $\mu^{+(2)}$ fields, and $Q^{(3)}$ and $I^{(3)}$ are caused by $P_0^{(3)}$ and $\mu^{+(3)}$. Using equations (\ref{eq:Q}) and (\ref{eq:I}), we impose the conservation of mass and charge for the reservoirs whose virtual total pressures and electro-chemical potential are both unknown. If either the $\pi_i$ or $\theta_i$ is unknown at some reservoir $i$, we impose only one of the equations (\ref{eq:Q}) and (\ref{eq:I}) and use the known value in lieu of the second equation. Generally, when pressure is known, the flow rate constraint is relaxed, and when electro-chemical potential is known, the current constraint is relaxed. Finally, this results in a linear system of equations coupling the entire network of pores, by which we can solve for reservoir quantities, $\pi_i$'s, and $\theta_i$'s. Then, we can substitute these values into equations (\ref{eq:p0}) and (\ref{eq:mu}) to obtain the $P_0$ and $\mu^+$ fields inside each pore.\\

Once we obtain $P_0$ and $\mu^+$ throughout the porous network, we can compute the temporal evolution of ion species using equations (\ref{eq:area_pos_ion}) and (\ref{eq:area_neg_ion}). As mentioned before, since we have assumed local electroneutrality inside each pore, we solve only the transport of negative ions using equation (\ref{eq:area_neg_ion}) and obtain the temporal evolution of positive ions by imposing cross-sectional  electroneutrality, $\overline{C^+} = \overline{C^-} + C_s(x)$.\\

The concentrations of internal reservoirs are also determined by writing the mass conservation of negative ions for those elements. By considering the internal reservoir as a computational control volume, we can write the mass conservation of negative ions in integral form as follows:
\begin{equation}
\int_{V_r}\frac{dC^-}{dt} dV_r - \int_{S_r}\vec{F}^-.\vec{dS} = 0,
\label{eq:res_vol}
\end{equation}  
where $V_r$ and $S_r$ are the reservoir volume and surface area respectively. Using area-averaged quantities, we rewrite equation (67) in the following form: 
\begin{equation}
\frac{d \overline{C^-}}{d t} = \frac{1}{V_r}\sum_{\substack{\text{Intersecting} \\ \text{pores}}} \overline{F_i^-}S_i .
\label{eq:res_dC}
\end{equation}
We use equation (\ref{eq:res_dC}) to compute temporal evolution of reservoir concentrations. In Section \ref{sec:summary}, we explain in detail our numerical approach including the spatial discretization and the time advancement scheme used to solve the governing equations.

\section{Summary of the model and solution algorithm} \label{sec:summary}
We derived two sets of equations under the assumption that all the pores in the network are thin and long (small aspect ratio). The first set of equations that are solved in the wall normal direction are summarized in Table (\ref{tbl:eqlb_eqn}). 
\begin{table}[H]
  \caption{Equilibrium equations to be solved in transverse directions}
  \label{tbl:eqlb_eqn}
  \centering
  \begin{tabular}{lll}
    \hline
      &\centering Poisson-Boltzmann eqn.  & \\
      \hline
          & Equation                                                                                    & B.C.\\
           & $\nabla^2_{\perp}\psi =\frac{1}{\lambda_0^2} \sinh(\psi)$   & $\frac{\partial \psi}{\partial x_{\perp}}\vert_{\text{wall}}=\sigma^*(x)$\\
           &                                                                                              &                                                 \\
           & $\frac{C^{\pm}}{C_0(x)}= \exp(\mp \psi)$                             &         \_                                        \\
           &                                                                                              &                                                 \\
        
      \hline
           & \centering Velocity profile                                                     &                                                  \\
           \hline
           &  $\nabla^2_{\perp}g^p = 1$                                                 &   $g^p\vert_{\text{wall}} = 0$       \\
           &                                                                                             &                                                   \\
           &  $\nabla^2_{\perp}g^e = - \nabla^2_{\perp}\psi$                 &   $g^e\vert_{\text{wall}} = 0$       \\
           &                                                                                             &                                                   \\
           &  $\nabla^2_{\perp}g^c = 2\exp(\psi)$                                   &   $g^c\vert_{\text{wall}} = 0$       \\
    \hline
  \end{tabular}
\end{table}
We solve these equations for a wide range of $\sigma^*$ and $\lambda^*$, where $\lambda_0$ is obtained only as an intermediate variable via iterations. We use these solutions to compute the area-averaged coefficients summarized in Table (\ref{tbl:coef}).

\begin{table}[H]
  \caption{Area-averaged coefficients}
  \label{tbl:coef}
  \centering
  \begin{tabular}{llll}
    \hline
      Symbol & Relation  & Symbol & Relation\\
      \hline
      $\bar{g}$                   &  $\frac{1}{S}\int_S \text{exp}(\psi)ds$      &   $\overline{g^{e-}} $    &  ${\frac{\overline{g^{e \prime}C^{-\prime}}}{2\vert \sigma^*\vert \lambda_D^2}}$      \\
      &  &  & \\
      $\overline{g^p}$        & $ \frac{1}{S}\int_S g^p ds$                      &   $\overline{g^{c-}}$     &  ${\frac{\overline{g^{c \prime}C^{-\prime}}}{2\vert \sigma^*\vert \lambda_D^2}}$         \\
       &  &  & \\
      $\overline{g^e} $       &  $\frac{1}{S}\int_S g^e ds$                     &   $\overline{g^{p+}} $     &   ${\frac{\overline{g^{p \prime}C^{+\prime}}}{2\vert \sigma^*\vert \lambda_D^2}}$        \\
       &  &  &\\
      $\overline{g^c} $       &  $\frac{1}{S}\int_S g^c ds$                           &   $\overline{g^{e+}}$  & ${\frac{\overline{g^{e \prime}C^{+\prime}}}{2\vert \sigma^*\vert \lambda_D^2}}$          \\
       &  &  &\\
      $\overline{g^{p-}}$  &  ${\frac{\overline{g^{p \prime}C^{-\prime}}}{2\vert \sigma^*\vert \lambda_D^2}} $                     &  $\overline{g^{c+}}$      &   ${\frac{\overline{g^{c \prime}C^{+\prime}}}{2\vert \sigma^*\vert \lambda_D^2}}$         \\
     \hline
    \end{tabular}
\end{table}
Once the area-averaged coefficients are tabulated and stored, we use these coefficients to solve the second set of time-dependent equations in the pore axial directions. These equations are summarized in Table (\ref{tbl:axial_eqn}). First, the equations representing the conservation of mass and charge are solved to determine $P_0$ and $\mu^+$ throughout the network. The mathematical procedure used to obtain the correct boundary conditions and the corresponding pressure and potentials fields were explained in Section \ref{sec:network}. We then use the area-integrated transport equation to calculate the temporal evolution of $\overline{C^-}$ inside pores and internal reservoirs. \\ 
\begin{table}[h]
  \caption{Area integrated equations to be solved in axial direction}
  \label{tbl:axial_eqn}
  \centering
  \begin{tabular}{lll}
    \hline
      &\centering Conservation of fluid flow  and ion current& \\
      \hline
      & Equation                                                                                    & B.C. - I.C.\\
      & $\frac{d}{d x}(A_1(x) \frac{d P_0}{d x}) + \frac{d}{d x}(B_1(x) \frac{d \mu^+}{d x})= \frac{d}{d x} (C_1(x)\frac{d C_0}{d x})$  & $\pi_L$, $\theta_L$\\
      &                                                                                                   &                                                 \\
      & $\frac{d}{d x}(A_2(x) \frac{d P_0}{d x}) + \frac{d}{d x}(B_2(x) \frac{d \mu^+}{d x})= \frac{d}{d x} (C_2(x)\frac{d C_0}{d x})$  & $\pi_R$, $\theta_R$\\
      &                                                                                                   &                                                 \\
        
      \hline
      & \centering Area integrated transport of negative ion                 &                                                  \\
       \hline
      &  $S\frac{\partial \overline{C^-}}{\partial t} +\frac{\partial}{\partial x} S\{\overline{uC^-} -2\frac{\overline{C^-}}{C_0} \frac{d C_0}{dx} + \overline{C^-}\frac{d \mu^+}{d x}\}=0$                                                         &   $\overline{C^-_L}$,  $\overline{C^-_R}$     \\
      &                                                                                                                     &   $\overline{C^-}(t=0)$                                                 \\
      \hline
  \end{tabular}
\end{table}

We now describe our numerical strategy to solve the aforementioned equations. In the development of our numerical model, we used the finite volume method. For the pore cross-sectional geometry, we considered two options: cylindrical pores and rectangular pores with wide and thin cross-sections, where the effects of sidewalls can be ignored. For these pores, the equilibrium solution can be represented as a 1D profile and thus, the transverse problem reduces to a 1D ODE. To solve these equations in the transverse directions, we used uniformly spaced cells either in the wall normal direction of a pore with rectangular cross-section or in the radial direction for a circular pore. The unknowns, $\psi$, $\frac{C^{\pm}}{C_0}$, $g^p$, $g^e$, and $g^c$ are located at cell centers. The Laplace operator in equilibrium equations are discretized with second order accuracy. In the Poisson-Boltzmann equation (\ref{eq:final_boltzmann}, \ref{eq:psi_eqn}), we solve for a correction, $\delta \psi$, via an iterative procedure.  As shown in \ref{app:poisson solver}, this iteration is crucial to properly handle the nonlinearity of the Poisson-Boltzmann equation and improve the stability and convergence rate of the linear solver. Moreover, since the tabulation parameters are $\sigma^*$ and $\lambda^*$, some iterations are required in order to find the corresponding value of $\lambda_0$, which is the parameter present in the Poisson-Boltzmann equation. Once we obtain the correct solution of $\psi$, we can solve for transverse dependent coefficients of velocity field ($g^e$ and $g^c$) using the implemented linear solver. To compute integral quantities present in the definition of area-averaged coefficients, we have used the second order rectangle/mid-point rule. We conducted a convergence study to verify the independence of the numerical solutions from the size of the computational cells. Once the area-averaged coefficients are computed over a wide range of $\sigma^*$ and $\lambda^*$ values, they are tabulated with respect to these parameters. Given the smooth dependence on $\sigma^*$ and $\lambda^*$, these input variables are spaced logarithmically, such that every factor of 10 in $\sigma^*$ and $\lambda^*$ is resolved by 10 points.\\

To solve the 1D transport equations (Table (\ref{tbl:axial_eqn})), we used uniformly spaced cells along the pore axes. Note that since the pore area cross-section can vary along the axis, the cell volumes may not be equal to each other.  All area-averaged coefficients and the driving potential fields, $P_0$ and $\mu^+$, and $\overline{C^-}$ are defined at cell centers, whereas we compute the area-averaged velocity and ion flux components at cell faces. At each time step, after updating $\overline{C^-}$ throughout the porous network, we extract the area-averaged coefficients by applying bi-linear interpolation in the developed table using the local values of $\sigma^*$ and $\lambda^*$ specified at cell centers. Then, we use second order differentiation and linear interpolation in order to approximate the derivatives and the area-averaged quantities at cell faces. For each reservoir, we consider one computational element whose volume, $\sigma^*$, and $\lambda_D$ are set to the values that belong to the largest adjacent cell. The inlet and outlet surface areas of each interface with intersecting pores are equal to the cross-section areas of the pores at their connecting points with the interface. This choice of volume brings better numerical stability when solving for $\overline{C^-}$ of internal reservoirs. The spatial variation of driving potentials, $P_0$ and $\mu^+$, are obtained by solving the conservation equations (\ref{eq:cont}) and (\ref{eq:current}) leading to a penta-diagonal matrix system that is solved by a direct linear solver for banded matrices. For the time integration of equation (\ref{eq:area_neg_ion}), we implemented a second order semi-implicit scheme to enhance the numerical stability of the developed model. The implicit time discretization of the area-integrated transport equation for negative ions is:
\begin{equation}
\frac{3\overline{C^-}^{(n+1)} - 4\overline{C^-}^{(n)}+ \overline{C^-}^{(n-1)}}{2\Delta t} = \frac{-1}{S} \frac{\partial}{\partial x}\{S\overline{F^-_x}^{(n+1)}\} + O(\Delta t^2), \label{eq:discrete_transport}
\end{equation}
where superscript $(n+1)$ refers to the quantities in the next time step. $(n)$ and $(n-1)$ correspond to the quantities at the current time step and the previous time step respectively. $\overline{F^-_x}^{(n+1)}$ is the area-averaged flux of negative ions at moment $(n+1)$, which includes the advection, diffusion, and electromigration effects as described in equation (\ref{eq:area_neg_ion}). The existence of stiff terms such as the diffusion term restricts the time step to small values, which can dramatically slow down the time-integration process and convergence to the steady state solution. To eliminate such limitations, we devised a fast implicit procedure, which is explained in detail in \ref{app:asymptotic} and \ref{app:implicit solver}. For initial condition, in most cases it coincides with the time when a driving force is applied. At the very first time step, one can assume equilibrium over the entire system (uniform $C_0$, $P_0$, and $\mu^+$). Therefore, we often choose initial conditions for $\overline{C^-}$ such that they correspond to uniform $C_0$.\\

To demonstrate the capabilities of the developed model, we present series of canonical and engineering problems in the next session, and show how our model predicts a wide range of phenomena relevant to electrokinetic porous networks.\\

\section{Numerical results}
\subsection{Osmosis phenomenon and its connection to pore size and surface charge}
\begin{figure}[h]
\makebox[\textwidth][c]{\includegraphics[width=0.6\textwidth]{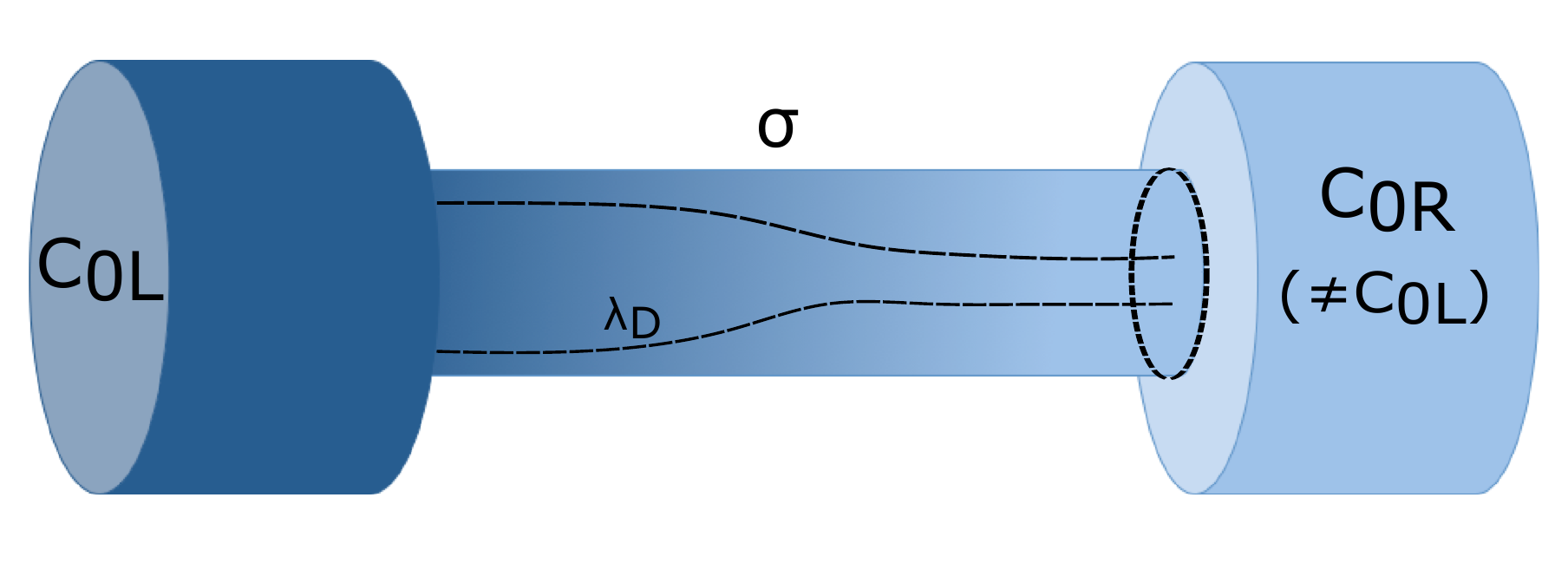}}
\caption{Schematic of a single pore with surface charge density of $\sigma$ connecting two closed reservoirs with different salt concentrations. The thickness of EDL varies across the system due to the existing salt concentration gradient. There is no externally imposed electric potential or pressure gradient. Dark blue refers to the zones with higher salt concentration and light blue region shows low salt concentration.}
\label{fig:osmosis_singlepore}
\end{figure}

\begin{figure}[h]
\makebox[\textwidth][c]{\includegraphics[width=0.55\textwidth]{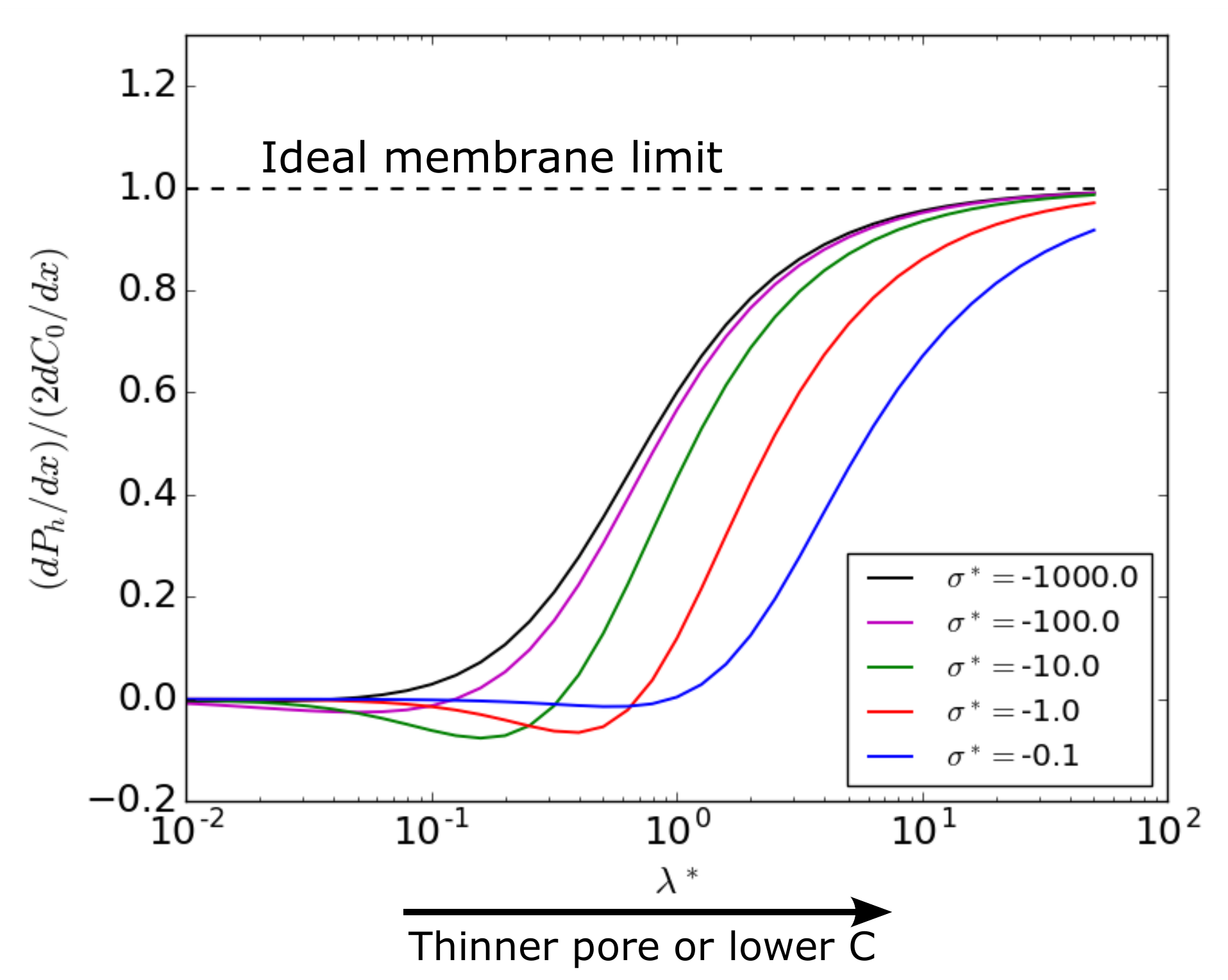}}
\caption{The ratio of virtual hydrodynamic pressure gradient and osmotic pressure gradient vs. $\lambda^*$ for different values of $\sigma^*$ and $\kappa=0.5$. At steady state, as pore EDLs become highly overlapped and pore surface charge density increases, the osmotic pressure is fully recovered.}
\label{fig:osmosispressure}
\end{figure}

Consider the schematic of a single pore whose ends lead to the closed reservoirs with known and unequal salt concentrations, as shown in Figure (\ref{fig:osmosis_singlepore}). The left reservoir contains higher salt concentration compared to the right reservoir. Given the dead-end flow condition, one expects the imposed diffusio-osmosis flow to be blocked by an induced pressure gradient. However, the system is also coupled to induce electric current via charge advection by both pressure driven and diffusio-osmotic flows, as described in equation (\ref{eq:cont}). Another required condition is that the net current through the system must be zero. Therefore, to predict the induced pressure difference, one needs to solve equations (\ref{eq:cont}) and (\ref{eq:current}) in a coupled fashion. Given zero flow and current, one can first integrate these equations to obtain equations for gradients of $P_0$ and $\mu^+$: 
\begin{equation}
A_1(x) \frac{d P_0}{d x} + B_1(x) \frac{d \mu^+}{d x} - C_1(x)\frac{d C_0}{d x} = 0,
\end{equation}
\begin{equation}
A_2(x) \frac{d P_0}{d x} + B_2(x) \frac{d \mu^+}{d x} - C_2(x)\frac{d C_0}{d x} = 0.
\end{equation}
Eliminating $\frac{d \mu^+}{d x}$ from the equations above, one can obtain $\frac{d P_0}{d x}$ in terms of $\frac{d C_0}{d x}$:
\begin{equation}
\frac{d P_0}{d x} = \frac{C_1 B_2 - C_2 B_1}{A_1 B_2 - A_2 B_1} \frac{d C_0}{d x}.
\end{equation}
Using equation (\ref{eq:finalP}), we can write the virtual total pressure as the superposition of virtual hydrodynamic pressure ($P_{h0}$) and virtual osmotic pressure ($2C_0$). The following relation is eventually obtained between the virtual hydrodynamic pressure gradient and the virtual osmotic pressure gradient:
\begin{equation}
\frac{d P_{h0}}{d x} = (\frac{1}{2} \frac{C_1 B_2 - C_2 B_1}{A_1 B_2 - A_2 B_1} + 1) \frac{d (2 C_0)}{d x},
\end{equation}
where the pre-factor is only a function of electro-hydrodynamic coupling parameter ($\kappa$) and tabulated area-averaged coefficients. Figure (\ref{fig:osmosispressure}) depicts the variation of this pre-factor with respect to $\lambda^*$ for different values of $\sigma^*$. Here we considered $\kappa=0.5$, which is relevant to aqueous electrolytes. The trend indicates the osmotic pressure is fully recovered as the pore EDLs get highly overlapped (thinner pore).\\

While in thermodynamics the same relation is obtained from macroscopic energy analysis for membranes with thin pores, the significance of the present result is that  it provides a description of the osmotic pressure via direct connection to the microscopic force balance relations. With this description, we not only capture the ideal membrane limit, but also quantify the non-ideal behavior when the pore size is finite. Additionally, the present result indicates a tradeoff  between pore size and pore charge, which can be utilized in designing perm-selective material. For example, Figure (\ref{fig:osmosispressure}) depicts that for highly charged pores, one can achieve roughly the same performance as in a low-charge pore, by choosing a pore size that is about 10 times thicker. This analysis provides quantitative guidelines for designing new membrane material that have lower viscous friction while maintaining other performance measures. 
\subsection{Deionization shock propagation through a micro-pore } \label{subsec:shock_propagation}
\begin{figure}[h]
\makebox[\textwidth][c]{\includegraphics[width=0.57\textwidth]{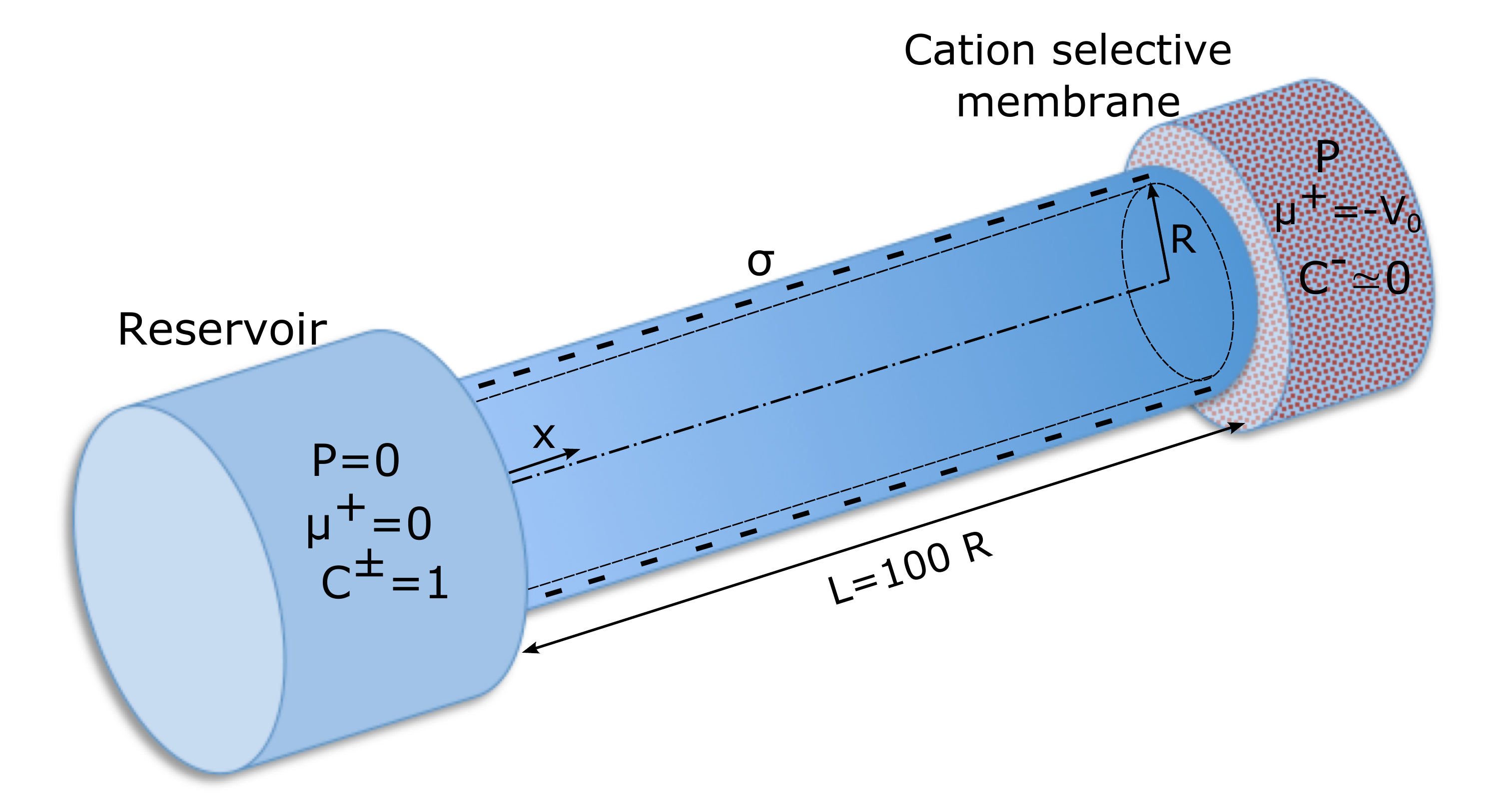}}
\caption{An axisymmetric normalized length micro-pore connecting a uniform reservoir to a cation selective membrane. The membrane enforces zero net flow and zero flux for anions.}
\label{fig:single_pore}
\end{figure}

To validate our numerical solution, we compared the numerical results from our reduced model with multi-dimensional direct numerical simulations of single micro-pore implemented by Nielsen et al. \cite{nielsen2014}. As shown in Figure (\ref{fig:single_pore}), the geometry of interest is a circular pore with normalized axial length. At the right boundary the pore is blocked by a cation selective membrane, which imposes zero area-averaged flux of negative ion species as well as zero net flow, $\bar{u}=0$. The electro-chemical potential is known at this end, however, the pressure is to be determined such that it results in the zero net flow through the system.  At the other side, the pore is connected to a large reservoir with known uniform salt concentration, pressure, and electro-chemical potential.  

In order to make a rational comparison to direct simulation, we looked at the results in Nielsen et al. obtained for the smallest value of aspect ratio ($R/L=0.01$). Their dimensionless quantities for the case we considered here were normalized Debye length $\lambda_{\text{ref}}/R=0.01$ and averaged charged density in the pore cross-section, $C_s=-0.1$. The equivalent non-dimensional values in our model are equal to $\lambda_D= \lambda_{\text{ref}}/h_p=\lambda_{\text{ref}}/(R/2) = 0.02$ and $\sigma^*=-C_s/(2\lambda_D^2)=-125$. The electrohydrodynamic coupling constant, $\kappa$ was set to 0.235 to match the corresponding parameter in the direct simulation. The direct simulation by Nielson et al. was conducted by solving steady state Poisson-Nernst-Planck-Stokes problem using the commercial multiphysics software COMSOL. In contrast, our mathematical model solves transient area-integrated transport equation and returns the temporal evolution of the system until it reaches steady state, when the deionization shock arrives close at the interface of the reservoir with uniform salt concentration. To resolve an initially thin shock front and avoid numerical oscillations in the concentration profile, we fully resolved the spatial domain with $n_x=800$ computational cells and chose $\Delta t = 10^{-4}$ for the time integration. We used the steady state quantity of the total electric current for different electro-chemical potential gradients applied to generate the I-V response of the system.\\ 

\begin{figure}[h]
\makebox[\textwidth][c]{\includegraphics[width=0.7\textwidth]{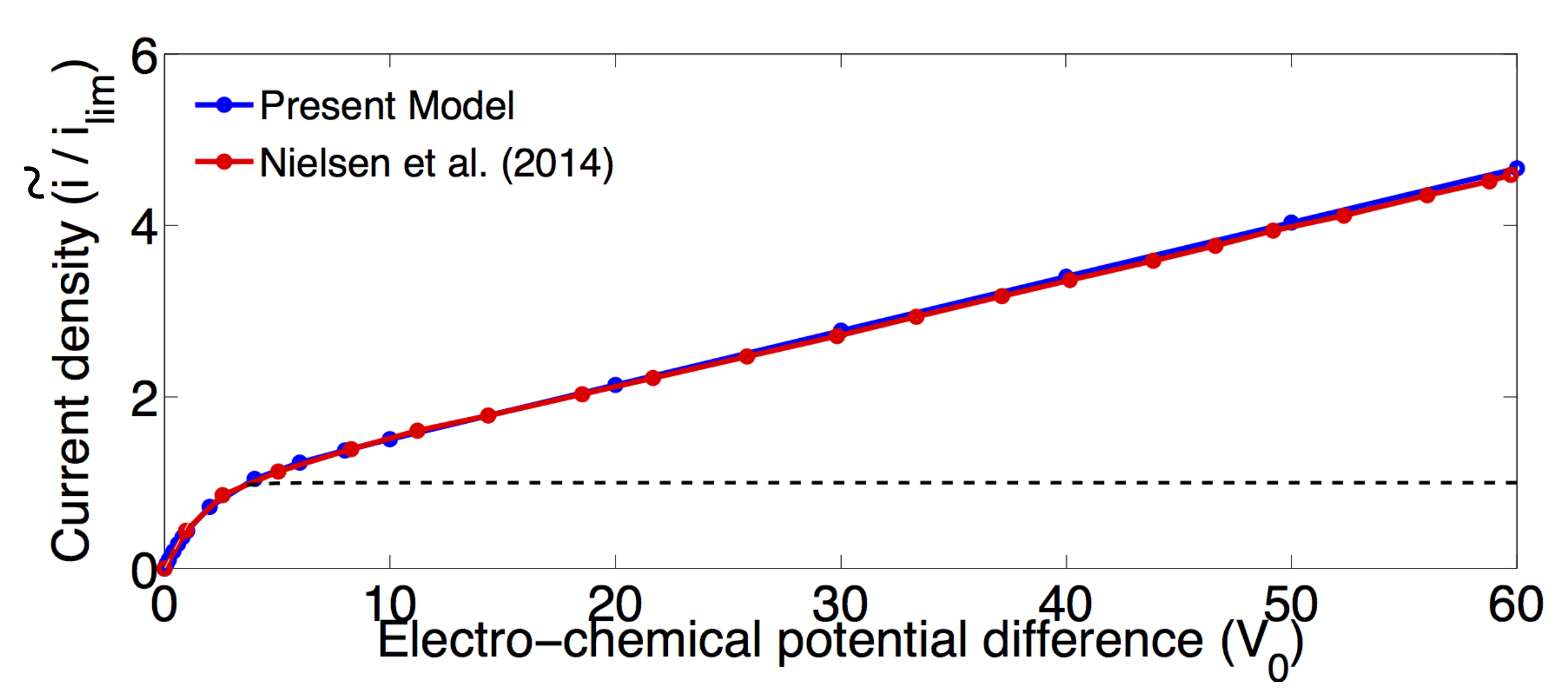}}
\caption{Normalized I-V characteristics for a cylindrical dead-end micro-pore with $\lambda_D=0.02$, $\sigma^*=-125$, and $\kappa=0.235$. The current densities are normalized by limiting current density, $i_{\text{lim}} = \frac{2zeDC_{\text{ref}}}{L}$, which is twice our model normalization factor. The dashed line represents the theoretical limiting current corresponding to solution with no surface charge effect and zero $\lambda_D$. }
\label{fig:iv}
\end{figure}

The I-V characteristics that we obtained for this geometry is depicted in Figure (\ref{fig:iv}) with the blue curve. The red curve represents the results from the full simulation by Nielsen et al. The curves are in good agreement over the entire range of applied voltages, corresponding to classically known ohmic regime $V<V_T$, and overlimiting regime. In this case, an overlimiting current with the slope of $di/dV=0.063$ is observed due to the surface conduction mechanism.

\begin{figure}[h]
\makebox[\textwidth][c]{\includegraphics[width=0.65\textwidth]{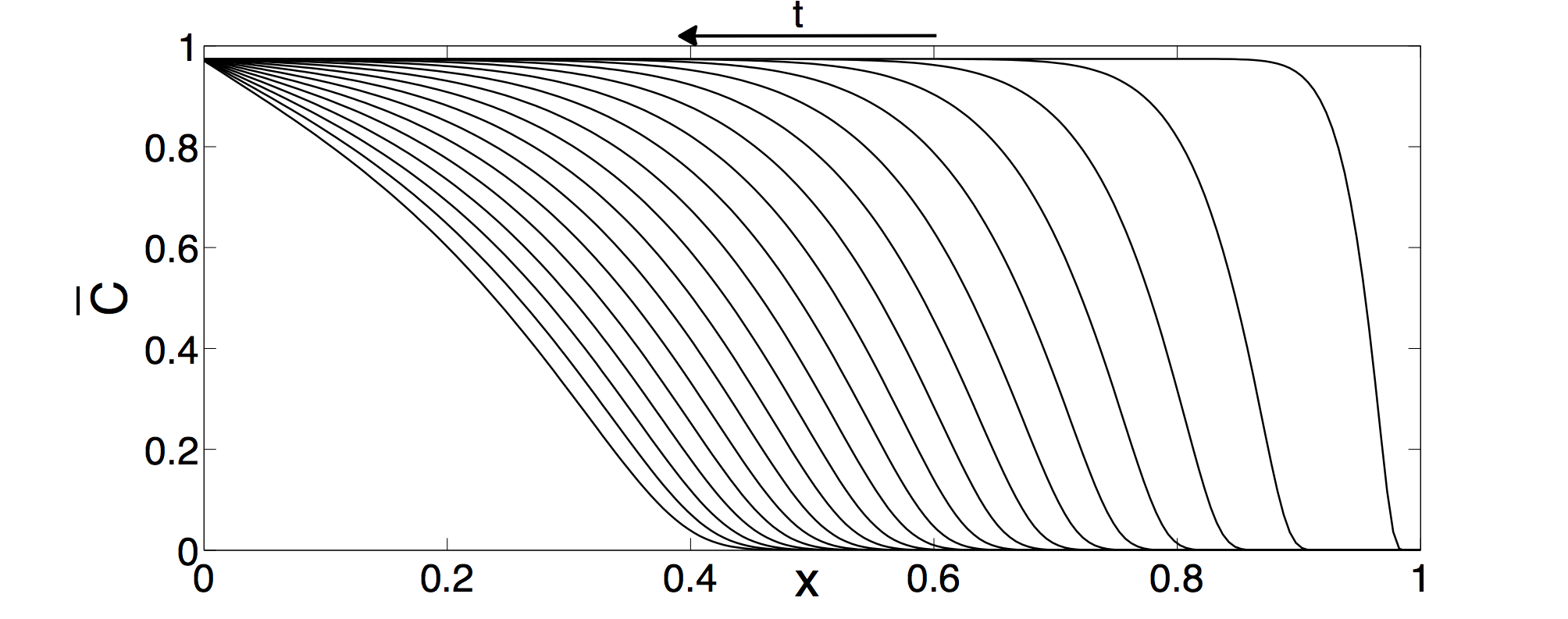}}
\caption{Propagation of a deionization shock in a micropore. Depicted is the area-averaged concentration profile versus x, which are uniformly sampled in time. The system parameters are  $\lambda_D=0.02$, $\sigma^*=-125$, and $\kappa=0.235$. The external potential gradient provoking the nonlinear dynamics is $V/V_T=60$.}
\label{fig:propag}
\end{figure}

\begin{figure}[h]
        \centering
        \begin{subfigure}[H]{0.57\textwidth}
                \includegraphics[width=\textwidth]{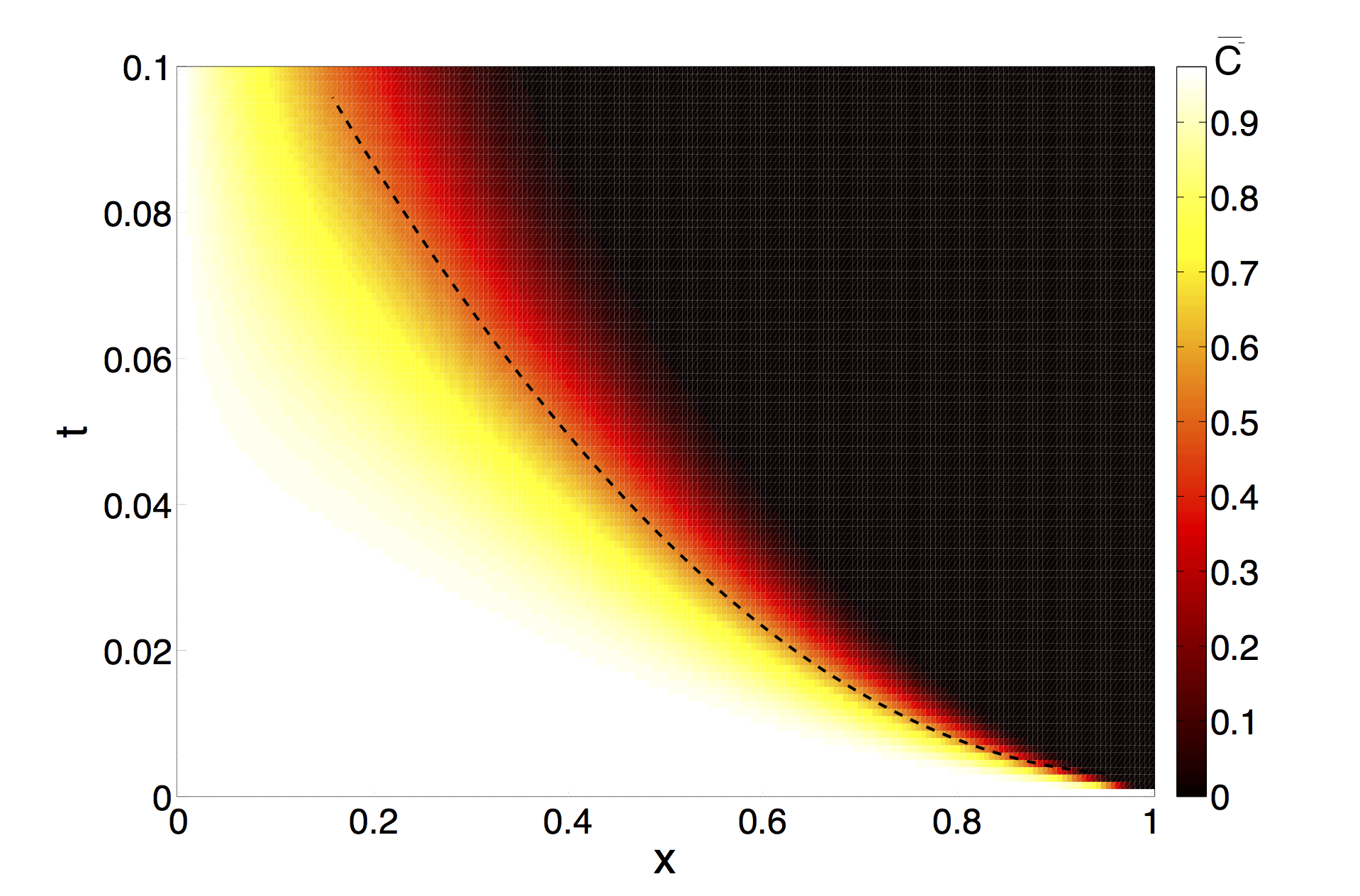}
                \caption{}
                \label{fig:space_temp_1}
        \end{subfigure}%
        ~ 
        \begin{subfigure}[H]{0.47\textwidth}
                \includegraphics[width=\textwidth]{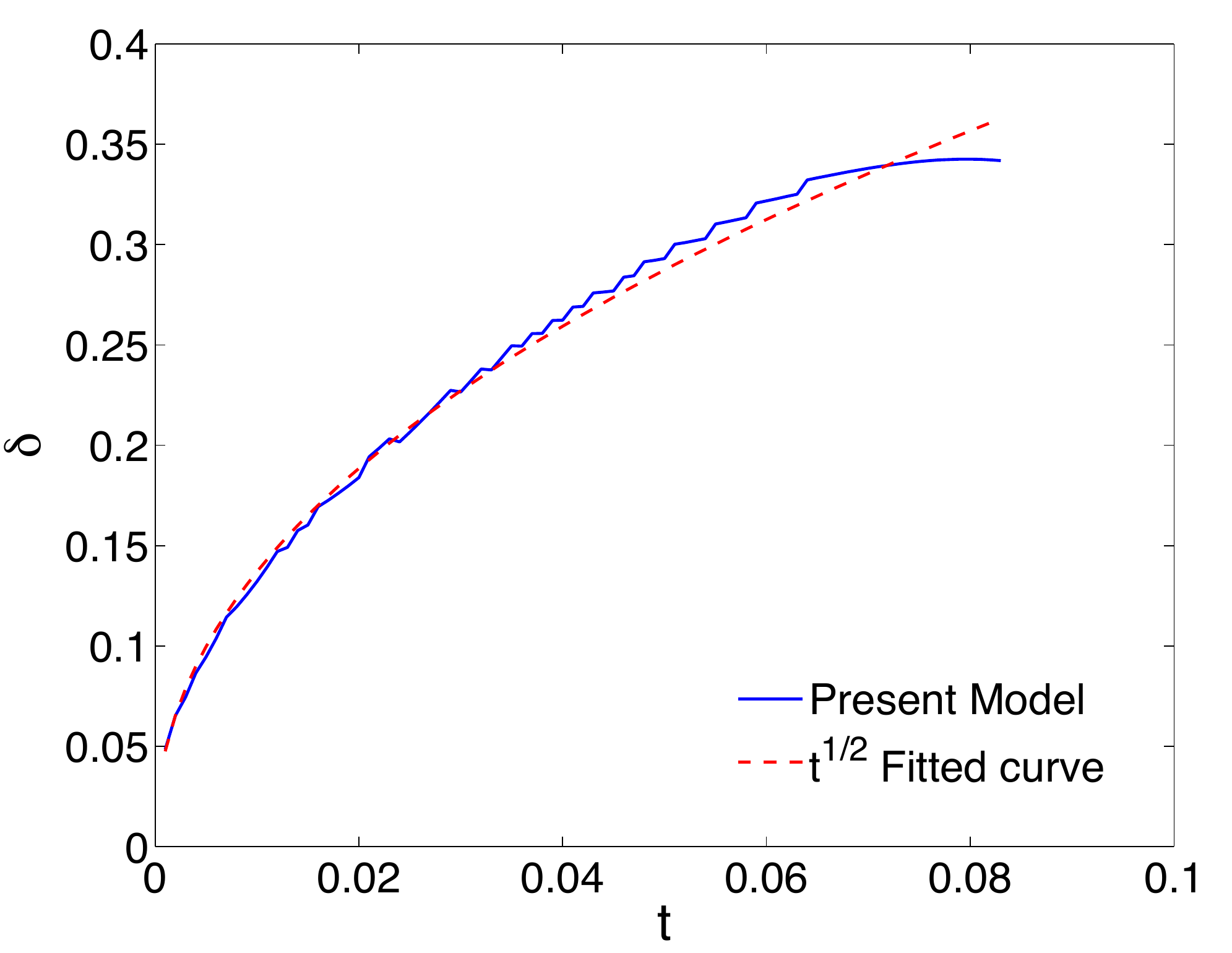}
                \caption{}
                \label{fig:space_temp_2}
        \end{subfigure}
        ~ 
         \caption{(a) Space-time evolution of area-averaged concentration of anions in a micropore. The system parameters are $\lambda_D=0.02$, $\sigma^*=-125$, and $\kappa=0.235$. (b) The variation of shock thickness vs. time and the fitted curve representing the $t^{1/2}$ scaling of the shock thickness.}
                \label{fig:deadend_temp}
 \end{figure}

Figure (\ref{fig:propag}) shows the transient solution of area-averaged anion concentration profile inside the micro-pore for which the applied voltage is $V/V_T=60$. Moreover, Figure (\ref{fig:space_temp_1}) presents the spatial and temporal evolution of negative ion concentration during the propagation of deionization shock. This figure confirms the $t^{1/2}$ scaling of shock distance from the membrane for early times under a constant voltage condition \cite{zangle2010}. The  $t^{1/2}$ scale has been shown by dashed line on this figure. We have shown the time variation of shock thickness until it reaches close to the end reservoir at steady state. The shock thickness is determined on each $\overline{C^-}$ vs. x curve as the distance between the intersections of the straight line that is tangent to $\overline{C^-}=0.5$ with lines $\overline{C^-}=0$ and $\overline{C^-}=1$. Once the shock thickness is computed, its time evolution was compared to the red curve representing the $t^{1/2}$ scaling in Figure (\ref{fig:space_temp_2}), which demonstrates the same growth rate as the one reported in the literature for early propagation times and\cite{manimartin2011}. 

\subsection{Microchannel-nanochannel-microchannel junction}
\begin{figure}[h]
\makebox[\textwidth][c]{\includegraphics[width=0.8\textwidth]{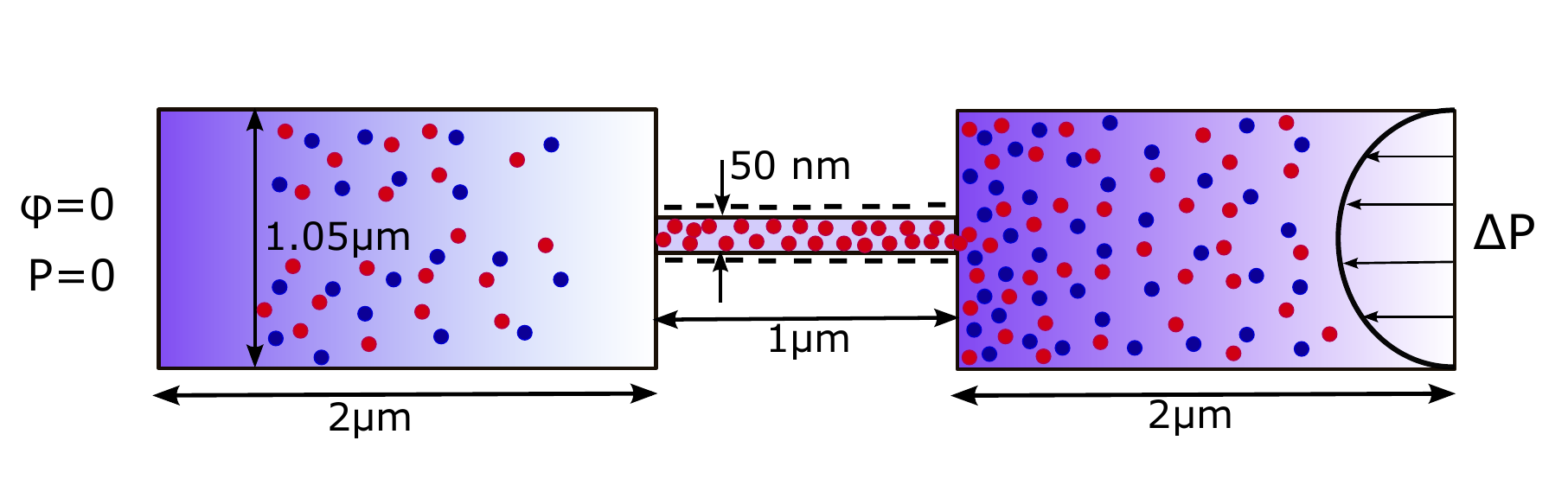}}
\caption{Schematic of a microfluidic pre-concentrator consisting of serial microchannel-nanochannel-microchannel. There is no charge on the microchannel surfaces, but the nanochannel possesses a fixed amount of negative charge on the surface. A constant pressure head is imposed, which leads to the development of streaming potential and the ICP phenomenon in the form of an enrichment zone in the right interface and a depletion zone in the left interface.}
\label{fig:mnm_schematic}
\end{figure}

In this section we present our results for a micro-nanofluidic pre-concentrator system operating based on hydrodynamic flow. As shown schematically in Figure (\ref{fig:mnm_schematic}), we consider a microchannel-nanochannel-microchannel system in which a negatively charged nano-channel bridges two uncharged micro-pores. We consider a specification adopted from the investigation of Wang et al.  \cite{wang2009} who performed full DNS of Poisson-Nernst-Planck-Stokes equations at steady state in multi-dimensions for this geometry. In their setting, the micro channels are $2 \mu m$ long and $1.05 \mu m$ thick. The nano channel is $1 \mu m$ long and $50 nm$ thick. In other words, their system involves elements that are not long and thin. This setting provides a useful scenario to assess the capabilities of our model when some of our derivation assumptions are relaxed. In our model, we obtained the area-averaged coefficients from the solutions to the Poisson-Boltzman system (Table (\ref{tbl:eqlb_eqn})) for a 2D geometry. This assumption is justified, given that the channel widths are much larger than their thickness.  The nanochannel includes a fixed surface charge density of $-2 mC/m^2$. The open ends of the microchannels  are connected  to the reservoirs filled with potassium chloride aqueous solution with the concentration of $0.1 mM$. According to Wang et al. \cite{wang2009}, we model the electrolyte using the following properties: $\varepsilon=7.08e-10 F/m$, z=1, $D=2e-9 m^2/s$ and, $\mu=8.9e-4Pa$ $s$. The system operates under a fixed pressure head $\Delta p = 0.1 atm$ imposed from the right end. The electric potential in the left reservoir is zero, but it is relaxed in the right reservoir as the induced streaming potential grows. To enforce this condition, we impose a net zero current since the system is open-loop. We let the system evolve until the streaming potential is fully established and the system reaches steady state. Below we present the results from our reduced order model and compare them against the DNS of the same pore system conducted by Wang et al.\\

\begin{figure}[h]
        \centering
        \begin{subfigure}[H]{0.53\textwidth}
                \includegraphics[width=\textwidth]{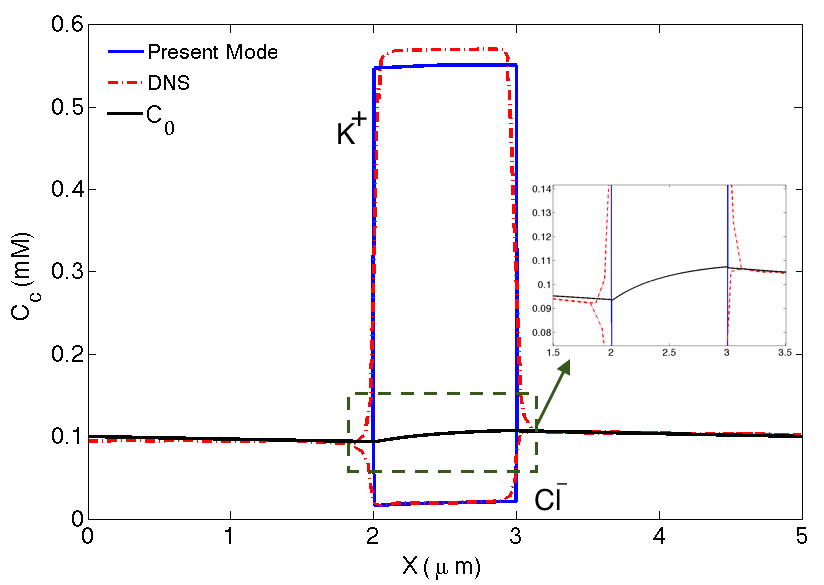}
                \caption{}
                \label{fig:concentration_x}
        \end{subfigure}%
        ~ 
        \begin{subfigure}[H]{0.52\textwidth}
                \includegraphics[width=\textwidth]{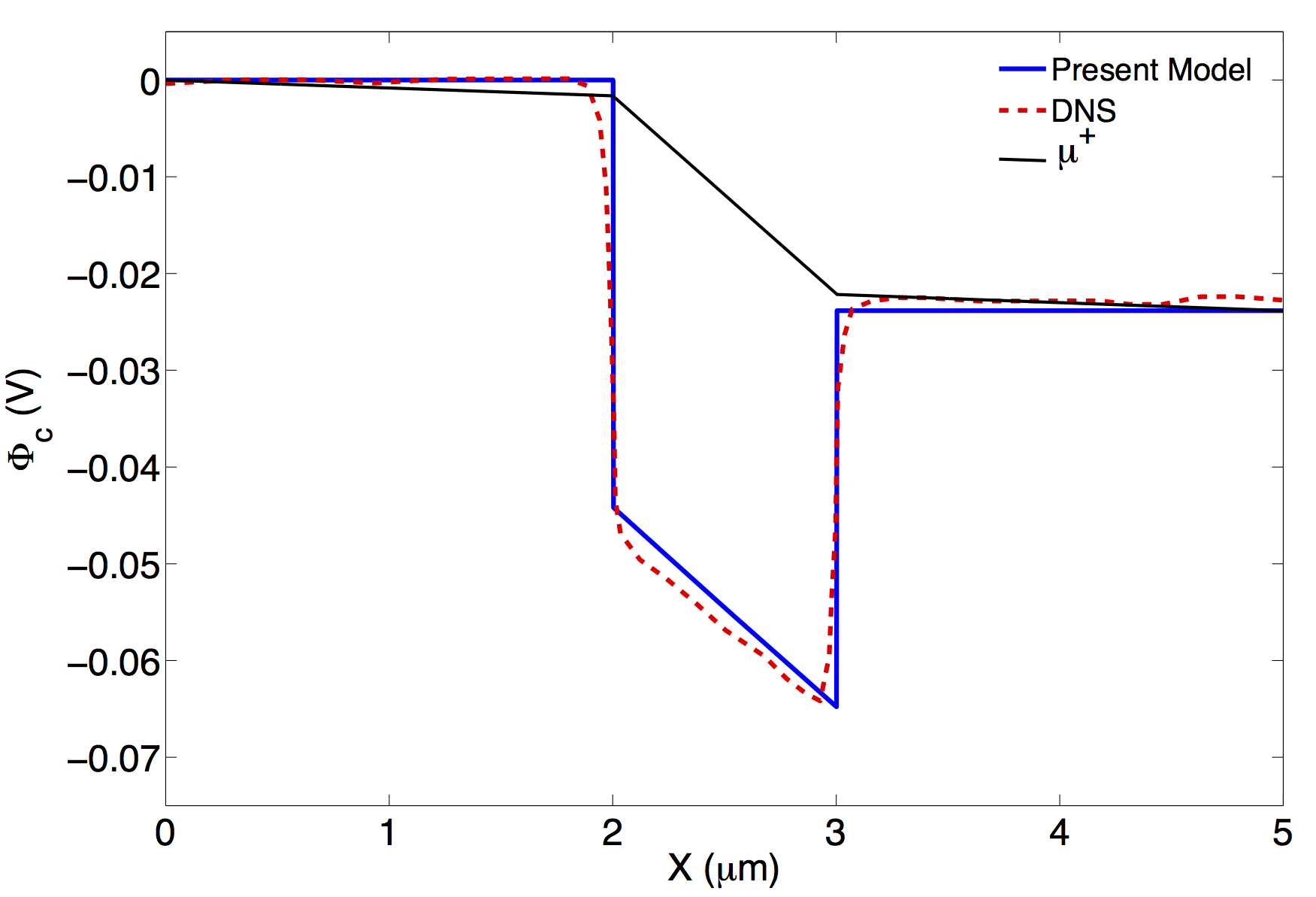}
                \caption{}
                \label{fig:potential_x}
        \end{subfigure}
        ~ 
         \caption{Spatial distribution of centerline ion concentration and electrostatic potential for a microchannel-nanochannel-microchannel system. (a) concentration of ions at the pore centerline versus x at steady state, computed by our numerical model (solid lines) as well as the direct simulation of Wang et al.  \cite{wang2009} (dashed lines). $C_0$ profile is continuous across the whole system, whereas the physical concentrations of ion species involve steep gradients at the interfaces of the nanochannel. (b) The centerline electric potential profile and the virtual electro-chemical potential along the pore axial direction.}
                \label{fig:deadend_temp}
 \end{figure}

According to equation (\ref{eq:final_boltzmann}), the centerline concentrations of species are equal to $C^{\pm} = C_0(x)\exp({\mp \psi_{\text{centerline}}})$, where $C_0(x)$ is given by the model.  $\psi_{\text{centerline}}(x)$ is the electrostatic potential that is obtained simply by the solution of the Poisson equation (\ref{eq:psi_eqn}) at the pore centerline. Thus, in addition to the 9 area-averaged coefficients, we also tabulated $\psi_{\text{centerline}}$ with respect to $\sigma^*$ and $\lambda^*$. Once we determine the values of $\psi_{\text{centerline}}$, we are able to compute the total electric potential at the centerline as well. Combining equations (\ref{eq:psi_def}) and (\ref{eq:finalMu}), one can obtain this potential as follows:
\begin{equation}
\phi_{\text{centerline}}(x) = \mu^+(x) - \ln(C_0(x)) + \psi_{\text{centerline}}(x),
\label{eq:phi_center}
\end{equation}
where $C_0$ and $\mu^+$ profiles are given by our model at steady state. Figure (\ref{fig:concentration_x}) compares the steady state concentration of ion species at the pore centerline obtained by our model against the DNS. Since there is no surface charge on the microchannel walls, the concentration of K$^+$ and Cl$^-$ are almost the same to sustain electroneutrality. However, as seen in Figure (\ref{fig:concentration_x}), in the nanochannel the concentration of  $K^+$ is higher than the concentration of Cl$^-$, which is necessary to screen the nanochannel negative surface charge. The high gap between the concentration of ion species is also related to the fact that there are highly overlapped EDLs formed inside the nanochannel, which prohibits the transmission of coions. Therefore, steep concentration gradients are observed at both interfaces of nanochannel with the microchannels.\\

The inset in Figure (\ref{fig:concentration_x}) magnifies the formation of ICP zones at the nanochannel interfaces. The negative ions carried by the hydrodynamic flow toward the right micro-nano interface are accumulated and do not move through the nanochannel due to its high negative surface charge. The concentration of positive ions also arises at this interface to maintain electroneutrality and hence, higher concentration of ions is built up at this interface. On the other hand, both ion species are carried away from the left interface leading to the formation of a low concentration zone. It should be noted that while there are high gradients in the ion concentrations at the centerline, the virtual concentration ($C_0$) is much closer to a uniform profile in the system. Our model and the DNS both show consistent results, which also match the underlying physics described above.\\

Figure (\ref{fig:potential_x}) shows the longitudinal variation of the electric potential at the centerline of the pores, which is computed by equation (\ref{eq:phi_center}). The potential is almost constant in the microchannels due to their low ohmic resistance compared to the nanochannel, but it varies almost linearly inside the nanochannel. While the virtual electro-chemical potential varies continuously in the system, the electric potential involves jumps at the micro-nano interfaces. These jumps lead to locally strong electric fields and significant electromigration fluxes, which are in balance with the large diffusion fluxes that result from the existing jump in the concentration profiles at the interfaces. In \cite{wang2009} these jumps were attributed to the Donnan potential. The Donnan model assumes uniform concentration of anions and cations across pore cross-section. Historically, this model has been assumed to be valid in the limit of thick EDL, $\lambda^* >> 1$. However, we point out that $|\sigma^*|= \frac{\text{z} e |\sigma| h_p}{k_B T \varepsilon} <<1$ is an additional necessary condition for the validity of the Donnan model. Here we provide a brief reasoning regarding the importance of the second condition. Given $\frac{\sigma}{\varepsilon}$ equals to the wall normal electric field, $\sigma^*$ represents the order of magnitude for the cross-sectional variation of electrostatic potential relative to the thermal voltage, specifically in the thick EDL limit. In other words, when $\lambda^* >> 1$ the variation of $\psi$ is on the order of $|\sigma^*|$. According to equation (\ref{eq:boltzmann_C}), this estimate implies that when $|\sigma^*|>1$, concentration profiles cannot be modeled by uniform profiles. Therefore, the validity of the Donnan model not only requires $\lambda^* >> 1$, but also it demands small values of $|\sigma^*|<<1$. In this case, $\sigma^*=-2.7$. Therefore, our model, which captures the cross-sectional profiles of ion concentration, can appropriately predict the area-averaged ion concentration in the nanochannel.\\

As indicated in the inset of Figure (\ref{fig:concentration_x}), the ICP effects are weak at the zones outside the nanochannel interfaces, as the overall magnitude of induced potential in the system are not large compared to the thermal voltage. The consistency of the results between our model and the DNS demonstrates that our model is able to capture the nonlinear response of the system with sufficient accuracy, even when the assumption of having low aspect ratio pores are relaxed for microchannels. 

\subsection{Flow recirculation in porous topologies involving loops}
\begin{figure}[H]
        \centering
        \begin{subfigure}[h]{0.42\textwidth}
                \includegraphics[width=\textwidth]{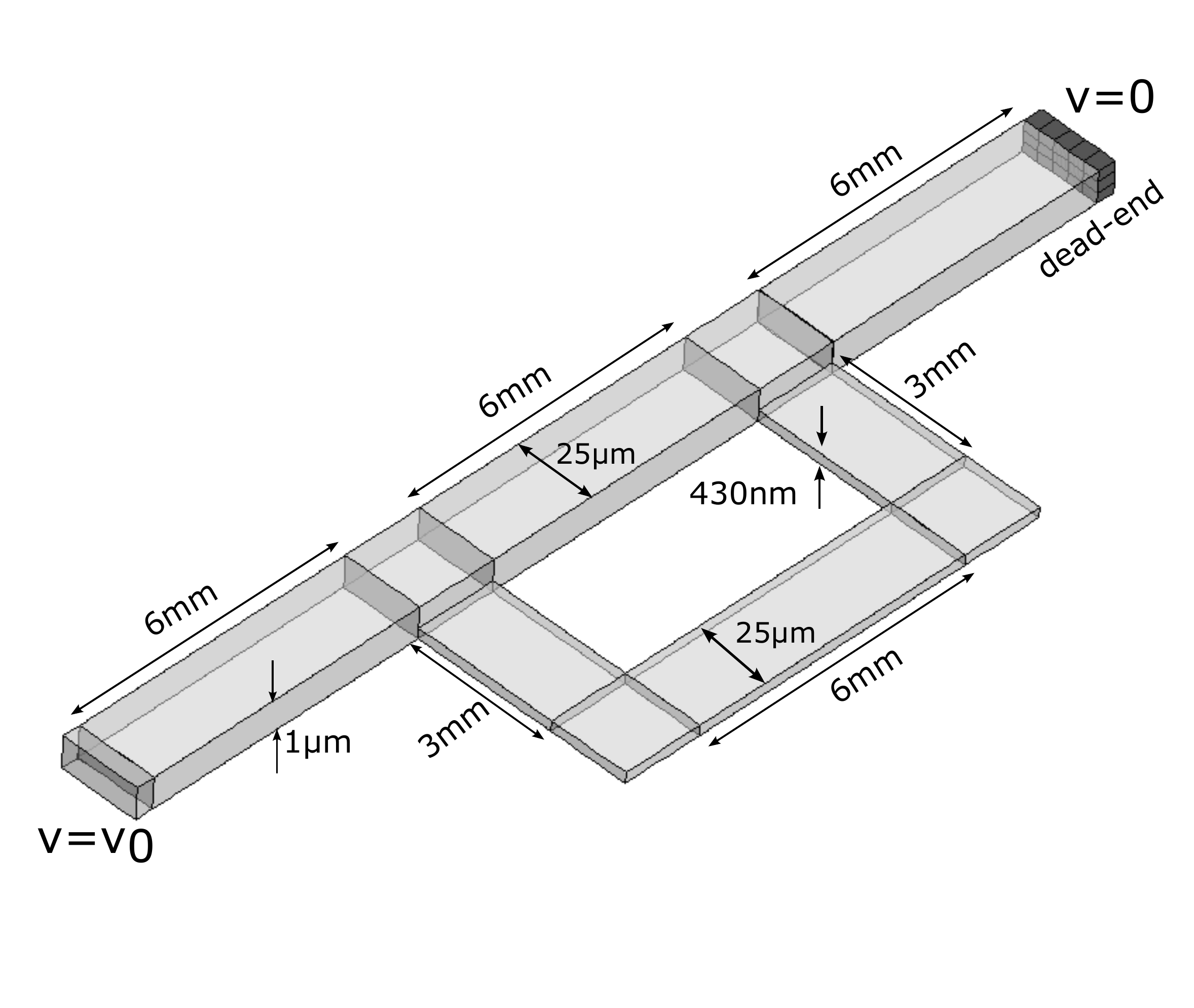}
                \caption{}
                \label{fig:circ_geometry}
        \end{subfigure}%
        ~ 
        \begin{subfigure}[h]{0.45\textwidth}
                \includegraphics[width=\textwidth]{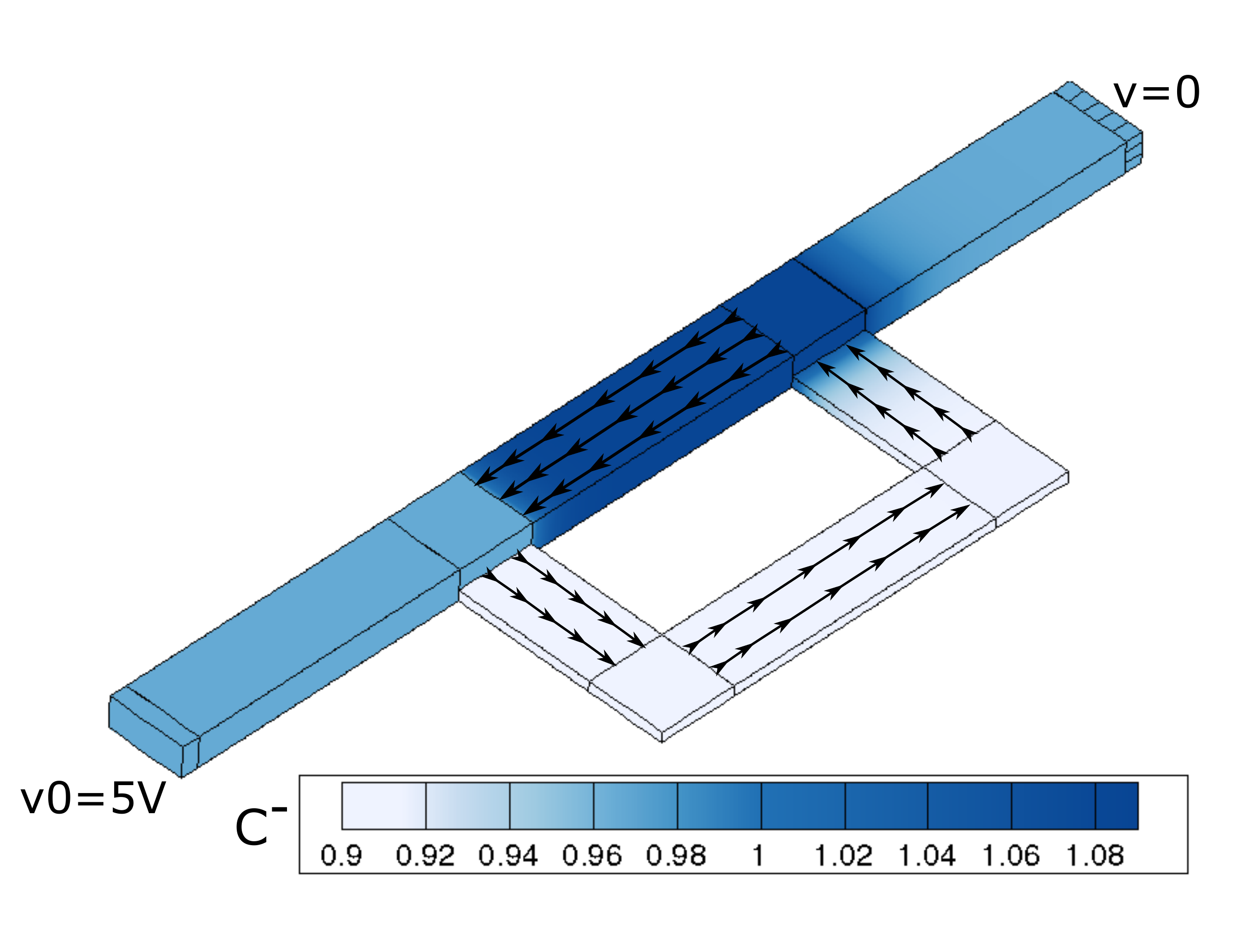}
                \caption{}
                \label{fig:circ_contours}
        \end{subfigure}
        ~ 
         \caption{(a) A network with two pores in parallel forming a loop subject to an external voltage. (b) contours of area-averaged anion concentration and the flow direction at steady state.}
                \label{fig:recirc_result}
 \end{figure}

Next, we consider a network involving a thin pore connecting in parallel to a thicker pore, shown in Figure (\ref{fig:circ_geometry}). We refer to these pores as nano-pore and micro-pore respectively, and consider a setting in which the micro-pore thickness is $1 \mu m$ and the nano-pore thickness is $430 nm$. We consider a uniform symmetric surface charge density of $\sigma=-13 mC/m^2$ throughout the system. At the two ends, the micro-pore is connected to two reservoirs with the same salt concentration of $1 mM$. The right end of the micro-pore is dead-end in order to avoid the net inflow and outflow in the system. An external potential difference of 5 V is applied across the micro-pore. To model this geometry, we partitioned the system into three microchannels and three nanochannels, as indicated in Figure (\ref{fig:circ_geometry}). For two internal reservoirs located at the interfaces of the micro-pores and nano-pores, the volume and the surface charge are equal to the values specified for the micro-pore computational cells. The other internal reservoirs have the same volumes and surface charge as the nano-pore computational elements, and effectively represent two arbitrary mid-elements of the longer nano-pore partitioned into three shorter nano-pores. For the results presented below, we used 200 computational cells for all the pores with the length $6 mm$, and 100 cells for the two nano-pores with the lengths $3 mm$. The time step used was $\Delta t = 5\times10^{-5}$, which allows the accurate computation of the system transient evolution.\\

Figure (\ref{fig:circ_contours}) presents the steady state contours of area-averaged concentration of anions as well as the flow direction in the interior section of the micro-pore and nano-pores under $\text{V}_0=5$ V. As observed, there is no big concentration gradient across the system, which is due to the fact that the pore sizes and electro-chemical properties are not substantially different. However, a significant recirculation net flow is formed in the loop zone. The explanation for this internal recirculation phenomenon is straightforward: the external electric field acts on the positively charged EDLs and induces an electroosmotic flow from left to right. Meanwhile, the dead-end boundary conditions results in a pressure rise in the right reservoir and thus the net flow in the system becomes zero. However, for the parallel internal pores, the micro-pore has a smaller hydraulic resistance compared to the nano-pore. As a result, for the combination of pressure difference and voltage difference imposed on these pores, the pressure effects win in the micro-pore, while electro-osmotic effects win in the nano-pore, and thus a counter-clockwise internal recirculation is induced in the system. All of these combined effects are captured quantitatively with the present reduced order model. The Peclet number associated with the recirculating flow rate in the loop is Pe=$\frac{\tilde{u}L}{D} = \frac{Q}{S_{\text{micro}}+S_{\text{nano}}}$= 6.3, indicating that such internal recirculations can significantly affect ion transport and mixing rate by surpassing molecular diffusion effects. These effects can be present as long as internal loops involve variation in pore sizes; any pore ratio other than 1 can generate such internal flows. A realistic porous medium containing a massive network of many pores offers numerous cites for inducing these recirculation zones. In a recent study, Deng et al. \cite{deng2013} attributed the mismatch between molecular diffusivity and effective diffusivity to such recirculation zones. However, the quantitative impact of this phenomenon on transport through porous media and its dependence on pore morphology is not well understood. The computational model presented here can be used as an effective tool for such investigations in the future.

\subsection{H-Junction}
\begin{figure}[h]
        \centering
        \begin{subfigure}[H]{0.38\textwidth}
                \includegraphics[width=\textwidth]{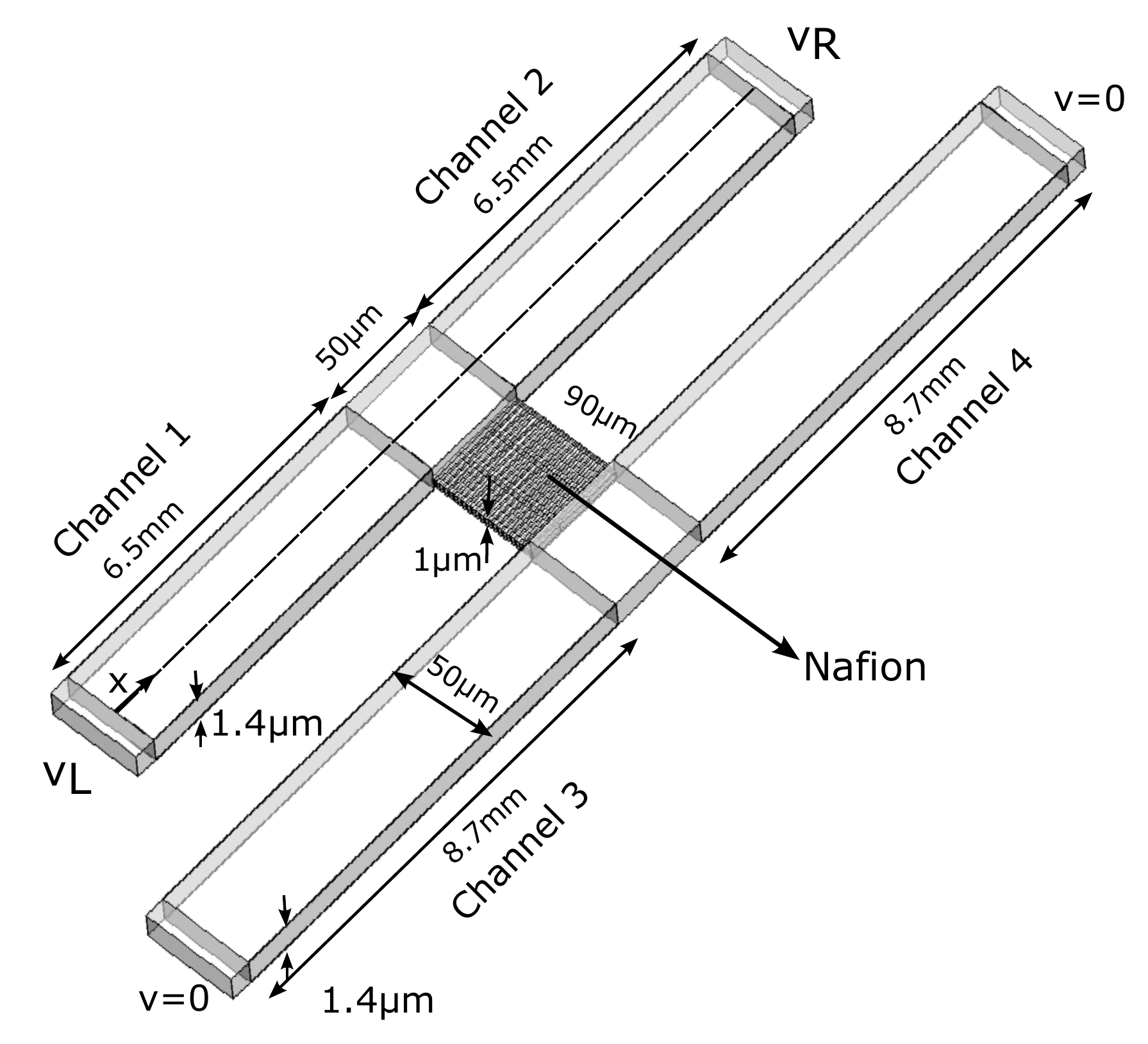}
                \caption{}
                \label{fig:h_geometry}
        \end{subfigure}%
        ~ 
        \begin{subfigure}[H]{0.39\textwidth}
                \includegraphics[width=\textwidth]{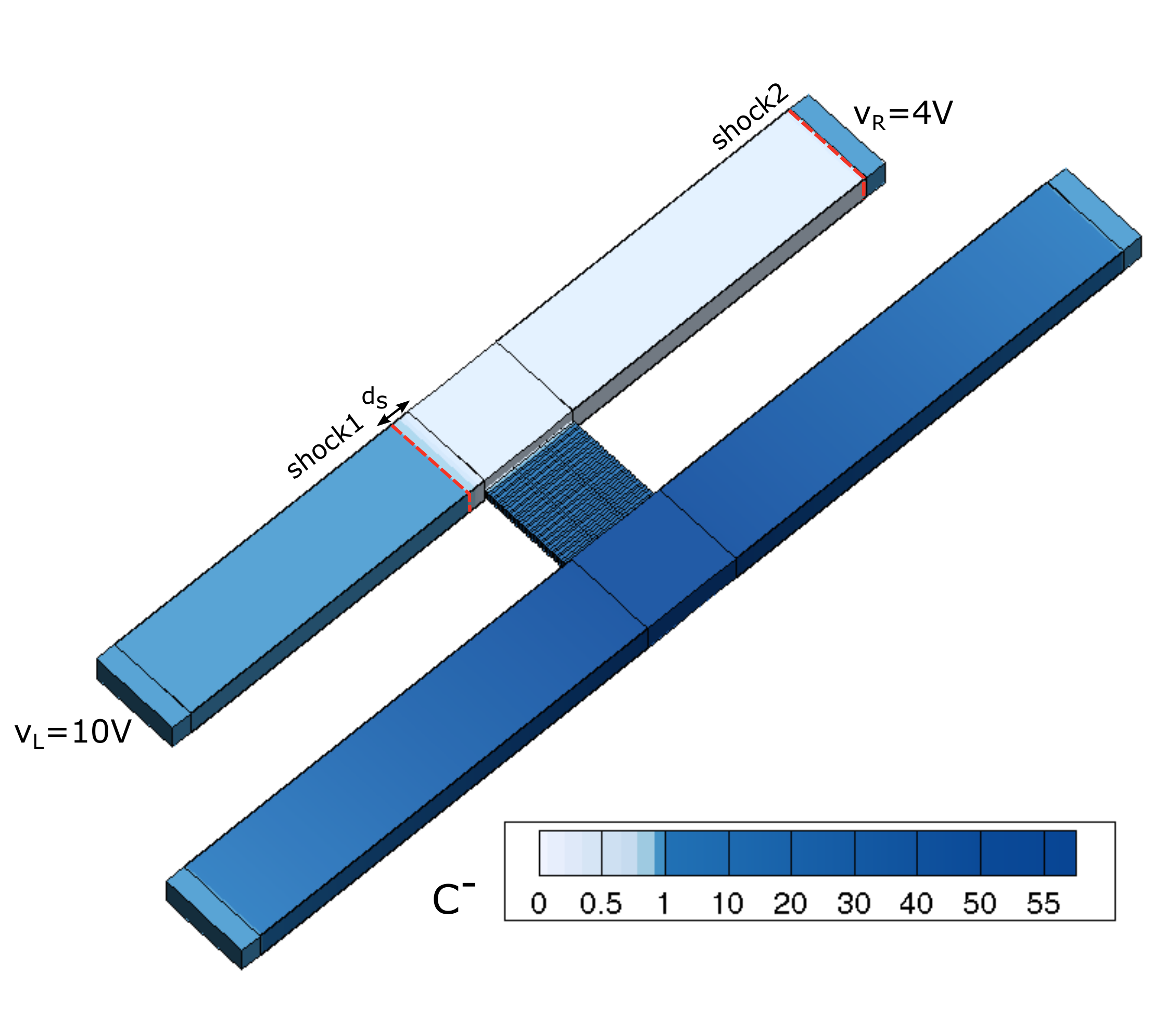}
                \caption{}
                \label{fig:h_contours}
        \end{subfigure}
        ~ 
         \caption{(a) Sketch of an H-junction microfluidic system consisting of two micro channels perpendicularly connected to a nafion with known porosity, nano-pore diameters, and surface charge density. (b) contours of area-averaged concentration of anions at steady state for $V_L=10 V$ and $V_R = 4 V$. The steady state shock locations are marked by red lines in Channel 1 and Channel 2. Shock 1, which propagates in the opposite direction of electro-osmotic flow, stops in a finite distance, $d_s$, from the nafion junction, whereas shock 2, which moves in the flow direction, propagates through the end of Channel 2.}
                \label{fig:hjunction}
 \end{figure}

The last model problem is an H-junction microfluidic network. This type of geometry has been used in numerous lab-on-a-chip experiments \cite{kim2010, kim2012, son2016} as a preconcentration device for detection of biological and chemical samples. As shown in Figure (\ref{fig:h_geometry}), two microchannels with different lengths are connected to a nafion channel, which we model as parallel replicas of a nano-scale pore. Previous experimental results show that a stationary interface (similar to deionization shocks) is formed in one of the microchannels indicated as Channel 1 in Figure (\ref{fig:h_geometry}), while the connecting microchannel indicated as Channel 2 experiences the full propagation of a depletion front. Prior to the present study, no mathematical analysis of this system has been performed in the context of deionization shocks. The question that still remains open is whether the experimentally observed interface is a deionization shock \cite{manimartin2011}, and if so, why this shock does not propagate away from the intersection, as observed in similar analyses \cite{mani2009, manimartin2011}. We show that our reduced order model captures these effects as observed in the experimental reports and confirms that the stationary interface is indeed a deionization shock. Secondly, inspired by our network model, we present an even simpler phenomenological model to explain why the deionization shock in this system remains stationary.\\

\begin{figure}[h]
\makebox[\textwidth][c]{\includegraphics[width=0.5\textwidth]{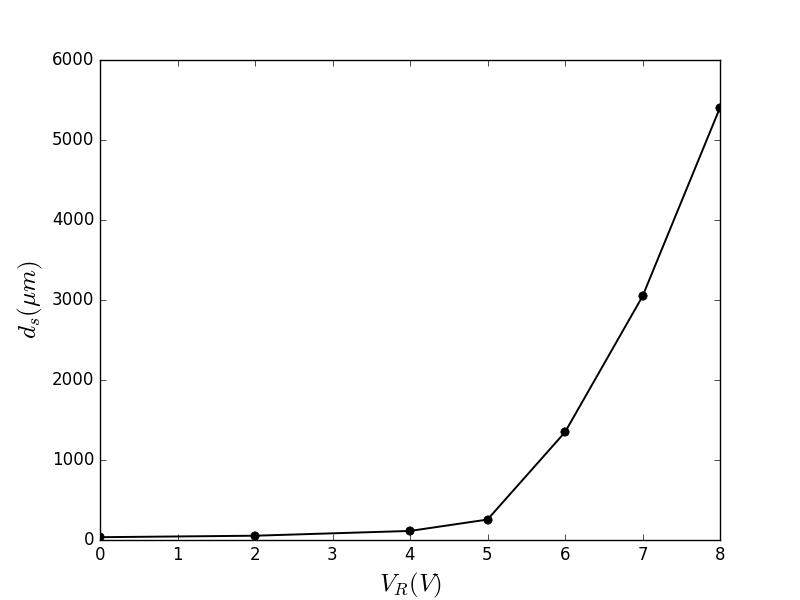}}
\caption{The shock distance from the nafion junction in Channel 1 for $V_L=10V$ and variable $V_R$. The distance increases monotonically as $V_R$ increases. For the case when $V_R = V_L$, shock1 and shock2 symmetrically propagate through the ends of Channel 1 and Channel2, leaving a completely depleted region behind them.}
\label{fig:shock distance}
\end{figure}

We model this system as a network of five pores, with two internal reservoirs and four terminal reservoirs. The two internal reservoirs are on the two sides of a cation-selective nafion element. Each microchannel is modeled as a single pore connecting one of the terminal reservoirs to the internal reservoir as shown in Figure (\ref{fig:h_geometry}). The system specifications are adopted from an experimental setting performed by Professor Kim's group in Seoul National University, Korea \cite{kimlab}. All microchannels have the same surface charge density, thickness, and width, which are $\sigma_m=-13 mC/m^2$,  $h = 1.4 \mu m$, and $w = 50 \mu m$ respectively. As shown in Figure (\ref{fig:h_geometry}), Channel 1 and Channel 2 are both $6.5 mm$ long, while Channel 3 and Channel 4 have lengths equal to $8.7 mm$. We consider a nafion element with volumetric porosity of $40\%$, which is $1 \mu m$ thin, $90 \mu m$ long, and $50 \mu m$ wide. We model this component as an element including circular nano-pores that all have the same diameters, $d_n=10 nm$ and surface charge density, $\sigma_n=-200 mC/m^2$. All terminal reservoirs contain the same salt concentration of $1 mM$ and are open to atmospheric pressure (i.e. $P_0=0$ for all terminal reservoirs). The terminal reservoirs that are connected to Channel 3 and 4 have zero potential, while different potential values are applied in the terminal reservoirs located at the two ends of Channel 1 and 2, whose potentials are indicated by $V_L$ and $V_R$ in Figure (\ref{fig:h_geometry}).  Since all reservoirs are physical components of microchannels, their thicknesses and surface charge densities match the quantities specified for microchannels' computational cells. Furthermore, the computational volumes of two internal reservoirs are set to their largest adjacent grid cell volumes, which are inside Channel 1 and Channel 3 respectively. We considered 3200 computational cells for each microchannel, and 30 elements for the nafion element. $\Delta t = 10^{-6}$ was employed to compute the results presented below.
 Figure (\ref{fig:h_contours}) is the steady state solution obtained after a long term time-advancement and represents the contours of area-averaged negative ion concentration. As in experimental results \cite{holtzel2007, kim2010, kim2012}, we predict a stationary interface near the nafion junction in Channel 1. Initially, two deionization shocks are formed near the junction and then they propagate away from the intersection towards the terminal reservoirs. This effect is very similar to the shock propagation phenomenon discussed in Section \ref{subsec:shock_propagation} for a simpler microchannel-membrane junction. However, here the shock in Channel 1 stops at a stable distance after a finite propagation length indicated as $d_s$ in Figure (\ref{fig:h_contours}). Moreover, in this figure, one can observe that there is an enrichment region occurring at the bottom interface of the nafion. Despite the non-symmetric propagation of depletion fronts, the enrichment fronts move in the bottom microchannels symmetrically, which is expected because the same boundary conditions are applied at the two ends of Channel 3  and Channel 4. Figure (\ref{fig:shock distance}) shows that, consistent with related experiments, there is a monotonic rise of $d_s$ by increasing $V_R$. While our model consistently predicts these trends, we next present a phenomenological model that better explains the observed behavior.

\begin{figure}[H]
\makebox[\textwidth][c]{\includegraphics[width=0.65\textwidth]{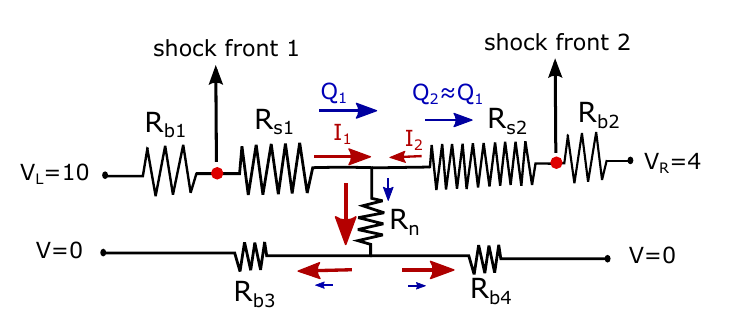}}
\caption{The circuit model of the H-junction system undergoing the propagation of deionization shocks in the top section. $R_{s1}$ and $R_{s2}$ are the amplified resistors by the expansion of the depleted region. In the bottom microchannel, the reduction of bulk resistance is observed as the accumulation of the ions occur. The sizes of the arrows indicating the relative magnitude of fluid flow rate (\textcolor{blue}{Q}) and the current (\textcolor{red} {I}) in different parts of the system; larger and thicker arrows refer to higher values of fluid and charge flow rates.}
\label{fig:ciruit}
\end{figure}

Next, we use ideas from circuit modeling to explain the dynamics associated with the propagation of deionization shocks in the H-junction considered. Figure (\ref{fig:ciruit}) presents a schematic of a circuit model at steady state. $R_{bi}$ refers to the electric resistance of channel $i$, which includes the combined effects of the bulk and surface charge. The depletion zone behind deionization shocks could be regarded as a large resistor and to distinguish its resistance from the general pore resistance, we have considered $R_{si}$. In this phenomenological model, we consider the nafion element to have a small electric resistance, given its significant surface to volume ratio with highly charged EDL. However, the hydraulic resistance of the nafion is assumed to be so large that almost no net flow can pass through this element.\\

Initially, no shock is propagated and Channel 1 and Channel 2 have low electric resistance indicated by $R_{b1}$, and $R_{b2}$, respectively. This leads to high electric currents in these two channels, $I_1$, and $I_2$. Mani and Bazant \cite{manimartin2011} developed an analysis of deionization shocks, and predicted that in the absence of net flow, deionization shock speed is proportional to electric current: 
 \begin{equation}
 \text{V}_{\text{shock}} \sim I .
\label{eq:shock speed no flow}
 \end{equation}
 For channels with negative surface charge, the shock propagates in the opposite direction of the current. As observed in Figure (\ref{fig:h_contours}) this leads to the propagation of shock fronts to the left in Channel 1 and to the right in Channel 2. In our system however, given the voltage difference $V_L>V_R$, a net electroosmotic flow is enforced from Channel 1 through Channel 2. A negligible portion of this flow passes through nafion element, but there is a net flow from left to right as indicated in Figure (\ref{fig:ciruit}). Using Hemholtz-Smoluchowski relation, the advective flow scales as $V_\text{adv} \sim \varepsilon\zeta(V_L-V_R)/\mu L$, where $\zeta$ represents the zeta potential associated with the EDLs. Combining this relation with equation (\ref{eq:shock speed no flow}) results in: 
\begin{equation}
 \text{V}_{\text{shock}} = \text{c}I + V_{\text{adv}},
 \label{eq:net shock speed}
 \end{equation}
where $c$ is a constant. On the contrary to Channel 2, which experiences a net flow in the direction of shock propagation, the advective flow in Channel 1 is against the direction of shock movement. In other words, in Channel 1 the two terms in the RHS of (\ref{eq:net shock speed}) compete with each other. Initially, the depletion zone behind deionization shocks are small, and the first term in (\ref{eq:net shock speed}) dominates, leading the shock to propagate to the left. However, as $\text{d}_\text{s}$ grows, the ohmic resistance $R_{s1}$ builds up and  decreases the electric current. Meanwhile, depletion leads to thicker EDLs and hence, larger $V_{\text{adv}}$ in the connected channels. These conditions shift in balance between the two terms, eventually leading to a situation that the second term in (\ref{eq:net shock speed}) becomes equal to the first term, when the deionization shock ceases further propagation. Based on this qualitative analysis, larger difference between $V_R$ and $V_L$ should lead to larger $V_{\text{adv}}$ and thus, a stationary shock closer to the intersection. As shown in Figure (\ref{fig:shock distance}), our model correctly captures these trends. 

\section{Conclusions}
We presented a multi-scale method that provides an efficient technique for accurate modeling of nonlinear electrokinetic phenomena in networks of micro-scale and nano-scale pores. Our model treats each pore as a one-dimensional element for which the coupled transient equations governing the interactions of fluid flow, charge transport, and ion transport are derived from an area-averaging procedure. Assuming equilibrium in cross-sectional directions, this model takes into account the EDL effect, which results in non-uniformity in cross-sectional profiles of flow and concentration. The effects of these non-uniformities on area-averaged fluxes are accounted for in terms of tabulated transport coefficients that can robustly handle all regimes of surface charge densities and EDL thicknesses. While similar models have been attempted in the past, the present model has unique advantages, which make it suitable for numerical simulations of porous networks: 
\begin{enumerate}
\item It covers the entire range of EDL to pore size ratios and does not impose any limitation on the thickness of EDL. This has been enabled by the tabulation of transport coefficients, which are otherwise not available analytically. Our table requires only two input parameters, $\sigma^*$ and $\lambda^*$, and the transport coefficients are smooth functions of these parameters. Therefore, the developed tables can be stored with a very low memory overhead.
\item The tabulated coefficients are bounded and remain non-singular for the entire range of $\sigma^*$ and $\lambda^*$. This has been achieved by proper choice of input parameters to the table and by properly normalizing each coefficient (for example the use of $C_s$ as opposed to $\overline{C^-}$ turns a $0\times \infty$ singularity to the product of finite numbers in the limit of low concentration).
\item  An asymptotic treatment is developed that automatically treats regions of zero concentration (i.e, behind strong deionization shocks) where the input $\lambda^*$ is outside of the table bounds.  
\item The discretization scheme of the model is fully conservative without any numerical leakage of ions. This property has been achieved by deriving the evolution equation for $\overline{C^-}$, which is a physically conserved quantity (as opposed to $C_0$, whose evolution does not map to the conservative quantity, $\overline{C^-}$, when computed discretely.) 
\item The model formulation is discretely well balanced and in the absence of any driving forces, equilibrium solution is fully recovered regardless of spatial mesh spacing. We enabled this property by utilizing virtual potentials as driving forces for fluxes. The virtual potentials are invariants of equilibrium solutions. 
\item The model is capable of simulating general network of pores that may contain discrete jumps in pore size, or multi-pore intersections. This property is gained by taking advantage of virtual driving potentials, which remain continuous across the system as well as proper coupling of conservation laws at pore intersections. 
\end{enumerate}

We evaluated our model performance for several model problems and discussed the key outcomes for each case. Our model has revealed to capture the osmotic pressure phenomenon that is induced between two reservoirs with different salt concentration when connected with thin pores. In addition, our model quantified non-ideal behavior when the pore size is finite. Additionally, we demonstrated that our model captures a wide range of phenomena associated with charged pores connecting with other pores or electrokinetic elements. These predictions were verified against multi-dimensional DNS, previously derived scaling laws, and/or validated against previously reported experiments. The studied examples include: nonlinear I-V curve for a micropore/membrane junction, deionization shock propagation in a micropore, streaming potential in a microchannel-nanochannel-microchannel junction, flow recirculation in a network with a loop connecting pores with different cross-sections, and stationary shock in a microfluidic H-junction.\\

The developed model provides an efficient framework for engineering analysis of various electrokinetic porous systems with applications in energy conversion, desalination, and lab-on-a-chip systems. In this paper, we demonstrated the results obtained for simple and small network topologies, even though this model is capable of modeling big networks of pores with arbitrary pore sizes and pore connections. A number of generalizations can be envisioned for the developed model. Given the modular nature of this model, many of these extensions are straightforward to implement. A wide range of capabilities can be achieved via only changing the tabulated coefficients. These include extensions to electrolyte systems that are non-symmetric, or EDL regimes that involve non-continuum effects  \cite{dirk2015, lee2015}. Another interesting extension would be to generalize the model to account for time-dependent surface charge, relevant to systems involving conducting pores such as supercapacitors \cite{mirzadeh2014}, or systems in which surface charge can be regulated by bulk chemistry \cite{mathias2012, mathias2014}. In these cases, the conservation laws must incorporate time dependent surface charge, and the tabulation strategy (input) may need to be changed depending on the constraints on the surface. Lastly, extensions to non-binary electrolytes allows tracing of multiple species within these systems.\\


\appendix
\section{Iterative Poisson solver} \label{app:poisson solver}
For clarity, we explain our numerical strategy for solving Poisson equation (equation (\ref{eq:psi_eqn}) ) in the cartesian coordinate for a 2D pore. The z-direction is assumed to be in the wall normal direction. The equation and surface charge boundary condition are as follows in this coordinate:
\begin{equation}
\frac{d^2 \psi}{d z^2} = \frac{1}{\lambda_0^2} \text{sinh}(\psi),
\label{eq:psiy}
\end{equation}
\begin{equation}
\frac{d \psi}{d z}|_{\text{wall}} = \sigma^*.
\label{eq:bc}
\end{equation}
Assume the parameters $\sigma^*$ and $\lambda_0$ are known. Before discretization, we substitute the following predictor-corrector format, $\psi = \psi^* + \delta \psi$ into equation (\ref{eq:psiy}) and apply the linearization technique to RHS:
\begin{equation}
\frac{d^2 \delta \psi}{d z^2}  + \frac{d^2 \psi^*}{d z^2} =  \frac{1}{\lambda_0^2}(\text{sinh}(\psi^*) + \delta \psi \text{cosh}(\psi^*)).
\end{equation}
An initial distribution of predictor $\psi^*$ is also set. We use uniformly spaced computational cells with thickness of $\Delta z$ and discretize Laplace operator with second order accuracy to obtain the final discretized equation:
\begin{equation}
\delta \psi_{i-1} + (-2-\frac{\Delta z^2}{\lambda_0^2}\text{cosh}(\psi^*_i))\delta \psi_i + \delta \psi_{i+1} = \frac{\Delta z^2}{\lambda_0^2}\sinh(\psi^*_i) - (\psi^*_{i-1} - 2\psi^*_i + \psi^*_{i+1}).
\label{eq:discrete_poisson}
\end{equation}
The discretized boundary condition is used to set the values of $\psi^*$ in the boundaries. Equation (\ref{eq:discrete_poisson}) results in a tri-diagonal system of equations, which are solved using Thomas algorithm. The linearization causes the left hand side matrix to be diagonally dominant, which ensures the convergance of the iterative procedure and also enhance the convergence rate. Each time after we obtain the corrector $\delta \psi$, we update $\psi^*$ by $\psi^* = \psi^* + \delta \psi$ and solve the system of equations derived from (\ref{eq:discrete_poisson})  using the new values of $\psi^*$. The iteration is terminated when the L2-norm of $\delta \psi$ is smaller than a threshold, which was set to be $1e-8$ in our computations. The calculation of area averaged coefficients is straightforward once the converged solution is obtained.\\

To develop the table based on $\sigma^*$ and $\lambda^*$, for each set of input parameters, we initialize $\bar{g}$ with 1 which results in the initial $\lambda_0=\lambda^*$. At the end of each iteration when the new $\psi^*$ is obtained and $\bar{g}$ is calculated, the value of $\lambda_0$ is re-set using equation (\ref{eq:lam_0_s}) and the system of equations are solved for new values of $\lambda_0$ and $\psi^*$. This iterative procedure eventually leads to the correct value of $\lambda_0$ in addition to improving the numerical stability. 

\section{Asymptotic treatment for regions of small concentration}\label{app:asymptotic}
In depletion regions behind strong deionization shocks, concentration of coions can physically become very small (i.e. $\overline{C^-} \sim10^{-10}$). In these zones, even though the second order discretization error is very small, it can lead to locally negative numerical concentrations. As long as the spatial coordinate is well resolved the magnitude of the unphysically negative concentration can be maintained to arbitrarily small values. Given that the EDLs contribute finite amount of counterions in depleted zones ($\overline{C^+} \sim C_s$), one can ensure that the error in the prediction of $\overline{C^-}$ does not affect the accuracy and positivity of conductivity fields (equal to $2\overline{C^-}+C_s$, in equation (\ref{eq:current})). Additionally, such numerical errors still lead to acceptably close to zero prediction of ion fluxes. The only problem with regions of negative concentration is that they interfere with the table reading step since the input to the table, $\lambda^*$, requires the computation of $\sqrt{\overline{C^-}}$. To resolve this issue, we formulated an asymptotic treatment that is utilized in the limit of very small positive values of concentration, or when the local concentartion turns into small negative value numerically. This treatment is constructed based on the asymptotic scaling of area-averaged coefficients with local $\overline{C^-} $, and relaxes the need of positive $\overline{C^-} $. According to Figure (\ref{fig:table_coeff}) the following scalings are obtained for each area-averaged coefficient in the limit of small $\overline{C^-} $ or large $\lambda^*$:
\begin{center}
$\bar{g} \sim \overline{C^-}^{1/2}$   ,   $\overline{g^e} \sim  \overline{C^-}^0$, and $\overline{g^c}  \sim  \overline{C^-}^{1/2}$ \\

$\overline{g^{p-}} \sim \overline{C^-}$, $\overline{g^{e-}} \sim \overline{C^-}$, $\overline{g^{c-}} \sim \overline{C^-}^{3/2}$\\

$\overline{g^{p+}} \sim \overline{C^-}^0$, $\overline{g^{e+}} \sim \overline{C^-}^0$, $\overline{g^{c+}} \sim \overline{C^-}^{1/2}$.
\end{center}

Using the mathematical relation between $C_0$ and $ \overline{C^-}$, one can see that $C_0 \sim  \overline{C^-}^{1/2}$. We rewrite the diffusion flux term in equation (\ref{eq:area_neg_ion}) as follows:
\begin{equation}
-2\frac{\overline{C^-}}{C_0}\frac{d C_0}{dx} = -\frac{\overline{C^-}}{C_0^2}\frac{d C_0^2}{dx} = \frac{-1}{\frac{\overline{C^-}}{\bar{g}^2}}\frac{d}{dx}(\frac{\overline{C^-}}{\bar{g}^2} \overline{C^-}) = \frac{-1}{\bar{f}}\frac{d}{dx}(\bar{f} \overline{C^-}).
\label{eq:diffusion treatment}
\end{equation}
In the new formulation of diffusion term, $\bar{f} = \frac{\overline{C^-}}{\bar{g}^2} $ is easily computed using tabulated $\bar{g}$. When $\overline{C^-}$ becomes small, $\bar{f}=\frac{\overline{C^-}}{\bar{g}^2} \sim  \overline{C^-}^0$, and the ratio becomes independent of concentration. For any given $\sigma^*$ we extrapolate the value of $\bar{f}$ for small concentration, based on the largest available $\lambda^*$ (smallest available concentration,  $\overline{C^-}_{\text{min}}$) in the table. This procedure still involves use of finite coefficients while bypassing the need for the computation of $\sqrt{\overline{C^-}}$.\\

We apply the same strategy on the terms containing $C_0$ in the conservation equations (\ref{eq:cont}) and (\ref{eq:current}). Thus, the RHS terms in these equations are written as follows:
\begin{equation}
 C_1\frac{d C_0}{d x} = - S \frac{\kappa}{2 \lambda_D^2} \overline{g^c} \frac{d C_0}{d x}= -S \frac{\kappa}{2 \lambda_D^2} \frac{g^c}{2C_0} \frac{d}{dx}(\bar{f} \overline{C^-})
\end{equation}

\begin{equation}
C_2 \frac{d C_0}{d x} = -S \{C_s\frac{\kappa}{2\lambda_D^2}(\overline{g^c} + \overline{g^{p-}} - \overline{g^{c-}}) + 2\bar{g} \} \frac{d C_0}{d x}=
\label{eq:current treatment}
\end{equation}
\begin{equation*}
-S \{ C_s \frac{\kappa}{2\lambda_D^2}(\frac{\overline{g^c}}{2C_0} + \frac{\overline{g^{c+}}}{2C_0} - \frac{\overline{g^{c-}}}{2C_0}) +\frac{1}{\bar{f}} \} \frac{d }{d x}(\bar{f} \overline{C^-}).
\end{equation*}
Similar to $\bar{f}$, $\frac{\overline{g^c}}{2C_0}$ and $\frac{\overline{g^{c+}}}{2C_0}$ scale with $\overline{C^-}^0$ in the limit of small concentration and thus they are computable using the quantities stored for the largest $\lambda^*$. However, $\frac{\overline{g^{c-}}}{2C_0} \sim \overline{C^-}$, which requires linear extrapolation to compute the value associated with the parameters beyond the table limits. Mathematically, this can be achieved by scaling the value obtained based on the largest $\lambda^*$ by factor of $\frac{ \overline{C^-}}{ \overline{C^-}_{\text{min}}}$. Although this treatment allows the negative concentration to emerge, it does not interfere with the physical prediction of the model as long as spatial resolution is sufficiently fine to resolve physical features. Namely, the finite ohmic resistance, electroosmotic flow, and advective current are computed accurately.\\

In LHS of equations (\ref{eq:cont}) and (\ref{eq:current}) the area-averaged coefficients which scale with $ \overline{C^-}^0$ are $\overline{g^e}$, $\overline{g^{p+}}$, and $\overline{g^{e+}}$. Thus, similar to $\bar{f}$, these coefficients are set to the tabulated quantities associated with the largest $\lambda^*$. However, $\overline{g^{p-}}$ and $\overline{g^{e-}}$ scale linearly with $\overline{C^-}$, and thus, the linear extrapolation is used to determine their values in the limit of zero concentration, when $\lambda^*$ goes beyond the table highest bound.\\

The treatment described by equations (\ref{eq:diffusion treatment}) to (\ref{eq:current treatment})  is exact for all ranges of concentration. Therefore, our spatial discretization uses the form presented above for the entire range of $C_0$. However, only when table inputs are out of bound, the extrapolations described above are used. We use the new formulation in Appendix 3 to explain our time-integration procedure.

\section{Semi-implicit solver}\label{app:implicit solver}
In this section, we describe the details of our second order semi-implicit time advancement scheme used to solve transport equation (\ref{eq:area_neg_ion}). The transport of area-averaged concentration for negative ions is driven by the following flux components:
\begin{equation}
\overline{F^-_x} =  \bar{u}\overline{C^-} +  \overline{u^{\prime}C^{-\prime}} - \frac{1}{\bar{f}}\frac{d}{dx}(\bar{f} \overline{C^-}) +  \overline{C^-}\frac{d\mu^+}{dx}, \label{eq:negflux}
\end{equation}
which consists of the advection, diffusion, and electromigration terms respectively. Since we solve for only negative ion concentration, for simplicity we drop the superscript $-$ and use $\bar{C}$ in our description of numerical strategy afterwards. Thus, the implict time discretization described by equation (\ref{eq:discrete_transport}) is simplified to:
\begin{equation}
\frac{3\bar{C}^{(n+1)} - 4\bar{C}^{(n)} + \bar{C}^{(n-1)}}{2\Delta t} = \frac{-1}{S} \frac{\partial}{\partial x}\{S\bar{F}_x^{(n+1)}\} + O(\Delta t^2). \label{eq:cbar_time_discrete}
\end{equation}

The implicit time advancement enables us to robustly treat the stiff diffusion terms, which otherwise would limit us to small time steps in regions where a quasi-steady phenomenon could be accurately captured by large time steps. To avoid costly calculations associated with the coupled nonlinear system described by (\ref{eq:cbar_time_discrete}), we devised an iterative procedure to solve this equation in a fashion similar to that applied by Karatay et al. \cite{karatay2015}. In an iteration loop we treat $\bar{C}$ portion of the flux terms implicitly and their remaining components are updated based on the latest available results. Once iterations converge, the full implicit solution to (\ref{eq:cbar_time_discrete}) is achieved. The terms treated implicitly include $\bar{u}\bar{C}$, $ \frac{1}{\bar{f}}\frac{d}{dx}(\bar{f} \bar{C})$, and $\bar{C}\frac{d\mu^+}{dx}$. The semi-implicit nature of our time-advancement procedure comes from the fact that the driving potentials,  $P_0$ and $\mu^+$ are to be computed at moment $(n)$ as all area-averaged pre-factors in conservation equations (\ref{eq:cont}) and (\ref{eq:current}) are obtained based on $\bar{C}^{(n)}$ quantities.\\

We introduce the delta notation of concentration as $\delta \bar{C} = \bar{C}^{(n+1)} - \bar{C}^{(n)}$. Using this notation, we apply linearization on the flux at moment $(n+1)$ and write it in the following form:
\begin{equation}
\overline{F_x}^{(n+1)} \simeq \overline{F_x}^{(n)} + \delta \overline{F_x},
\label{eq:f_next}
\end{equation}
where $\overline{F_x}^{(n)}$ refers to the flux quantities at time $t^{(n)}$, and $\delta \overline{F_x}$ contains the terms treated implicitly:
\begin{equation}
\delta \overline{F_x} (\delta \bar{C}) =  \bar{u}^{(n)}\delta \bar{C} + \frac{1}{\bar{f}}\frac{d}{dx}(\bar{f} \delta \bar{C}) + \delta \bar{C} \frac{d \mu^{+(n)}}{dx}. \label{eq:delta_f}
\end{equation}
Using the introduced delta notation and equations (\ref{eq:f_next}) and (\ref{eq:delta_f}), we rewrite equation (\ref{eq:cbar_time_discrete}) as follows: 
\begin{equation}
\frac{3\delta \bar{C}}{2\Delta t}  + \frac{1}{S} \frac{\partial}{\partial x}\{ S \delta \overline{F_x}(\delta \bar{C})\} \simeq  \frac{\bar{C}^{(n)} - \bar{C}^{(n-1)}}{2\Delta t} + \frac{-1}{S} \frac{\partial}{\partial x}\{ S\overline{F_x}^{(n)}\}. \label{eq:del_c_del_flux}
\end{equation}
Equation (\ref{eq:del_c_del_flux}) can be solved by inverting the left-hand side operator and computing $\delta \bar{C}$. At this form, we achieve first-order temporal accuracy, since only a portion of the flux terms is computed at time (n+1). However, it is possible to increase the order of accuracy by employing an iterative procedure and using a series of intermediate solutions $\bar{C}^*$, which eventually matches $\bar{C}^{(n+1)}$. We define the intermediate correction as $\delta \bar{C}^* = \bar{C}^{*+1} - \bar{C}^{*}$ and replace $\bar{C}^{(n+1)}$ in equation (\ref{eq:cbar_time_discrete}) with $\bar{C}^{*+1} = \delta \bar{C}^* + \bar{C}^{*}$. The final equation used for iteration is as follows:
\begin{equation}
\frac{3\delta \bar{C}^*}{2\Delta t} + \frac{1}{S}\frac{\partial}{\partial x} \{S\delta \overline{F_x}^*(\delta \bar{C^*}) \} = \frac{-3\bar{C}^* + 4 \bar{C}^{(n)} - \bar{C}^{(n-1)}}{2\Delta t} - \frac{1}{S} \frac{\partial}{\partial x}\{ S\overline{F_x}^*(\bar{C^*}) \}, \label{eq:itr_transport}
\end{equation} 
where $\overline{F_x}^*$ and $\delta \overline{F_x}^*$ are evaluated based on the most recent concentration computed, $\bar{C}^*$. With only one iteration per time step when $\bar{C}^* = \bar{C}^{(n)}$ initially, equation (\ref{eq:itr_transport}) is exactly equal to (\ref{eq:del_c_del_flux}) and the method reduces to first accuracy. However, with two or higher iterations per time step  the scheme becomes second order in time and is shown to be highly robust \cite{karatay2015}.\\

Equation (\ref{eq:itr_transport}) is solved inside each pore, and thus appropriate boundary conditions are required for $\delta \bar{C}^*$. To determine the correct boundary conditions, we employ the same trick we used to solve the conservation equations (\ref{eq:cont}) and (\ref{eq:current}) to obtain the virtual total pressure and electro-chemical potential fields. For each pore in the network, we solve equation (\ref{eq:itr_transport}) three times for different boundary conditions and source terms summarized below:
\begin{table}[H]
  \caption{three B.C. conditions and source terms in RHS used to solve equation (\ref{eq:itr_transport}).}
  \label{tbl:deltaC_solns}
  \centering
  \begin{tabular}{llll}
    \hline
    Case  & Left B.C.  & Right B.C. & RHS\\
    \hline
    1        & $\delta \bar{C}^*_{L} = 0$   & $\delta \bar{C}^*_{R}=  0$   & $\text{RHS} =\frac{-3\bar{C}^* + 4 \bar{C}^{(n)} - \bar{C}^{(n-1)}}{2\Delta t} - \frac{1}{S} \frac{\partial}{\partial x}\{ S\overline{F_x}^* \}$\\
              &                                           &                                            &                                                                                     \\
    2        & $\delta \bar{C}^*_{L}= 1$   & $\delta \bar{C}^*_{R} = 0$   &  $\text{RHS} = 0$\\
              &                                           &                                                    &                                                          \\
    3        & $\delta \bar{C}^*_{L} = 0$   & $\delta \bar{C}^*_{R} = 1 $  &  $ \text{RHS} =0$\\
    \hline
  \end{tabular}
\end{table}
We combine three solutions that are obtained for these cases to find the concentration correction ($\delta \bar{C}^*$) throughout the network:
\begin{equation}
\delta \bar{C}^* = \delta \bar{C}^{*(1)} + \delta \bar{C}^*_{L} \delta \bar{C}^{*(2)} +  \delta \bar{C}^*_{R} \delta \bar{C}^{*(3)},
\label{eq:delta_c}
\end{equation}
where $ \delta \bar{C}^{*(i)}$ are the solutions obtained for case 1, 2, and 3. The boundary values, $\delta \bar{C}^*_{L}$ and $ \delta \bar{C}^*_{R}$, are obtained by enforcing equation (\ref{eq:itr_transport}) for internal reservoir. For terminal reservoirs whose concentration values are known, we impose $\delta \bar{C}^*=0$. After solving for concentration correction in the reservoirs, we compute this variable throughout the network using equation (\ref{eq:delta_c}) and the new concentration field is obtained by $\bar{C}^{*+1} = \delta \bar{C}^* + \bar{C}^{*}$. This procedure is applied iteratively until $\delta \bar{C}^*\simeq 0$, which implies that $\bar{C}^{(n+1)} \simeq \bar{C}^{*}$. We run our simulation with three iterations to ensure that the method is second order accurate and the solution is sufficiently robust. 

\end{document}